\documentclass[preprint2]{aastex}
\usepackage{epsfig}
\usepackage[fleqn]{amsmath}
\usepackage{mathptmx}
\usepackage{graphicx}
\usepackage{bm}
\shorttitle{Construction of second-order} \shortauthors{Hu et al.}

\begin{document}


\title{Construction of second-order six-dimensional Hamiltonian-conserving scheme}


\author{Shiyang Hu$^{1}$, Xin Wu$^{1,2,3, \dag}$, Enwei Liang$^{1,3}$}
\affil{1. School of Physical Science and Technology, Guangxi
University,
Nanning 530004, China \\
2. School of Mathematics, Physics and Statistics $\&$
Center of Application and Research of Computational Physics, Shanghai
University of Engineering Science, Shanghai 201620, China \\
3. Guangxi Key Laboratory for Relativistic Astrophysics, Guangxi
University, Nanning 530004, China} \email{$\dag$ Corresponding
Author wuxin$\_$1134@sina.com (X. W.); %}
2312639147@qq.com (S. H.), lew@gxu.edu.cn (E. L.)}


\begin{abstract}

It is shown analytically that the energy-conserving implicit
nonsymplectic scheme of  Bacchini, Ripperda, Chen and Sironi
provides a first-order accuracy to numerical solutions of a
six-dimensional conservative Hamiltonian system. Because of this,
a new second-order energy-conserving implicit scheme is proposed.
Numerical simulations of Galactic model hosting a BL Lacertae
object and magnetized rotating black hole background support these
analytical results. The new method with appropriate time steps is
used to explore the effects of varying the parameters on the
presence of chaos in the two physical models. Chaos easily occurs
in the Galactic model as the mass of the nucleus, the internal
perturbation parameter, and the anisotropy of the potential of the
elliptical galaxy increase. The dynamics of charged particles
around the magnetized Kerr spacetime is easily chaotic for larger
energies of the particles, smaller initial angular momenta of  the
particles, and stronger magnetic fields. The chaotic properties
are not necessarily weakened when the black hole spin increases.
The new method can be used for any six-dimensional Hamiltonian
problems, including globally hyperbolic spacetimes with readily
available (3+1) split coordinates.

\end{abstract}


\emph{Unified Astronomy Thesaurus concepts}:  Black hole physics
(159); Computational methods (1965); Computational astronomy
(293); Celestial mechanics (211); Galaxy dynamics (591)



\section{Introduction}

Although the Schwarzschild black hole and the Kerr-Newman black
hole are integrable, their analytical solutions are too difficult
to be explicitly expressed in terms of elementary functions. These
spacetimes become nonintegrable in general and then have no
analytical solutions when external  magnetic fields are included.
Numerical integration schemes are good tools to treat these
problems. Low-order explicit Runge-Kutta
integrators without adaptive step-size control, such as a
conventional fourth-order  Runge-Kutta scheme, are applicable for
not only such light-like and time-like geodesics or nongeodesics
in the general theory of relativity (Bronzwaer et al. 2018, 2020;
Wang et al. 2021a, 2021b, 2021c; Wu et al. 2021), but also other
Hamiltonian and nonHamiltonian problems, e.g., the solar system
dynamics, extrasolar planets, and galaxy models (Carlberg $\&$
Innanen 1987; Caranicolas 1984, 1993; Zotos 2011). They should
yield accurate and reliable numerical solutions in short
integration times. However, they would have unphysical energy
drifts over long integration times and then provide unreliable
results. On the contrary, high-order explicit
Runge-Kutta-Fehlberg (RKF) methods with adaptive step-size control
can yield higher  precision numerical solutions, and are very
useful and important in the long-term dynamics of many
astrophysical problems, for example, eccentric multi-body orbits
and N-body problems. Such high-order  methods require more
expensive computations compared to the low-order Runge-Kutta
integrators. 

Conservation of energy along a trajectory is important in
long-term numerical simulations. It is an intrinsic property of
conservative Hamiltonian dynamics. Checking the energy accuracy is
often used to test the performance of a numerical integrator
because high energy accuracy would bring high-precision solutions
in many situations although high energy accuracy does not always
lead to high-precision solutions for any cases. The growth speed
of errors in the solutions is governed by the relative errors in
the individual Keplerian energies in two-body problems, perturbed
two-body problems and $N$-body problems, therefore, energy
conservation or suppressing the growth of individual Keplerian
energy errors is helpful to weaken the Lyapunov's instability
influence on the accuracy of numerical integration, and results in
more stable orbital motions (Avdyushev 2003). Because the
frequencies of periodic motions depend on the energies in general,
suppressing the accumulation of energy errors plays an important
role in the capability of accurately grasping periodic motions
over long time runs. The lower-order explicit
Runge-Kutta methods combined with manifold corrections conserve
one or more first integrals (or slowly-varying quantities) like
energy (Nacozy 1971; Fukushima 2003; Wu et al. 2007; Ma et al.
2008; Wang et al. 2016, 2018; Deng et al. 2020), and are suitable
for simulating the long-term dynamics. In this way, some
non-geometric numerical integrators such as the Runge-Kutta family
methods can be reformed as a class of geometric integration
methods (Hairer et al. 2006).

Symplectic integrators are also a class of
geometric integration methods. They have a major advantage over
the low-order Runge-Kutta methods with manifold corrections and
high-order RKF methods in long-term integrations. This is because
they have very nice long-time properties, like bounding of energy
error, maintenance of phase space volume, conservation of first
integrals, conservation of symplectic structure, in some instances
time-symmetry/reversibility etc. Symplectic integrations of
non-separable Hamiltonian systems are generally implicit,
requiring more expensive numerical iterations compared to explicit
methods. Some examples are the second-order implicit symplectic
midpoint rule (Brown 2006), implicit schemes with adaptive step
size control (Seyrich $\&$ Lukes-Gerakopoulos 2012), and explicit
and implicit combined symplectic schemes (Preto $\&$ Saha 2009;
Kop\'{a}\v{c}ek et al. 2010; Lubich et al. 2010; Zhong et al.
2010; Mei et al. 2013a, 2013b). There are extended phase-space
explicit symplectic-like or symplectic methods (Liu et al. 2016;
Luo et al. 2017; Li $\&$ Wu 2017; Christian $\&$ Chan 2021; Pan et
al. 2021), and explicit symplectic algorithms (Wang et al. 2021a,
2021b, 2021c; Wu et al. 2021). These low-order symplectic or
symplectic-like integrations have been successfully applied to
Hamiltonian systems describing the motions of particles in curved
spacetimes and the motions of spinning compact binaries. Notably,
conservations of first integrals such as the associated
Hamiltonian in these symplectic integrations do not mean that such
first integrals are conserved exactly, but mean that the
integrals' errors are bounded in time. In fact, the symplectic
integrations conserve modified Hamiltonians rather than the real
Hamiltonians of the considered differential equations.

There have been a class of numerical schemes that conserve energy
to machine precision (Chorin et al. 1978; Feng $\&$ Qin 2009), and
are even more accurate in the energy accuracy than symplectic
integrators. Qin (1987) constructed an exact energy-conserving
method for a four-dimensional system by Hamiltonian differencing.
The discretization of each component of the Hamiltonian gradient
is the average of four Hamiltonian difference terms. The method is
implicit nonsymplectic, and gives second-order accuracy to
numerical solutions. Such integrator was also given by Itoh $\&$
Abe (1988). The construction of energy-conserving  schemes based
on Hamiltonian Formulations is more complex as the dimension of
Hamiltonians increases. As an extension, a more complex
energy-conserving scheme for a six-dimensional Hamiltonian system
was proposed by Bacchini et al. (2018), and is suitable  for the
numerical integration of time-like (massive particles) and null
(photons) geodesics in any given 3+1 split spacetime. The
Hamiltonian energy-conserving method was further applied to
simulate test particle trajectories in general relativistic
magnetohydrodynamic simulations (Bacchini et al. 2019). Following
this idea, Hu et al. (2019) introduced a Hamiltonian
energy-conserving method to eight-dimensional problems. More
recently, a second-order energy-conserving scheme was specifically
designed for ten-dimensional Hamiltonian problems (Hu et al.
2021), and can be used for post-Newtonian Hamiltonian systems of
spinning compact binaries (Wu $\&$ Xie 2010; Wu et al. 2015; Huang
et al. 2016).

In the present paper, we demonstrate that the
energy-conserving scheme of Bacchini et al. (2018) is in actuality
first order accurate. We construct a new second-order
energy-conserving  scheme for six-dimensional Hamiltonian
problems. This is one of the main aims in this paper. Another aim
is the application of the newly proposed energy-conserving  scheme
to the dynamics of two six-dimensional systems.

The rest of this paper is organized as follows. In Section 2, we
analytically show that the energy-conserving  scheme of Bacchini
et al. (2018) is indeed exactly energy-conserving, and yields a
first-order accuracy to the numerical solutions.  A new
second-order energy-conserving  scheme is introduced. In Section
3, a Galactic model hosting a BL Lacertae object is used to test
the performance of the scheme of Bacchini et al and the newly
proposed method. The dynamics of the Galactic model is
investigated. Instead, the dynamics of charged particles moving
around a rotating black hole in an external magnetic field is used
as a test model in Section 4. Finally, the main results are
concluded in Section 5. Three Appendixes are used to list the
discrete forms of the related energy-conserving schemes.

\section{Reconstructing an energy-conserving scheme for a six-dimensional Hamiltonian system}

We theoretically show that the energy-conserving scheme of
Bacchini et al. (2018) yields a first-order accuracy rather than a
second-order  accuracy to numerical solutions of a six-dimensional
Hamiltonian system. Then, we introduce a new  energy-conserving
method, which makes the numerical solutions  accurate to second
order.

\subsection{Accuracy of numerical solutions for the energy-conserving scheme of Bacchini et al.}

Set $\bm{q}=(q_1,q_2,q_3)$ as generalized coordinates and
$\bm{p}=(p_1,p_2,p_3)$ as conjugate momenta. Consider a
six-dimensional conservative Hamiltonian system
\begin{equation}
  H(\bm{q},\bm{p})=H(q_1,q_2,q_3,p_1,p_2,p_3).
\end{equation}
This Hamiltonian has the canonical equations
\begin{eqnarray}
  \dot{\bm{q}} &=& \frac{\partial {\emph{H}}}{\partial {\bm{p}}},
  \\
  \dot{\bm{p}} &=& -\frac{\partial {\emph{H}}}{\partial {\bm{q}}}.
\end{eqnarray}

Take $h=t_{n+1}-t_{n}$ as an interval between  time $t_{n}$
corresponding to an $n$th step and time $t_{n+1}$ corresponding to
an $(n+1)$th step, i.e., a time step. In terms of the Taylor's
formula, the solutions from point $(q_{1}^{n}$, $q_{2}^{n}$,
$q_{3}^{n}$, $p_{1}^{n}$, $p_{2}^{n}$, $p_{3}^{n})$ advancing
towards time $h$ are expressed as
\begin{eqnarray}
q_{i}^{n+1} &=& q_{i}^{n}+ h \dot{q}_{i} +\frac{h^2}{2}
\ddot{q}_{i}+ \mathcal{O}(h^3) \nonumber \\
&=& q_{i}^{n}+ h\frac{\partial H^n}{\partial
p_i}+\frac{h^2}{2}\sum^{3}_{j=1} (\frac{\partial^2 H^n}{\partial
q_j\partial p_i}\dot{q}_j \nonumber \\ && +\frac{\partial^2
H^n}{\partial p_j\partial p_i}\dot{p}_j )+\mathcal{O}(h^3) \nonumber \\
&=& q_{i}^{n}+ h\frac{\partial H^n}{\partial
p_i}+\frac{h}{2}\sum^{3}_{j=1} (\frac{\partial^2 H^n}{\partial
q_j\partial p_i}\Delta q_j \nonumber \\ && +\frac{\partial^2
H^n}{\partial p_j\partial p_i}\Delta p_j )+\mathcal{O}(h^3),
\\
p_{i}^{n+1} &=& p_{i}^{n}+ h \dot{p}_{i} +\frac{h^2}{2}
\ddot{p}_{i}+ \mathcal{O}(h^3) \nonumber \\
&=& p_{i}^{n}-h\frac{\partial H^n}{\partial
q_i}-\frac{h^2}{2}\sum^{3}_{j=1} (\frac{\partial^2 H^n}{\partial
q_j\partial q_i}\dot{q}_j \nonumber \\ && +\frac{\partial^2
H^n}{\partial p_j\partial q_i}\dot{p}_j )+\mathcal{O}(h^3) \nonumber \\
&=& p_{i}^{n}- h\frac{\partial H^n}{\partial
q_i}-\frac{h}{2}\sum^{3}_{j=1} (\frac{\partial^2 H^n}{\partial
q_j\partial q_i}\Delta q_j \nonumber \\ && +\frac{\partial^2
H^n}{\partial p_j\partial q_i}\Delta p_j )+\mathcal{O}(h^3),
\end{eqnarray}
where $i=1,2,3$, $H^n=H(q_{1}^{n}$, $q_{2}^{n}$, $q_{3}^{n}$,
$p_{1}^{n}$, $p_{2}^{n}$, $p_{3}^{n})$, $\Delta
q_j=q_{j}^{n+1}-q_{j}^{n}$, $\Delta p_j=p_{j}^{n+1}-p_{j}^{n}$,
$\dot{q}_j=\Delta q_j/h$, and $\dot{p}_j=\Delta p_j/h$. Clearly,
$\Delta q_j\sim h$ and $\Delta p_j\sim h$. The solutions based on
the Taylor's formula in Equations (4) and (5) are explicitly
given, and are accurate to the order of $h^2$, i.e., the second
order.

On the other hand, the derivatives in Equations (2) and (3) can be
discretized. A simple discrete method is listed in Appendix A.
Using Equations (A1)-(A6), we easily derive the relation
\begin{eqnarray}\label{hami}
&&H(q_{1}^{n+1},q_{2}^{n+1},q_{3}^{n+1},
p_{1}^{n+1},p_{2}^{n+1},p_{3}^{n+1}) \nonumber \\ &&=
H(q_{1}^{n},q_{2}^{n},q_{3}^{n},p_{1}^{n},p_{2}^{n},p_{3}^{n}).
\end{eqnarray}
This shows that the solutions $(q_{1}^{n+1}$, $q_{2}^{n+1}$,
$q_{3}^{n+1}$, $p_{1}^{n+1}$, $p_{2}^{n+1}$, $p_{3}^{n+1})$
determined by the solutions $(q_{1}^{n}$, $q_{2}^{n}$,
$q_{3}^{n}$, $p_{1}^{n}$, $p_{2}^{n}$, $p_{3}^{n})$ can exactly
preserve the Hamiltonian (1), i.e., energy if the Hamiltonian
denotes an energy. Expanding the Hamiltonian in the right-hand
side of Equation (A1) at point $(q_{1}^{n}$, $q_{2}^{n}$,
$q_{3}^{n}$, $p_{1}^{n}$, $p_{2}^{n}$, $p_{3}^{n})$ in terms of
Taylor's formula, we have
\begin{eqnarray}
q_{1}^{n+1} &=& q_{1}^{n}+ h\frac{\partial H^n}{\partial
p_1}+\frac{h}{2}\Delta p_1\frac{\partial^2 H^n}{\partial p^2_1}
\nonumber \\ &&
 +\mathcal{O} [h(\Delta p_1)^2].
\end{eqnarray}
Here, the solutions $(q_{j}^{n+1}$, $p_{j}^{n+1})$ in Equations
(4) and (5) are labelled as $(q_{jT}$, $p_{jT})$, and the
solutions $(q_{j}^{n+1}$, $p_{j}^{n+1})$ in Equations (A1)-(A6)
with Equation (7) are named as $(q_{jA}$, $p_{jA})$. The
difference between $q_{1T}$ and $q_{1A}$ is estimated by
\begin{eqnarray}
q_{1T}-q_{1A} &=& \frac{h}{2}(\sum^{3}_{j=1} \Delta q_j
\frac{\partial }{\partial q_j} +\sum^{3}_{k=2} \Delta p_k
\frac{\partial }{\partial p_k}) \frac{\partial H^n}{\partial p_1}
\nonumber
\\ && +\mathcal{O}(h^3)\sim \mathcal{O}(h^2).
\end{eqnarray}
Note that $\Delta q_j\sim h$ and $\Delta p_j\sim h$ are considered
in the right-hand side. Similarly, $q_{kT}-q_{kA}\sim
\mathcal{O}(h^2)$ and $p_{jT}-p_{jA}\sim \mathcal{O}(h^2)$. Thus,
Equations (A1)-(A6) are an implicit, nonsymplectic, exact
energy-conserving scheme, which provides a first-order accuracy to
the numerical solutions.

Bacchini et al. (2018) gave a more complex  discrete method to the
derivatives in Equations (2) and (3), described in Appendix B.
Each Hamiltonian gradient is replaced with the average of six
Hamiltonian difference terms. Equations (B1)-(B6) still satisfy
Equation (6), therefore, they are also an implicit, nonsymplectic,
exact energy-conserving scheme without doubt. Bacchini et al.
claimed that Equations (B1)-(B6) can provide a second-order
accuracy to the numerical solutions. However, we confirm that
Equations (B1)-(B6) have a first-order accuracy only, as Equations
(A1)-(A6) do. In order to show this result, we mark the solution
$q_{1}^{n+1}$ in Equations (B7) as $q_{1B}$, and estimate the
difference between the solutions $q_{1T}$ and $q_{1B}$ as follows:
\begin{eqnarray}
  q_{1T}-q_{1B} & = & -\frac{h}{6}[ \frac{\partial}{\partial q_{3}}\Delta q_{3}
  + \frac{\partial}{\partial p_{3}}\Delta p_{3}- \frac{\partial}{\partial q_{2}}\Delta q_{2} \nonumber \\
  && - \frac{\partial}{\partial p_{2}}\Delta p_{2} ] \frac{\partial H^n}{\partial
  p_{1}}+ \mathcal{O}(h^{3})  \nonumber \\
  &\sim& \mathcal{O}(h^{2}).
\end{eqnarray}
In such a similar way, we also have $q_{kT}-q_{kB}\sim
\mathcal{O}(h^2)$ and $p_{jT}-p_{jB}\sim\mathcal{O}(h^2)$, where
$q_{kB}$ and $p_{jB}$ correspond to the solutions in Equations
(B2)-(B6). In other words, the truncation errors of the solutions
given in Equations (B1)-(B6) are the order of $\mathcal{O}(h^2)$.
This fact sufficiently confirms that the solutions determined by
Equations (B1)-(B6) are accurate to first order rather than second
order. In fact, the energy-conserving method for eight-dimensional
Hamiltonian systems proposed by Hu et al. (2019) is also accurate
to first order.

\subsection{New energy-conserving scheme with second-order accuracy}

Following the works of Bacchini et al. (2018) and Hu et al.
(2021), we propose a new energy-conserving scheme for the
six-dimensional Hamiltonian problem with dynamical Equations (2)
and (3). It is a discrete form with respect to the two phase-space
points at times $t_n$ and $t_{n+1}$, described in Appendix C.
Equations (C1)-(C6) are energy-conserving because they exactly
satisfy Equation (6). When they are expanded, they become
Equations (4) and (5). Therefore, their solutions $(q_{jC}^{n+1}$,
$p_{jC}^{n+1})$ minus the solutions $(q_{jT}$, $p_{jT})$ obtained
from the Taylor's formula are $q_{jC}-q_{jT}\sim \mathcal{O}(h^3)$
and $p_{jC}-p_{jT}\sim \mathcal{O}(h^3)$. Namely, the truncation
errors of the solutions given in Equations (C1)-(C6) are the order
of $h^3$. This indicates that Method $M_C$ provides a second order
accuracy to the numerical solutions.

As is illustrated, a new second-order energy-conserving method for
eight-dimensional Hamiltonian systems will be presented and
considered as an erratum of Hu et al. (2019). Of curse, the
energy-conserving method for ten-dimensional Hamiltonian problems
(Hu et al. 2021) has the second order accuracy to the numerical
solutions without question.

In the later discussions, we consider two physical problems to
test the numerical performance of the existing energy-conserving
scheme in Equations (B1)-(B6) and the newly proposed
energy-conserving scheme in Equations (C1)-(C6). We also focus on
the dynamics of the considered  problems.

\section{Galactic model}

The center of the galaxy is the gathering place
of compact object. The study of spacetime properties and emission
spectra around these celestial bodies would be helpful to
understand the formation and evolution of the galaxies. In this
section, a three-dimensional galaxy model hosting a BL Lacertae
object (Zotos 2012a, 2012b, 2013, 2014) is used to test the
performance of the related algorithms. Then, the dynamics of
galaxy model is investigated.

\subsection{Description of Galactic model}

The Galactic model considered by Zotos (2012a, 2012b, 2013, 2014)
is expressed as
\begin{equation}\label{G1}
H=\frac{1}{2} (p^{2}_x+p^{2}_y+p^{2}_z)+V_{G}(x,y,z)+V_{B}(x,y,z).
\end{equation}
$V_{G}$ is a host elliptical galaxy with the logarithmic potential
\begin{equation}\label{G2}
V_{G}(x,y,z)=\frac{v_{0}^{2}}{2} \ln(x^{2}+\alpha y^{2}+
bz^{2}-\lambda x^{3}+c_{b}^{2}).
\end{equation}
$c_{b}$ is a bulge of radius of the elliptical galaxy, and $v_{0}$
is a parameter for the consistency of the galactic units. $\alpha$
is associated to the flattening of the galaxy along the $y$ axis,
and $b$ describes the flattening of the galaxy along the $z$ axis.
$\lambda\ll1$ corresponds to an internal perturbation. $V_{B}$
relates to the description of BL Lac objects as a relatively rare
subclasses of Active Galactic Nuclei (AGN) at the nucleus of the
elliptical galaxy. It is described by a spherically symmetric
Plummer potential
\begin{equation}\label{G3}
V_{B}(x,y,z)=-\frac{GM_{n}}{\sqrt{x^{2}+y^{2}+z^{2}+c_{n}^2}}.
\end{equation}
$G$ represents the gravitational constant. $M_n$ denotes the mass
of the nucleus, and $c_n$ is the scale length of the nucleus.

For convenience, Equations (A1)-(A6), Equations (B1)-(B6), and
Equations (C1)-(C6) are respectively labelled as Method A ($M_A$),
Method B ($M_B$), and Method C ($M_C$). For comparison,
second-order Runge-Kutta method (RK2) and second-order explicit
symplectic method (S2) are independently used to solve the system
(10). An eighth- and ninth-order Runge-Kutta-Fehlberg integrator
[RKF89] with adaptive step sizes is used to provide
higher-precision reference solutions. The related units and
parameters are specified as follows. The distance unit is $1$kpc,
and the speed unit is 9.77813 \textrm{km/s}. In this case, time
unit is 1kpc/9.77813 \textrm{km/s}=$10^{8}$ years. The mass and
energy units are $2.325 \times 10^{7}\textrm{M}_{\bigodot}$ and
95.6118 (\textrm{km/s})$^{2}$, respectively. The parameters are
taken as $v_{0}=15.3403565$, $c_{n}=0.25$, and $c_{b}=1.5$.

\subsection{Numerical evaluations}

At first, we take parameter $\lambda=0$ and  select two different
orbits whose initial conditions are $H=450$, $b=1$, $y=p_x=p_z=0$,
$x=3$, and $z=0.1$. Parameters $\alpha=1$ and $M_{n}=10$ for Orbit
1, while $\alpha=0.1$ and $M_{n}=400$ for Orbit 2. The initial
values of $p_y>0$ of the two orbits are given by Equation (10).
The dimensionless time-step takes $h=10^{4}/10^{8}$, which
corresponds to a real physical time of $10^{4}$ years. Figures 1
(a) and (b) plot the relative Hamiltonian errors when the five
algorithms act on  Orbits 1 and 2 to $10^{10}$ years
(corresponding to $10^{6}$ steps of integrations). Methods
$M_{A}$, $M_{B}$ and $M_{C}$ have approximately same energy
errors. These errors are relatively small and slowly grow up to an
order of  $10^{-13}$ due to roundoff errors. The three integrators
should be energy-conserving if the roundoff errors are negligible.
RK2 remains bounded in errors for Orbit 1 in Figure 1(a), but does
not for Orbit 2 in Figure 1(b). S2 always gives bounded errors to
Orbits 1 and 2 because of its symplecticity. RK2 exhibits the
largest errors, whereas $M_{A}$, $M_{B}$ and $M_{C}$ yield the
smallest errors. In particular, the Hamiltonian errors for
$M_{A}$, $M_{B}$ and $M_{C}$ seem to be independent of the choice
of Orbit 1 or Orbit 2. However, the position errors for the new
method $M_{C}$ in Figures 1 (c) and (d) are almost the same as
those for the second-order methods RK2 and S2, and the position
errors for Method $M_{B}$ are approximate to those for Method
$M_{A}$. The position errors for $M_{C}$, RK2 and S2 are smaller
than those for $M_A$ and $M_B$.

The dependence of relative Hamiltonian error $\Delta H/H$ on
Hamiltonian $H$ in Figures 2 (a)-(c) shows that the Hamiltonian
errors among the three schemes $M_A$, $M_B$ and $M_C$ have no
dramatic differences, and are several orders of magnitude smaller
than S2 or RK2 for different choices of mass parameter $M_n$. On
the other hand, the absolute position errors for $M_A$ are almost
consistent with those for $M_B$, but are two or three orders of
magnitude larger than for the methods S2 and $M_C$ in Figures 2
(d)-(f). These results are still supported in Figure 3 that
describes the dependence of the relative Hamiltonian error (or the
absolute position error) on the mass parameter $M_n$ for different
choices of parameter $\lambda$. The error trends with an increase
of time step in Figure 4 describe that the relative Hamiltonian
errors for RK2 are slightly larger than those for S2, but are
relatively larger than those for anyone of Methods $M_A$, $M_B$
and $M_C$. The relative Hamiltonian errors increase with an
increase of time step for RK2 and S2, whereas are independent of
any choice of time steps for $M_A$, $M_B$ and $M_C$. On the other
hand, the absolute position errors grow with an increase of
dimensionless time step $h>1/100000$ for the five methods. In
particular, the rules of the growth of absolute position error
with time step $h$ for the these algorithms are $\Delta r\propto
h$ for $M_A$ and $M_B$, and $\Delta r\propto h^2$ for $M_C$, S2
and RK2. Clearly, $M_C$ with S2 has the smallest position errors
for dimensionless time steps $10^{-5}\leq h\leq 10^{-3}$.

In short, the numerical results in Figures 1-4 have sufficiently
conformed that the schemes $M_A$, $M_B$ and $M_C$ can conserve
energy if the roundoff errors are neglected. They are also greatly
superior to the second-order symplectic method S2 in conservation
of energy. However, the three  energy-conserving schemes are
different in accuracy of the solutions. $M_A$ and $M_B$ have
almost the same position errors. $M_C$ with S2 has, too.
Particular for dimensionless time steps $10^{-5}\leq h\leq
10^{-3}$, $M_C$ and S2 give  the best accuracy to the solutions.
In other words, the numerical tests have supported that  $M_B$
yields a first-order accuracy to the numerical solutions and $M_C$
has a second-order accuracy.

\subsection{Dynamics of orbits}

Considering that Method $M_C$ with an optimal time step (such as
$h=10^{-4}$) shows better performance in conservation of energy
and accuracy of solutions, we apply it to give some insight into
the dynamical behavior of orbits. Seen from Figures 5 (a) and (b),
the two orbits seem to have distinct three-dimensional
configurations. Orbit 1 in Figure 5(a) seems to be periodic or
quasi-periodic, but Orbit 2 in Figure 2(b) seems to be chaotic.
These results are not shown through the method of
Poincar\'{e}-sections/maps because the phase space has six
dimensions, but can be confirmed by fast Lyapunov indicators
(FLIs) in Figures 5 (c) and (d). Here, the FLIs are calculated in
terms of the two-particle method (Wu et al. 2006). Different time
rates of the growth of FLIs are used to identify the regular or
chaotic behavior. A bounded orbit is ordered when its FLI grows
algebraically with time, but chaotic if its FLI increases
exponentially. Based on this criterion, the regularity of Orbit 1
and the chaoticity of Orbit 2 are clearly identified. The results
of Method $M_{C}$ are consistent with those of the high-precision
algorithm RKF89.

Using the technique of FLIs, we trace the effects of varying the
parameters on the occurrence of chaos. The initial conditions are
still those in Figure 1, and parameters $H=400$, $\alpha=1.6$, and
$b=0.8$ are always fixed. Mass $M_{n}$ is given several values,
and $\lambda$ ranges from 0 to 0.03  with an interval of $3 \times
10^{-4}$. For each value of $\lambda$, the FLI is obtained after
$3 \times 10^{5}$ integration steps. In this way, the relation
between the FLI and $\lambda$ is described in Figure 6. 5 is found
to be a threshold of FLIs between order and chaos. The values of
$\lambda$ with $FLI\geq 5$ correspond to the chaoticity, whereas
the values of $\lambda$ with $FLI< 5$ indicate the regularity.
Figure 6 relates to the description of finding chaos by scanning a
space of parameter $\lambda$; in fact, this figure establishes a
correspondence between parameter $\lambda$ and chaos or order. It
is shown clearly in Figure 6 that a transition from order to chaos
easily occurs as parameter $\lambda$ increases. If $\lambda$ is
given, the presence of chaos also becomes easier with an increase
of mass $M_{n}$. Particularly for a larger value of $M_{n}$ in
Figures 6 (h) and (i), all values of $\lambda$ indicate chaos.
These dynamical results of order and chaos obtained from Method
$M_{C}$ are in perfect agreement with those given by RKF89.
Scanning a two-dimensional space of parameters $\alpha$ and $M_n$
in Figure 7 shows that the occurrence of chaos is difficult when
the values of $\alpha$ are in the neighbourhood of 1, but it is
easy when the values of $\alpha$ are far away from 1.  The effect
of varying the parameter $b$ on the presence of chaos should be
similar to that of varying the parameter $\alpha$. The effects of
varying the parameters on the presence of chaos in the present
paper are the same as those of Zotos (2014).

An explanation to the dependence of chaos on the parameters is
given here. A larger value of perturbation parameter $\lambda$ or
mass $M_{n}$ means strengthening the gravity of BL Lac objects in
Equations (10)-(12), therefore, there is a more chance of chaos.
For $\alpha\approx 1$ in Equation (11), the potential of the
elliptical galaxy tends to the isotropy with respect to the three
axes in the case of $b=1$. However,  the potential destroys the
isotropy for $\alpha$ far away from 1. As a result, chaos is
easily present.

\section{Magnetized rotating black hole}

By analyzing the motion of charged particles
around a black hole, one can understand the spacetime properties
around the black hole and test the general theory of relativity.
In addition, the motion of charged particles reflects the
evolution of the accretion disk around the black hole, and is
useful to understand the accretion process of the black hole.
Because of this, the dynamics of charged particles moving around a
rotating black hole in an external magnetic field was considered
in the work of Kop\'{a}\v{c}ek $\&$ Karas (2014). Now, the
dynamical model is used to evaluate the above-mentioned
algorithms. The dynamics of charged particles is further
surveyed.

\subsection{Dynamical model}

 In Boyer-Lindquist coordinates $x^{\mu}=(t,r,\theta,\varphi)$, the Kerr metric
is expressed as (Misner et al. 1973)
\begin{eqnarray}\label{KN}
\textrm{d}s^{2} &=& \frac{\Sigma}{\Delta} \textrm{d}r^{2}+\Sigma \textrm{d}\theta^{2}
-\frac{\Delta}{\Sigma}[\textrm{d}t-a\textrm{sin}\theta \textrm{d}\varphi]^{2} \nonumber \\
&& +\frac{\textrm{sin}^{2}\theta}{\Sigma}
[(r^{2}+a^{2})\textrm{d}\varphi-a\textrm{d}t]^{2},
\end{eqnarray}
where $a$ stands for the spin parameter (i.e., the specific
angular momentum) of the Kerr black hole with mass $M$, $\Sigma$
and $\Delta$ are written as follows:
\begin{eqnarray}
  \Delta &=& r^{2}-2Mr+a^2, \\
  \Sigma &=& r^{2}+a^{2}\textrm{cos}^{2}\theta.
\end{eqnarray}
Gravitational constant $G$ and speed of light $c$ take geometrized
units: $c=G=1$.

Kop\'{a}\u{c}ek $\&$ Karas (2014) assumed that the rotating black
hole has a non-zero electric charge $Q$. Although such a black
hole is the Kerr-Newman black hole in this case, the
electromagnetic field generated by the black hole's electric
charge is so weak that it does not affect the spacetime and plays
an important role in the motion of charged particles around the
black hole. Because of this, the metric still remains unaltered by
$Q$. They also considered that the rotating Kerr black hole is
immersed in an asymptotically uniform magnetic field with the
vector potential:
\begin{eqnarray}
  A_{t} &=& \frac{aB_{z}Mr}{\Sigma} (1+\textrm{cos}^{2}\theta) -aB_{z}-\frac{Qr}{\Sigma} \nonumber \\
  && +\frac{aMB_{x}\textrm{sin}\theta \textrm{cos}\theta}{\Sigma}
  (r\textrm{cos}\psi-a\textrm{sin}\psi), \\
  A_{r} &=& -B_{x}(r-M)\textrm{cos}\theta \textrm{sin}\theta
  \textrm{sin}\psi, \\
  A_{\theta} &=& -B_{x}(r^{2}\textrm{cos}^{2}\theta-Mr\textrm{cos}2\theta+a^{2}\textrm{cos}2\theta)\textrm{sin}\psi \nonumber \\
  && -aB_{x}(r\textrm{sin}^{2}\theta+M\textrm{cos}^{2}\theta)\textrm{cos}\psi,
\end{eqnarray}
\begin{eqnarray}
  A_{\varphi} &=& B_{z}\sin^{2}\theta \left[ \frac{1}{2}(r^{2}+a^{2})-\frac{a^{2}Mr}{\Sigma}(1+\cos^{2}\theta) \right] \nonumber \\
  && -B_{x}\sin\theta \cos\theta \left[ \Delta \cos\psi+\frac{(r^{2}+a^{2})M}{\Sigma} \right. \nonumber \\
  && \left. \cdot(r\cos\psi-a\sin\psi)\right]+\frac{Qar\sin^{2}\theta}{\Sigma}.
\end{eqnarray}
This magnetic field was derived by Wald (1974) and generalized by
Bi\v{c}\'{a}k $\&$ Jani\v{s} (1985). Here, $B_x$ and $B_z$ are
constant magnetic parameters, $\psi$ and $r_{\pm}$ read as
\begin{eqnarray}
\psi &=& \varphi+\frac{a}{r_{+}-r_{-}}\ln \frac{r-r_{+}}{r-r_{-}},
\\ r_{\pm} &=& M \pm \sqrt{M^{2}-a^{2}}.
\end{eqnarray}

The dynamics of a test particle with charge $q$ and mass $m$
moving around the black hole with external magnetic field can be
described by the following super-Hamiltonian
\begin{equation}\label{SH}
  \mathcal{H} = \frac{1}{2m} g^{\mu \nu}
  (p_{\mu}-qA_{\mu})(p_{\nu}-qA_{\nu}),
\end{equation}
where the particle's generalized momenta $p_{\mu}=m g_{\mu
\nu}\dot{x}^{\nu}+qA_{\mu}$. Its canonical equations are
\begin{eqnarray}
 \dot{t} &=&\frac{\partial \mathcal{H}}{\partial p_{t}} \nonumber \\
 &=& g^{tt} (p_{t}-qA_{t})+ g^{t\varphi}(p_{\varphi}-qA_{\varphi}), \\
\dot{p_{t}} &=& -\frac{\partial \mathcal{H}}{\partial t}=0; \\
  \dot{r}&=&\frac{\partial \mathcal{H}}{\partial p_{r}}, ~~~~\dot{\theta}=\frac{\partial\mathcal{H}}{\partial p_{\theta}},
  ~~~~\dot{\varphi}=\frac{\partial \mathcal{H}}{\partial
  p_{\varphi}}, \\
\dot{p_{r}} &=& -\frac{\partial \mathcal{H}}{\partial r}, ~
\dot{p_{\theta}} = -\frac{\partial \mathcal{H}}{\partial \theta},
~\dot{p_{\varphi}}=-\frac{\partial \mathcal{H}}{\partial \varphi}.
\end{eqnarray}
Equation (24) shows that the conjugate momentum $p_{t}$ is a
constant of motion and is related to energy $E$ of the test
particle, namely, $p_{t}=-E$. Another constant is
\begin{equation}\label{SH1}
  \mathcal{H} = -\frac{m}{2}.
\end{equation}
Other constants like the particle's angular momentum are no longer
present due to the magnetic field governed by parameter $B_x$
breaking axial symmetry. In this sense, the super-Hamiltonian is a
six-dimensional nonintegrable system, whose evolution is dominated
by Equations (25) and (26).

For simplicity, scale transformations are used as dimensionless
operations to the super-Hamiltonian. The operations are as
follows: $r\rightarrow rM$, $t\rightarrow tM$, $\tau\rightarrow
\tau M$, $a\rightarrow aM$, $Q\rightarrow QM$, $E\rightarrow Em$,
$p_r\rightarrow mp_r$, $p_{\varphi}\rightarrow mMp_{\varphi}$,
$p_{\theta}\rightarrow mMp_{\theta}$, $q\rightarrow mq$,
$B_x\rightarrow B_x/M$, $B_z\rightarrow B_z/M$ and
$\mathcal{H}\rightarrow m\mathcal{H}$. As a result, $M\rightarrow
1$ in the above expressions, and the black hole's angular momentum
satisfies $|a|\leq 1$. In addition, $\mathcal{H} = -1/2$ and
$|Q|\leq 1$.

\subsection{Numerical investigations}

Let us consider two orbits with same initial values
$p_{r}=p_{\varphi}=0$ and charge $q=1$. Orbit I has other initial
conditions $r=3.9$, $\theta=1.15$, $\varphi=0$ and $L=
p_{\varphi}=6$, and parameters $a=0.9$, $Q=1$, $E=1.61$,
$B_{x}=0.001$ and $B_{z}=1$. Orbit II has other initial conditions
$r=5$, $\theta=1$, $\varphi=\pi/3$ and $L= 5.6$, and parameters
$a=0.8$, $Q=0.5$, $E=1.325$, $B_{x}=0.007$ and $B_{z}=0.7$. The
time step is $h=0.01$. The second-order explicit symplectic
algorithm (S2) is replaced by the second-order midpoint implicit
symplectic method (IS2). Figure 8 supports that the three
energy-conserving schemes have same good effects on Hamiltonian
conservations, compared with the methods RK2 and IS2. However,
Method $M_C$ is basically the same as the second-order methods IS2
and RK2, and Method $M_B$ is similar to the first-order method
$M_A$ in accuracy of the solutions. This shows again that $M_B$
gives a first-order accuracy to the numerical solutions, and $M_C$
possesses a second-order accuracy. These results are also
confirmed in Figure 9 that describes the dependence of the errors
on the time steps. Good choices of time steps are from $h\sim
10^{-3}$ to $h\sim 10^{-2}$.

Now, Method  $M_C$ with the appropriate time step $h= 10^{-2}$ is
applied to study the long-term evolution of orbits.  Orbits I and
II have different three-dimensional configurations in Figures 10
(a) and (b). The FLIs in Figures 10 (c) and (d) indicate the
regularity of Orbit I and  the chaoticity of Orbit II. The results
of the FLIs for Method $M_C$  are completely consistent with those
for RKF89. The effects of varying energies $E$ on the FLIs in
Figures 11 (a)-(d) show that chaos easily occurs as the energy $E$
increases. In addition, an increase of $B_{x}$ with $B_{z}=1$
causes a smaller energy to induce chaos. The result is consistent
with that of Kop\'{a}\v{c}ek $\&$ Karas (2014).
Here, the FLI for each value of $E$ is obtained
after $5 \times 10^{5}$ integration steps. The energies with FLIs
$<$ 5 correspond to the regularity, but those with FLIs $\geq$ 5
indicate the onset of strong chaos. Chaos occurs for $E\geq 1.544$
with $B_{x}=0.001$ [Figure 11(a)], $E\geq 1.54$ with $B_{x}=0.005$
[Figure 11(b)], $E\geq 1.536$ with $B_{x}=0.01$ [Figure 11(c)],
and $E\geq 1.52$ with $B_{x}=0.05$ [Figure 11(d)]. To clearly show
the dependence of the orbital dynamical behavior on a variation of
magnetic field parameter $B_x$, we plot Figures 11 (e) and (f),
where another magnetic field parameter $B_z=1$ is fixed. Chaos
occurs for $B_x\geq 0.03416$ with $E=1.5697$ [Figure 11(e)], and
$B_x\geq 0.01486$ with $E=1.576$ [Figure 11(f)]. This shows that
magnetic field parameter $B_x$ has a critical value, which makes
the dynamics transit from order to chaos. In fact, this value is
closely related to $E$. These results clearly describe the
above-mentioned dependence of the dynamical transition from order
to chaos with an increase of energy or magnetic field parameter
$B_{x}$. 

One of the results concluded from Figure 11 is
that chaos can occur for a smaller energy as  $B_{x}$ with
$B_{z}=1$ increases. What about the dynamical transition from
order to chaos with a variation of $B_{z}$? Figure 12(a) answers
this question. Here, we take the initial conditions $r=3.5$,
$\theta=1$ and $\varphi=0$ and the parameters $B_{x}=0$, $a=0.8$,
$E=1.48$, $L=5$ and $Q=1$, and let  $B_{z}$ range from 0.6 to 0.9
with an interval of 0.002. Regular regions of $B_{z}$ are [0.612,
0.654] and [0.756, 0.9], and chaotic regions of $B_{z}$ are [0.6,
0.61] and [0.656, 0.696]. Clearly, a dynamical transition from
order to chaos occurs  in the vicinity of $B_{z}=0.65$. The FLIs
for three values of $B_{z}$ in Figure 12(b) explicitly show that
the regularity exists for $B_{z}=0.65$, and chaos for
$B_{z}=0.656$ is weaker than for $B_{z}=0.658$. The Poincar\'{e}
sections in  Figure 12(c) display that  $B_{z}=0.65$ corresponds
to three islands of the regularity (i.e., a resonance), whereas
$B_{z}=0.658$ leads to losing the three islands and a number of
discrete points filled with a small area, i.e., the chaotic
behavior. The orbit for $B_{z}=0.656$, located in a sepratrix
location between the regular islands and the chaotic layers, seems
to be three islands of the regularity, but consists of three thin
chaotic layers. That is, $B_{z}$=0.65, 0.656 and 0.658 correspond
to order, weak chaos, and strong chaos, respectively. This
chaoticity in Figures 10-12 is due to external perturbations from
the magnetic field forces governed by parameters $B_{x}$ and/or
$B_{z}$. Given $B_{x}=B_{z}=0$, Equation (22) is the Kerr-Newman
black hole (13), which is integrable and nonchaotic. However, the
magnetic field forces for $B_{x}\neq0$ and/or $B_{z}\neq0$ cause
Equation (22) to be nonintegrable. When the magnetic field forces
are small, the motions are mainly dominated by the gravity from
the black hole and are still regular Kolmogorov-Arnold-Moser (KAM)
tori. In spite of the regularity, these tori are twisted by the
external perturbations. As the external  perturbations get
stronger, some tori are destroyed and departures from stability
and resonances appear. When the black hole's gravity basically
matches with the magnetic field forces, chaos occurs. This is just
an example shown in Figure 12(c). Now, we explain why the
technique of Poincar\'{e}-sections/maps can be used in Figure 12
but cannot in Figures 10 and 11. For the case of $B_{x}=0$ in
Figure 12,  the angular momentum $L=g_{t\varphi}\dot{t}
+g_{\varphi\varphi}\dot{\varphi}+qA_{\varphi}$ is conserved. Thus,
all orbits are restricted to a 4-dimensional phase space made of
$r$, $\theta$, $p_r$ and $p_{\theta}$, and the technique of
Poincar\'{e}-sections/maps can work well. For the case of
$B_{x}\neq0$ in Figures 10 and 11, $L$ varies with time.
Therefore, all motions are in the 6-dimensional phase space made
of $r$, $\theta$, $\varphi$, $p_r$, $p_{\theta}$, and
$p_{\varphi}$, and the technique of Poincar\'{e}-sections/maps is
not suitable for this case. Because $B_x$ destroys the
conservation of the angular momentum $L$ but $B_z$ does not, $B_x$
and $B_z$ exert different influences on the chaotic behavior. That
is, an increase of $B_x$ breaks the axial symmetry and strengthens
the chaotic properties in Figures 11 (e) and (f), but that of
$B_z$ does not always strengthen in Figure 12(a). 

For $B_{x}\neq0$ and $B_{z}\neq0$, Figure 13 shows that a smaller
initial angular momentum $L$ of the particle easily yields chaos.
The strength of chaos is not always enhanced  or weakened as the
black hole's angular momentum $a$ increases. In fact, chaos is
stronger for $0.6\leq a\leq 0.64$ with $5.4\leq L\leq 5.6$, but is
absent for $0.76\leq a\leq 0.8$ with any initial angular momenta
$L$ in Figure 13(a). Chaos is stronger for $0.6\leq a\leq 0.68$
with $5.4\leq L\leq 5.7$, but is absent for $0.78\leq a\leq 0.8$
with any initial angular momenta $L$ in Figure 13(b). Although
chaos is present for all values of $a$ and $L$ considered in
Figure 13(c), stronger chaos occurs for $0.64\leq a\leq 0.76$. The
result is not completely consistent with that of Takahashi $\&$
Koyama (2009) on the black hole spin weakening the chaotic
properties. This is due to different combinations of initial
conditions and other dynamical parameters. There is no universal
rule on the relation between the dynamical transition and the
black hole spin (Sun et al. 2021).

\section{Summary}

It is shown analytically that the existing six-dimensional
Hamiltonian-conserving algorithm of Bacchini et al. (2018) does
not possess a second-order accuracy to numerical solutions, but
has a first-order accuracy only. A new second-order
six-dimensional Hamiltonian-conserving scheme is proposed. Taking
the Galactic model hosting a BL Lacertae object and the dynamics
of charged particles moving around a rotating black hole in an
external magnetic field as test models, we numerically confirm
that the existing method of Bacchini et al and the newly proposed
scheme are energy-conserving, but have different performances in
accuracy of the numerical solutions. The numerical solutions are
accurate to first order for the former scheme, but to second order
for the latter method.

The new energy-conserving method combined with appropriate time
steps is used to explore the effects of varying the parameters on
the presence of chaos in the two physical models. Chaos easily
occurs in the Galactic model as the mass of the nucleus, the
internal perturbation parameter, and the anisotropy of the
potential of the elliptical galaxy increase. Larger energies of
the particles, smaller initial initial angular momenta of  the
particles, and stronger magnetic fields (that are mainly governed
by parameter $B_x$) are helpful to induce chaos in the magnetized
Kerr spacetime. The chaotic properties are not necessarily
weakened when  the black hole spin increases.

The new scheme has no symplecticity. However, it
is time reversibility as a property of some particular symplectic
problems, and it has also excellent energy conservation. Because
of the time reversibility, the new method including the particular
symplectic problems is suitable for tracing the  origin and
evolution of some celestial objects. The new second-order
energy-conserving method involving a second order symplectic
integrator with appropriately smaller steps can achieve similar
accuracies of  high-order symplectic partitioned Runge-Kutta
methods with correspondingly larger time steps. It can also take
less computational cost, compared to the high-order methods. Based
on these facts, a second-order symplectic integrator is often used
as very long-time integrations of celestial objects, for example,
$N$-body problems in the solar system (Wisdom $\&$ Holman 1991).
The new scheme can be used for any six-dimensional Hamiltonian
problems, including globally hyperbolic spacetimes with readily
available (3+1) split coordinates. These spacetimes are, e.g., the
Kerr metric and the Kerr black hole with external magnetic fields.

\section*{Acknowledgments}

The authors are very grateful to a referee for
valuable comments and suggestions. This research has been
supported by the National Natural Science Foundation of China
[Grant Nos. 11973020 (C0035736), 11533004, 11663005, 11533003, and
11851304], the Special Funding for Guangxi Distinguished
Professors (2017AD22006), and the National Natural Science
Foundation of Guangxi (Nos. 2018GXNSFGA281007 and 2019JJD110006).

\appendix
\section*{Appendix}

\section{Simple discrete method of the derivatives}

\begin{eqnarray}\label{3a}
\frac{q_{1}^{n+1}-q_{1}^{n}}{h} &=&
\frac{1}{p_{1}^{n+1}-p_{1}^{n}}
\left[H(q_{1}^{n},q_{2}^{n},q_{3}^{n},p_{1}^{n+1},p_{2}^{n},p_{3}^{n})
\right. \left.
-H(q_{1}^{n},q_{2}^{n},q_{3}^{n},p_{1}^{n},p_{2}^{n},p_{3}^{n})
\right],\\
 \frac{q_{2}^{n+1}-q_{2}^{n}}{h} &=&
\frac{1}{p_{2}^{n+1}-p_{2}^{n}}\left[H(q_{1}^{n},q_{2}^{n},
q_{3}^{n},p_{1}^{n+1},p_{2}^{n+1},p_{3}^{n}) \right. \left.
-H(q_{1}^{n},q_{2}^{n},q_{3}^{n},p_{1}^{n+1},p_{2}^{n},p_{3}^{n})
\right],
\\
\frac{q_{3}^{n+1}-q_{3}^{n}}{h} &=&
\frac{1}{p_{3}^{n+1}-p_{3}^{n}}
\left[H(q_{1}^{n},q_{2}^{n},q_{3}^{n},p_{1}^{n+1},p_{2}^{n+1},p_{3}^{n+1})
\right. \left.
-H(q_{1}^{n},q_{2}^{n},q_{3}^{n},p_{1}^{n+1},p_{2}^{n+1},p_{3}^{n})
\right],
\\
\frac{p_{1}^{n+1}-p_{1}^{n}}{h} &=& -
\frac{1}{q_{1}^{n+1}-q_{1}^{n}}
\left[H(q_{1}^{n+1},q_{2}^{n},q_{3}^{n},p_{1}^{n+1},p_{2}^{n+1},p_{3}^{n+1})
\right. \left.
-H(q_{1}^{n},q_{2}^{n},q_{3}^{n},p_{1}^{n+1},p_{2}^{n+1},p_{3}^{n+1})
\right],
\\
\frac{p_{2}^{n+1}-p_{2}^{n}}{h} &=&
-\frac{1}{q_{2}^{n+1}-q_{2}^{n}}
\left[H(q_{1}^{n+1},q_{2}^{n+1},q_{3}^{n},p_{1}^{n+1},p_{2}^{n+1},p_{3}^{n+1})
\right. \left.
-H(q_{1}^{n+1},q_{2}^{n},q_{3}^{n},p_{1}^{n+1},p_{2}^{n+1},p_{3}^{n+1})
\right],
\\
\frac{p_{3}^{n+1}-p_{3}^{n}}{h} &=&
-\frac{1}{q_{3}^{n+1}-q_{3}^{n}}\left[H(q_{1}^{n+1},q_{2}^{n+1},q_{3}^{n+1},p_{1}^{n+1},p_{2}^{n+1},p_{3}^{n+1})
\right. \left.
-H(q_{1}^{n+1},q_{2}^{n+1},q_{3}^{n},p_{1}^{n+1},p_{2}^{n+1},p_{3}^{n+1})
\right].
\end{eqnarray}

\section{Existing complex discrete method of the derivatives}

The discrete equations (39)-(44) with respect to Equations (2) and
(3) in the work of Bacchini et al. (2018) are written as follows:
\begin{eqnarray}\label{6}
\frac{q_{1}^{n+1}-q_{1}^{n}}{h} &=&
\frac{1}{6(p_{1}^{n+1}-p_{1}^{n})} \{
[H(q_{1}^{n+1},q_{2}^{n},q_{3}^{n},p_{1}^{n+1},p_{2}^{n},p_{3}^{n})
-H(q_{1}^{n+1},q_{2}^{n},q_{3}^{n},p_{1}^{n},p_{2}^{n},p_{3}^{n}) ]  \nonumber \\
&& + [H(q_{1}^{n+1},q_{2}^{n+1},q_{3}^{n+1},p_{1}^{n+1},p_{2}^{n+1},
p_{3}^{n+1})-H(q_{1}^{n+1},q_{2}^{n+1},q_{3}^{n+1},p_{1}^{n},p_{2}^{n+1},p_{3}^{n+1})] \nonumber \\
&& + [H(q_{1}^{n},q_{2}^{n+1},q_{3}^{n+1},p_{1}^{n+1},p_{2}^{n+1},p_{3}^{n+1})
-H(q_{1}^{n},q_{2}^{n+1},q_{3}^{n+1},p_{1}^{n},p_{2}^{n+1},p_{3}^{n+1})] \nonumber \\
&& + [H(q_{1}^{n+1},q_{2}^{n},q_{3}^{n+1},p_{1}^{n+1},p_{2}^{n},p_{3}^{n+1})
-H(q_{1}^{n+1},q_{2}^{n},q_{3}^{n+1},p_{1}^{n},p_{2}^{n},p_{3}^{n+1})] \nonumber \\
&& + [H(q_{1}^{n},q_{2}^{n},q_{3}^{n},p_{1}^{n+1},p_{2}^{n},p_{3}^{n})
-H(q_{1}^{n},q_{2}^{n},q_{3}^{n},p_{1}^{n},p_{2}^{n},p_{3}^{n})] \nonumber \\
&& + [H(q_{1}^{n},q_{2}^{n},q_{3}^{n+1},
p_{1}^{n+1},p_{2}^{n},p_{3}^{n+1})
-H(q_{1}^{n},q_{2}^{n},q_{3}^{n+1},p_{1}^{n},p_{2}^{n},p_{3}^{n+1})]
\},
\end{eqnarray}
\begin{eqnarray}\label{7}
\frac{q_{2}^{n+1}-q_{2}^{n}}{h} &=&
\frac{1}{6(p_{2}^{n+1}-p_{2}^{n})} \{
[H(q_{1}^{n+1},q_{2}^{n+1},q_{3}^{n},p_{1}^{n+1},p_{2}^{n+1},p_{3}^{n})
-H(q_{1}^{n+1},q_{2}^{n+1},q_{3}^{n},p_{1}^{n+1},p_{2}^{n},p_{3}^{n})] \nonumber \\
&& +[H(q_{1}^{n},q_{2}^{n+1},q_{3}^{n},p_{1}^{n},p_{2}^{n+1},p_{3}^{n})-H(q_{1}^{n},q_{2}^{n+1},q_{3}^{n},p_{1}^{n},p_{2}^{n},p_{3}^{n})] \nonumber \\
&& +[H(q_{1}^{n},q_{2}^{n},q_{3}^{n},p_{1}^{n},p_{2}^{n+1},p_{3}^{n})-H(q_{1}^{n},q_{2}^{n},q_{3}^{n},p_{1}^{n},p_{2}^{n},p_{3}^{n})] \nonumber \\
&& +[H(q_{1}^{n+1},q_{2}^{n+1},q_{3}^{n+1},p_{1}^{n+1},p_{2}^{n+1},p_{3}^{n+1})
-H(q_{1}^{n+1},q_{2}^{n+1},q_{3}^{n+1},p_{1}^{n+1},p_{2}^{n},p_{3}^{n+1})] \nonumber \\
&& +[H(q_{1}^{n+1},q_{2}^{n},q_{3}^{n},p_{1}^{n+1},p_{2}^{n+1},p_{3}^{n})-H(q_{1}^{n+1},q_{2}^{n},q_{3}^{n},p_{1}^{n+1},p_{2}^{n},p_{3}^{n})] \nonumber \\
&&+[H(q_{1}^{n+1},q_{2}^{n},q_{3}^{n+1},p_{1}^{n+1},p_{2}^{n+1},p_{3}^{n+1})
-H(q_{1}^{n+1},q_{2}^{n},q_{3}^{n+1},p_{1}^{n+1},p_{2}^{n},p_{3}^{n+1})]
\},
\end{eqnarray}
\begin{eqnarray}\label{8}
\frac{q_{3}^{n+1}-q_{3}^{n}}{h} &=&
\frac{1}{6(p_{3}^{n+1}-p_{3}^{n})} \{ [
H(q_{1}^{n+1},q_{2}^{n+1},q_{3}^{n+1},p_{1}^{n+1},p_{2}^{n+1},p_{3}^{n+1})
-H(q_{1}^{n+1},q_{2}^{n+1},q_{3}^{n+1},p_{1}^{n+1},p_{2}^{n+1},p_{3}^{n})].  \nonumber \\
&& +[H(q_{1}^{n},q_{2}^{n+1},q_{3}^{n+1},p_{1}^{n},p_{2}^{n+1},p_{3}^{n+1}) -H(q_{1}^{n},q_{2}^{n+1},q_{3}^{n+1},p_{1}^{n},p_{2}^{n+1},p_{3}^{n})] \nonumber \\
&& +[H(q_{1}^{n},q_{2}^{n+1},q_{3}^{n},p_{1}^{n},p_{2}^{n+1},p_{3}^{n+1}) -H(q_{1}^{n},q_{2}^{n+1},q_{3}^{n},p_{1}^{n},p_{2}^{n+1},p_{3}^{n})] \nonumber \\
&& +[H(q_{1}^{n},q_{2}^{n},q_{3}^{n+1},p_{1}^{n},p_{2}^{n},p_{3}^{n+1})-H(q_{1}^{n},q_{2}^{n},q_{3}^{n+1},p_{1}^{n},p_{2}^{n},p_{3}^{n})] \nonumber \\
&& +[H(q_{1}^{n+1},q_{2}^{n+1},q_{3}^{n},p_{1}^{n+1},p_{2}^{n+1},p_{3}^{n+1})-H(q_{1}^{n+1},q_{2}^{n+1},q_{3}^{n},p_{1}^{n+1},p_{2}^{n+1},p_{3}^{n})] \nonumber \\
&& +[H(q_{1}^{n},q_{2}^{n},q_{3}^{n},
p_{1}^{n},p_{2}^{n},p_{3}^{n+1})
-H(q_{1}^{n},q_{2}^{n},q_{3}^{n},p_{1}^{n},p_{2}^{n},p_{3}^{n})]
\},
\end{eqnarray}
\begin{eqnarray}\label{9}
  \frac{p_{1}^{n+1}-p_{1}^{n}}{h} &=& -\frac{1}{6(q_{1}^{n+1}-q_{1}^{n})} \{ [ H(q_{1}^{n+1},q_{2}^{n},q_{3}^{n},p_{1}^{n},p_{2}^{n},p_{3}^{n})
  -H(q_{1}^{n},q_{2}^{n},q_{3}^{n},p_{1}^{n},p_{2}^{n},p_{3}^{n})] \nonumber \\
&& +[H(q_{1}^{n+1},q_{2}^{n+1},q_{3}^{n+1},p_{1}^{n},p_{2}^{n+1},p_{3}^{n+1})-H(q_{1}^{n},q_{2}^{n+1},q_{3}^{n+1},p_{1}^{n},p_{2}^{n+1},p_{3}^{n+1})] \nonumber \\
&& +[H(q_{1}^{n+1},q_{2}^{n+1},q_{3}^{n+1},p_{1}^{n+1},p_{2}^{n+1},p_{3}^{n+1})-H(q_{1}^{n},q_{2}^{n+1},q_{3}^{n+1},p_{1}^{n+1},p_{2}^{n+1},p_{3}^{n+1})] \nonumber \\
&& +[H(q_{1}^{n+1},q_{2}^{n},q_{3}^{n+1},p_{1}^{n},p_{2}^{n},p_{3}^{n+1})-H(q_{1}^{n},q_{2}^{n},q_{3}^{n+1},p_{1}^{n},p_{2}^{n},p_{3}^{n+1})] \nonumber \\
&& +[H(q_{1}^{n+1},q_{2}^{n},q_{3}^{n},p_{1}^{n+1},p_{2}^{n},p_{3}^{n})-H(q_{1}^{n},q_{2}^{n},q_{3}^{n},p_{1}^{n+1},p_{2}^{n},p_{3}^{n})] \nonumber \\
&& +[H(q_{1}^{n+1},q_{2}^{n},q_{3}^{n+1},
p_{1}^{n+1},p_{2}^{n},p_{3}^{n+1})
-H(q_{1}^{n},q_{2}^{n},q_{3}^{n+1},p_{1}^{n+1},p_{2}^{n},p_{3}^{n+1})]
\},
\end{eqnarray}
\begin{eqnarray}\label{10}
  \frac{p_{2}^{n+1}-p_{2}^{n}}{h} &=& -\frac{1}{6(q_{2}^{n+1}-q_{2}^{n})} \{ [ H(q_{1}^{n+1},q_{2}^{n+1},q_{3}^{n},p_{1}^{n+1},p_{2}^{n},p_{3}^{n})
  -H(q_{1}^{n+1},q_{2}^{n},q_{3}^{n},p_{1}^{n+1},p_{2}^{n},p_{3}^{n})] \nonumber \\
&& +[H(q_{1}^{n},q_{2}^{n+1},q_{3}^{n},p_{1}^{n},p_{2}^{n},p_{3}^{n})-H(q_{1}^{n},q_{2}^{n},q_{3}^{n},p_{1}^{n},p_{2}^{n},p_{3}^{n})] \nonumber \\
&& +[H(q_{1}^{n},q_{2}^{n+1},q_{3}^{n},p_{1}^{n},p_{2}^{n+1},p_{3}^{n})-H(q_{1}^{n},q_{2}^{n},q_{3}^{n},p_{1}^{n},p_{2}^{n+1},p_{3}^{n})] \nonumber \\
&& +[H(q_{1}^{n+1},q_{2}^{n+1},q_{3}^{n+1},p_{1}^{n+1},p_{2}^{n},p_{3}^{n+1})-H(q_{1}^{n+1},q_{2}^{n},q_{3}^{n+1},p_{1}^{n+1},p_{2}^{n},p_{3}^{n+1})] \nonumber \\
&& +[H(q_{1}^{n+1},q_{2}^{n+1},q_{3}^{n},p_{1}^{n+1},p_{2}^{n+1},p_{3}^{n})-H(q_{1}^{n+1},q_{2}^{n},q_{3}^{n},p_{1}^{n+1},p_{2}^{n+1},p_{3}^{n})] \nonumber \\
&& +[H(q_{1}^{n+1},q_{2}^{n+1},q_{3}^{n+1},p_{1}^{n+1},
p_{2}^{n+1},p_{3}^{n+1})-H(q_{1}^{n+1},q_{2}^{n},q_{3}^{n+1},p_{1}^{n+1},
p_{2}^{n+1},p_{3}^{n+1})] \},
\end{eqnarray}
\begin{eqnarray}\label{11}
\frac{p_{3}^{n+1}-p_{3}^{n}}{h} &=&
-\frac{1}{6(q_{3}^{n+1}-q_{3}^{n})} \{ [
H(q_{1}^{n+1},q_{2}^{n+1},q_{3}^{n+1},p_{1}^{n+1},p_{2}^{n+1},p_{3}^{n})-H(q_{1}^{n+1},q_{2}^{n+1},q_{3}^{n},p_{1}^{n+1}, p_{2}^{n+1},p_{3}^{n})] \nonumber \\
&& + [H(q_{1}^{n},q_{2}^{n+1},q_{3}^{n+1},p_{1}^{n},p_{2}^{n+1},p_{3}^{n})-H(q_{1}^{n},q_{2}^{n+1},q_{3}^{n},p_{1}^{n},p_{2}^{n+1},p_{3}^{n})] \nonumber \\
&& + [H(q_{1}^{n},q_{2}^{n+1},q_{3}^{n+1},p_{1}^{n},p_{2}^{n+1},p_{3}^{n+1})-H(q_{1}^{n},q_{2}^{n+1},q_{3}^{n},p_{1}^{n},p_{2}^{n+1},p_{3}^{n+1})] \nonumber \\
&& + [H(q_{1}^{n},q_{2}^{n},q_{3}^{n+1},p_{1}^{n},p_{2}^{n},p_{3}^{n})-H(q_{1}^{n},q_{2}^{n},q_{3}^{n},p_{1}^{n},p_{2}^{n},p_{3}^{n})] \nonumber \\
&& + [H(q_{1}^{n+1},q_{2}^{n+1},q_{3}^{n+1},p_{1}^{n+1},p_{2}^{n+1},p_{3}^{n+1})-H(q_{1}^{n+1},q_{2}^{n+1},q_{3}^{n},p_{1}^{n+1},p_{2}^{n+1},p_{3}^{n+1})] \nonumber \\
&& + [H(q_{1}^{n},q_{2}^{n},q_{3}^{n+1},
p_{1}^{n},p_{2}^{n},p_{3}^{n+1})
-H(q_{1}^{n},q_{2}^{n},q_{3}^{n},p_{1}^{n},p_{2}^{n},p_{3}^{n+1})]
\}.
\end{eqnarray}

As we adjust Equation (A1) to Equation (7), we apply the Taylor
expansion to Equation (B1) and obtain
\begin{eqnarray}\label{26}
  q_{1}^{n+1} & = & q_{1}^{n}+ \frac{h}{6} \left\{ (\frac{\partial H^n}{\partial p_{1}}+\frac{1}{2} \frac{\partial ^{2} H^n}{\partial p_{1}^{2}}\Delta p_{1})
  + \left[\frac{\partial}{\partial p_{1}}H(q_{1}^{n+1},q_{2}^{n},q_{3}^{n},p_{1}^{n},p_{2}^{n},p_{3}^{n})
  +\frac{1}{2} \frac{\partial ^{2} H^n}{\partial p_{1}^{2}}\Delta p_{1} \right] \right. \nonumber \\
  && + \left[\frac{\partial}{\partial p_{1}}H(q_{1}^{n+1},q_{2}^{n+1},q_{3}^{n+1},p_{1}^{n},p_{2}^{n+1},p_{3}^{n+1})
  +\frac{1}{2} \frac{\partial ^{2} H^n}{\partial p_{1}^{2}}\Delta p_{1} \right] \nonumber \\
  && + \left[\frac{\partial}{\partial p_{1}}H(q_{1}^{n},q_{2}^{n+1},q_{3}^{n+1},p_{1}^{n},p_{2}^{n+1},p_{3}^{n+1})
  +\frac{1}{2} \frac{\partial ^{2} H^n}{\partial p_{1}^{2}}\Delta p_{1} \right] \nonumber \\
  && + \left[\frac{\partial}{\partial p_{1}}H(q_{1}^{n+1},q_{2}^{n},q_{3}^{n+1},p_{1}^{n},p_{2}^{n},p_{3}^{n+1})
  +\frac{1}{2} \frac{\partial ^{2} H^n}{\partial p_{1}^{2}}\Delta p_{1} \right] \nonumber \\
  && \left. + \left[\frac{\partial}{\partial p_{1}}H(q_{1}^{n},q_{2}^{n},q_{3}^{n+1},p_{1}^{n},p_{2}^{n},p_{3}^{n+1})
  +\frac{1}{2} \frac{\partial ^{2} H^n}{\partial p_{1}^{2}}\Delta p_{1} \right] \right\} \nonumber \\
  & = & q_{1}^{n}+ h \frac{\partial H^n}{\partial p_{1}}+\frac{h}{2} \frac{\partial ^{2} H^n}{\partial p_{1}^{2}}\Delta p_{1}
  +\frac{h}{6} \left[ \frac{\partial ^{2}}{\partial p_{1} \partial q_{1}}\Delta q_{1}+ (\frac{\partial ^{2}}{\partial p_{1} \partial q_{1}}\Delta q_{1}
  +\frac{\partial ^{2}}{\partial p_{1} \partial q_{2}}\Delta q_{2}
   \right. \nonumber \\
  && +\frac{\partial ^{2}}{\partial p_{1} \partial q_{3}}\Delta q_{3} +\frac{\partial ^{2}}{\partial p_{1} \partial p_{2}}\Delta p_{2} +\frac{\partial ^{2}}{\partial p_{1} \partial p_{3}}\Delta p_{3})
  + (\frac{\partial ^{2}}{\partial p_{1} \partial q_{2}}\Delta q_{2}
  +\frac{\partial ^{2}}{\partial p_{1} \partial q_{3}}\Delta q_{3}  \nonumber \\
  && +\frac{\partial ^{2}}{\partial p_{1}
  \partial p_{2}}\Delta p_{2}+\frac{\partial ^{2}}{\partial p_{1} \partial p_{3}}\Delta p_{3})
  + (\frac{\partial ^{2}}{\partial p_{1} \partial q_{1}}\Delta q_{1}
  +\frac{\partial ^{2}}{\partial p_{1} \partial q_{3}}\Delta q_{3}+\frac{\partial ^{2}}{\partial p_{1} \partial p_{3}}\Delta p_{3})
   \nonumber \\
  && \left. +(\frac{\partial ^{2}}{\partial p_{1} \partial q_{3}}\Delta q_{3} +\frac{\partial ^{2}}{\partial p_{1} \partial p_{3}}\Delta p_{3}) \right] H^n \nonumber \\
  & = & q_{1}^{n}+ h\frac{\partial H^n}{\partial p_{1}}+\frac{h}{2}\sum^{3}_{j=1}(\frac{\partial }{\partial p_{j}}\Delta p_{j}
  +\frac{\partial}{\partial q_{j}}\Delta q_{j})\frac{\partial H^n}{\partial  p_{1}}\nonumber \\
  && +\frac{h}{6}\left[ \frac{\partial}{\partial q_{3}}\Delta q_{3}
  + \frac{\partial}{\partial p_{3}}\Delta p_{3}- \frac{\partial}{\partial q_{2}}\Delta q_{2}- \frac{\partial}{\partial p_{2}}\Delta p_{2} \right] \frac{\partial H^n}{\partial
  p_{1}}+ \mathcal{O}(h^{3}).
\end{eqnarray}

\section{New complex discrete method of the derivatives}

\begin{eqnarray}\label{17}
  \frac{q_{1}^{n+1}-q_{1}^{n}}{h} &=& \frac{1}{6(p_{1}^{n+1}-p_{1}^{n})} \{ [H(q_{1}^{n},q_{2}^{n},q_{3}^{n},p_{1}^{n+1},p_{2}^{n},p_{3}^{n})
  -H(q_{1}^{n},q_{2}^{n},q_{3}^{n},p_{1}^{n},p_{2}^{n},p_{3}^{n})]  \nonumber \\
 && +[H(q_{1}^{n+1},q_{2}^{n},q_{3}^{n},p_{1}^{n+1},p_{2}^{n},p_{3}^{n})-H(q_{1}^{n+1},q_{2}^{n},q_{3}^{n},p_{1}^{n},p_{2}^{n},p_{3}^{n})] \nonumber \\
 && +[H(q_{1}^{n},q_{2}^{n},q_{3}^{n+1},p_{1}^{n+1},p_{2}^{n},p_{3}^{n+1})-H(q_{1}^{n},q_{2}^{n},q_{3}^{n+1},p_{1}^{n},p_{2}^{n},p_{3}^{n+1})] \nonumber \\
 && +[H(q_{1}^{n+1},q_{2}^{n+1},q_{3}^{n},p_{1}^{n+1},p_{2}^{n+1},p_{3}^{n})-H(q_{1}^{n+1},q_{2}^{n+1},q_{3}^{n},p_{1}^{n},p_{2}^{n+1},p_{3}^{n})] \nonumber \\
 && +[H(q_{1}^{n},q_{2}^{n+1},q_{3}^{n+1},p_{1}^{n+1},p_{2}^{n+1},p_{3}^{n+1})-H(q_{1}^{n},q_{2}^{n+1},q_{3}^{n+1},p_{1}^{n},p_{2}^{n+1},p_{3}^{n+1})] \nonumber \\
 && +[H(q_{1}^{n+1},q_{2}^{n+1},q_{3}^{n+1},p_{1}^{n+1},p_{2}^{n+1},p_{3}^{n+1})-H(q_{1}^{n+1},q_{2}^{n+1},q_{3}^{n+1},p_{1}^{n},p_{2}^{n+1},p_{3}^{n+1})] \},
\end{eqnarray}
\begin{eqnarray}\label{18}
  \frac{q_{2}^{n+1}-q_{2}^{n}}{h} &=& \frac{1}{6(p_{2}^{n+1}-p_{2}^{n})} \{ [H(q_{1}^{n},q_{2}^{n},q_{3}^{n},p_{1}^{n},p_{2}^{n+1},p_{3}^{n})
  -H(q_{1}^{n},q_{2}^{n},q_{3}^{n},p_{1}^{n},p_{2}^{n},p_{3}^{n})] \nonumber \\
 && +[H(q_{1}^{n},q_{2}^{n+1},q_{3}^{n},p_{1}^{n},p_{2}^{n+1},p_{3}^{n})-H(q_{1}^{n},q_{2}^{n+1},q_{3}^{n},p_{1}^{n},p_{2}^{n},p_{3}^{n})] \nonumber \\
 && +[H(q_{1}^{n+1},q_{2}^{n},q_{3}^{n},p_{1}^{n+1},p_{2}^{n+1},p_{3}^{n})-H(q_{1}^{n+1},q_{2}^{n},q_{3}^{n},p_{1}^{n+1},p_{2}^{n},p_{3}^{n})] \nonumber \\
 && +[H(q_{1}^{n},q_{2}^{n+1},q_{3}^{n+1},p_{1}^{n},p_{2}^{n+1},p_{3}^{n+1})-H(q_{1}^{n},q_{2}^{n+1},q_{3}^{n+1},p_{1}^{n},p_{2}^{n},p_{3}^{n+1})] \nonumber \\
 && +[H(q_{1}^{n+1},q_{2}^{n},q_{3}^{n+1},p_{1}^{n+1},p_{2}^{n+1},p_{3}^{n+1})-H(q_{1}^{n+1},q_{2}^{n},q_{3}^{n+1},p_{1}^{n+1},p_{2}^{n},p_{3}^{n+1})] \nonumber \\
 && +[H(q_{1}^{n+1},q_{2}^{n+1},q_{3}^{n+1},p_{1}^{n+1},p_{2}^{n+1},p_{3}^{n+1})-H(q_{1}^{n+1},q_{2}^{n+1},q_{3}^{n+1},p_{1}^{n+1},p_{2}^{n},p_{3}^{n+1})] \},
\end{eqnarray}
\begin{eqnarray}\label{19}
  \frac{q_{3}^{n+1}-q_{3}^{n}}{h} &=& \frac{1}{6(p_{3}^{n+1}-p_{3}^{n})} \{ [H(q_{1}^{n},q_{2}^{n},q_{3}^{n},p_{1}^{n},p_{2}^{n},p_{3}^{n+1})
  -H(q_{1}^{n},q_{2}^{n},q_{3}^{n},p_{1}^{n},p_{2}^{n},p_{3}^{n})] \nonumber \\
 && +[H(q_{1}^{n},q_{2}^{n},q_{3}^{n+1},p_{1}^{n},p_{2}^{n},p_{3}^{n+1})-H(q_{1}^{n},q_{2}^{n},q_{3}^{n+1},p_{1}^{n},p_{2}^{n},p_{3}^{n})] \nonumber \\
 && +[H(q_{1}^{n},q_{2}^{n+1},q_{3}^{n},p_{1}^{n},p_{2}^{n+1},p_{3}^{n+1})-H(q_{1}^{n},q_{2}^{n+1},q_{3}^{n},p_{1}^{n},p_{2}^{n+1},p_{3}^{n})] \nonumber \\
 && +[H(q_{1}^{n+1},q_{2}^{n},q_{3}^{n+1},p_{1}^{n+1},p_{2}^{n},p_{3}^{n+1})-H(q_{1}^{n+1},q_{2}^{n},q_{3}^{n+1},p_{1}^{n+1},p_{2}^{n},p_{3}^{n})] \nonumber \\
 && +[H(q_{1}^{n+1},q_{2}^{n+1},q_{3}^{n},p_{1}^{n+1},p_{2}^{n+1},p_{3}^{n+1})-H(q_{1}^{n+1},q_{2}^{n+1},q_{3}^{n},p_{1}^{n+1},p_{2}^{n+1},p_{3}^{n})] \nonumber \\
 && +[H(q_{1}^{n+1},q_{2}^{n+1},q_{3}^{n+1},p_{1}^{n+1},p_{2}^{n+1},p_{3}^{n+1})-H(q_{1}^{n+1},q_{2}^{n+1},q_{3}^{n+1},p_{1}^{n+1},p_{2}^{n+1},p_{3}^{n})] \},
\end{eqnarray}
\begin{eqnarray}\label{20}
  \frac{p_{1}^{n+1}-p_{1}^{n}}{h} &=& -\frac{1}{6(q_{1}^{n+1}-q_{1}^{n})} \{[ H(q_{1}^{n+1},q_{2}^{n},q_{3}^{n},p_{1}^{n},p_{2}^{n},p_{3}^{n})
  -H(q_{1}^{n},q_{2}^{n},q_{3}^{n},p_{1}^{n},p_{2}^{n},p_{3}^{n})] \nonumber \\
 && +[H(q_{1}^{n+1},q_{2}^{n},q_{3}^{n},p_{1}^{n+1},p_{2}^{n},p_{3}^{n})-H(q_{1}^{n},q_{2}^{n},q_{3}^{n},p_{1}^{n+1},p_{2}^{n},p_{3}^{n})] \nonumber \\
 && +[H(q_{1}^{n+1},q_{2}^{n+1},q_{3}^{n},p_{1}^{n},p_{2}^{n+1},p_{3}^{n})-H(q_{1}^{n},q_{2}^{n+1},q_{3}^{n},p_{1}^{n},p_{2}^{n+1},p_{3}^{n})] \nonumber \\
 && +[H(q_{1}^{n+1},q_{2}^{n},q_{3}^{n+1},p_{1}^{n+1},p_{2}^{n},p_{3}^{n+1})-H(q_{1}^{n},q_{2}^{n},q_{3}^{n+1},p_{1}^{n+1},p_{2}^{n},p_{3}^{n+1})] \nonumber \\
 && +[H(q_{1}^{n+1},q_{2}^{n+1},q_{3}^{n+1},p_{1}^{n},p_{2}^{n+1},p_{3}^{n+1})-H(q_{1}^{n},q_{2}^{n+1},q_{3}^{n+1},p_{1}^{n},p_{2}^{n+1},p_{3}^{n+1})] \nonumber \\
 && +[H(q_{1}^{n+1},q_{2}^{n+1},q_{3}^{n+1},p_{1}^{n+1},p_{2}^{n+1},p_{3}^{n+1})-H(q_{1}^{n},q_{2}^{n+1},q_{3}^{n+1},p_{1}^{n+1},p_{2}^{n+1},p_{3}^{n+1})] \},
\end{eqnarray}
\begin{eqnarray}\label{21}
  \frac{p_{2}^{n+1}-p_{2}^{n}}{h} &=& -\frac{1}{6(q_{2}^{n+1}-q_{2}^{n})} \{[ H(q_{1}^{n},q_{2}^{n+1},q_{3}^{n},p_{1}^{n},p_{2}^{n},p_{3}^{n})
  -H(q_{1}^{n},q_{2}^{n},q_{3}^{n},p_{1}^{n},p_{2}^{n},p_{3}^{n})] \nonumber \\
 && +[H(q_{1}^{n},q_{2}^{n+1},q_{3}^{n},p_{1}^{n},p_{2}^{n+1},p_{3}^{n})-H(q_{1}^{n},q_{2}^{n},q_{3}^{n},p_{1}^{n},p_{2}^{n+1},p_{3}^{n})] \nonumber \\
 && +[H(q_{1}^{n},q_{2}^{n+1},q_{3}^{n+1},p_{1}^{n},p_{2}^{n},p_{3}^{n+1})-H(q_{1}^{n},q_{2}^{n},q_{3}^{n+1},p_{1}^{n},p_{2}^{n},p_{3}^{n+1})] \nonumber \\
 && +[H(q_{1}^{n+1},q_{2}^{n+1},q_{3}^{n},p_{1}^{n+1},p_{2}^{n+1},p_{3}^{n})-H(q_{1}^{n+1},q_{2}^{n},q_{3}^{n},p_{1}^{n+1},p_{2}^{n+1},p_{3}^{n})] \nonumber \\
 && +[H(q_{1}^{n+1},q_{2}^{n+1},q_{3}^{n+1},p_{1}^{n+1},p_{2}^{n},p_{3}^{n+1})-H(q_{1}^{n+1},q_{2}^{n},q_{3}^{n+1},p_{1}^{n+1},p_{2}^{n},p_{3}^{n+1})] \nonumber \\
 && +H(q_{1}^{n+1},q_{2}^{n+1},q_{3}^{n+1},p_{1}^{n+1},p_{2}^{n+1},p_{3}^{n+1})-H(q_{1}^{n+1},q_{2}^{n},q_{3}^{n+1},p_{1}^{n+1},p_{2}^{n+1},p_{3}^{n+1})] \},
\end{eqnarray}
\begin{eqnarray}\label{22}
\frac{p_{3}^{n+1}-p_{3}^{n}}{h} &=&
-\frac{1}{6(q_{3}^{n+1}-q_{3}^{n})} \{[
H(q_{1}^{n},q_{2}^{n},q_{3}^{n+1},p_{1}^{n},p_{2}^{n},p_{3}^{n})
-H(q_{1}^{n},q_{2}^{n},q_{3}^{n},p_{1}^{n},p_{2}^{n},p_{3}^{n})] \nonumber \\
 &&+ [H(q_{1}^{n},q_{2}^{n},q_{3}^{n+1},p_{1}^{n},p_{2}^{n},p_{3}^{n+1})-H(q_{1}^{n},q_{2}^{n},q_{3}^{n},p_{1}^{n},p_{2}^{n},p_{3}^{n+1})] \nonumber \\
 &&+ [H(q_{1}^{n+1},q_{2}^{n},q_{3}^{n+1},p_{1}^{n+1},p_{2}^{n},p_{3}^{n})-H(q_{1}^{n+1},q_{2}^{n},q_{3}^{n},p_{1}^{n+1},p_{2}^{n},p_{3}^{n})] \nonumber \\
 &&+ [H(q_{1}^{n},q_{2}^{n+1},q_{3}^{n+1},p_{1}^{n},p_{2}^{n+1},p_{3}^{n+1})-H(q_{1}^{n},q_{2}^{n+1},q_{3}^{n},p_{1}^{n},p_{2}^{n+1},p_{3}^{n+1})] \nonumber \\
 &&+ [H(q_{1}^{n+1},q_{2}^{n+1},q_{3}^{n+1},p_{1}^{n+1},p_{2}^{n+1},p_{3}^{n})-H(q_{1}^{n+1},q_{2}^{n+1},q_{3}^{n},p_{1}^{n+1},p_{2}^{n+1},p_{3}^{n})] \nonumber \\
 && + H(q_{1}^{n+1},q_{2}^{n+1},q_{3}^{n+1},p_{1}^{n+1},p_{2}^{n+1},p_{3}^{n+1})-H(q_{1}^{n+1},q_{2}^{n+1},q_{3}^{n},p_{1}^{n+1},p_{2}^{n+1},p_{3}^{n+1})] \}.
\end{eqnarray}

\newpage

\begin{figure*}%[tbph]
\center{
\includegraphics[scale=0.3]{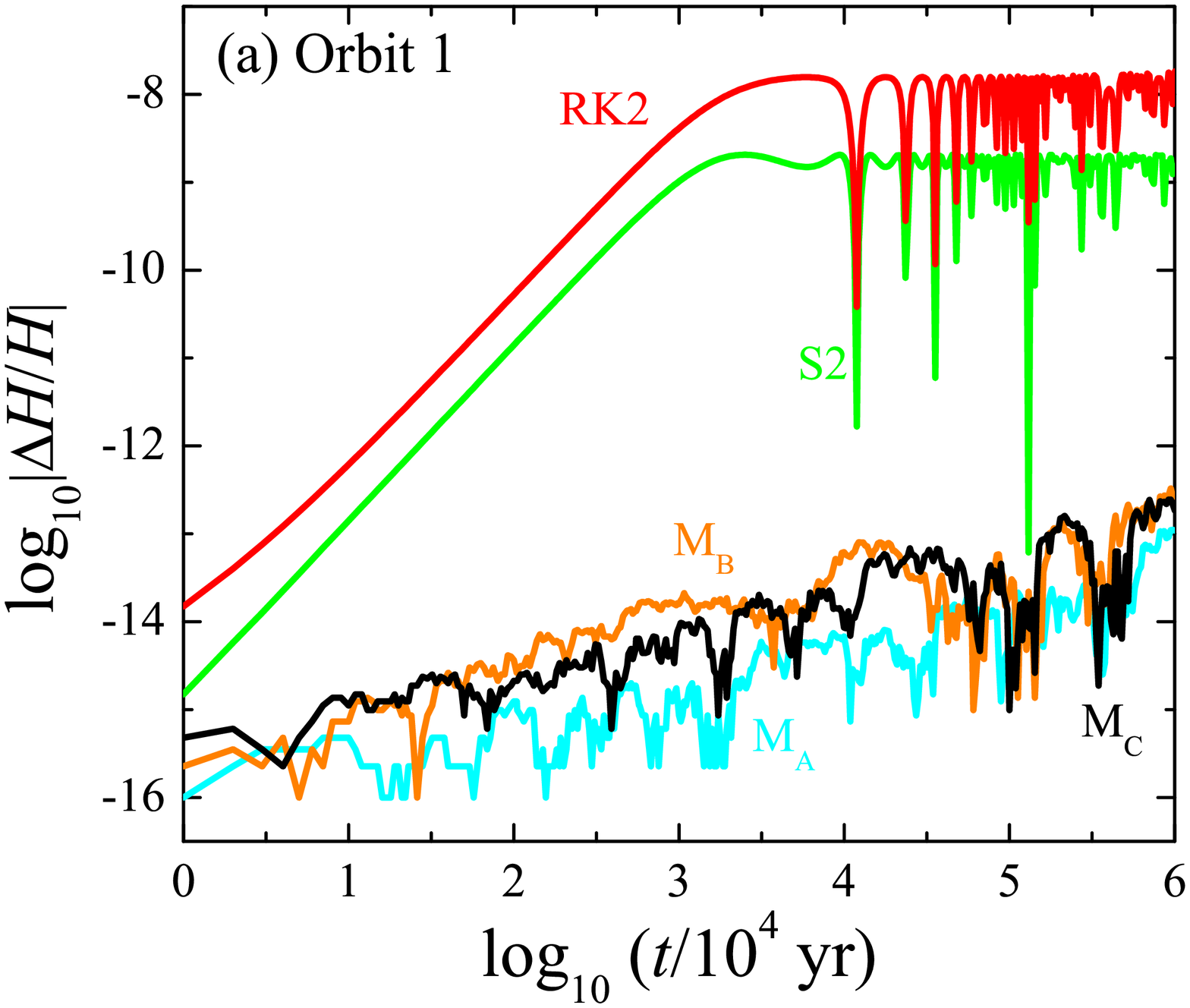}
\includegraphics[scale=0.3]{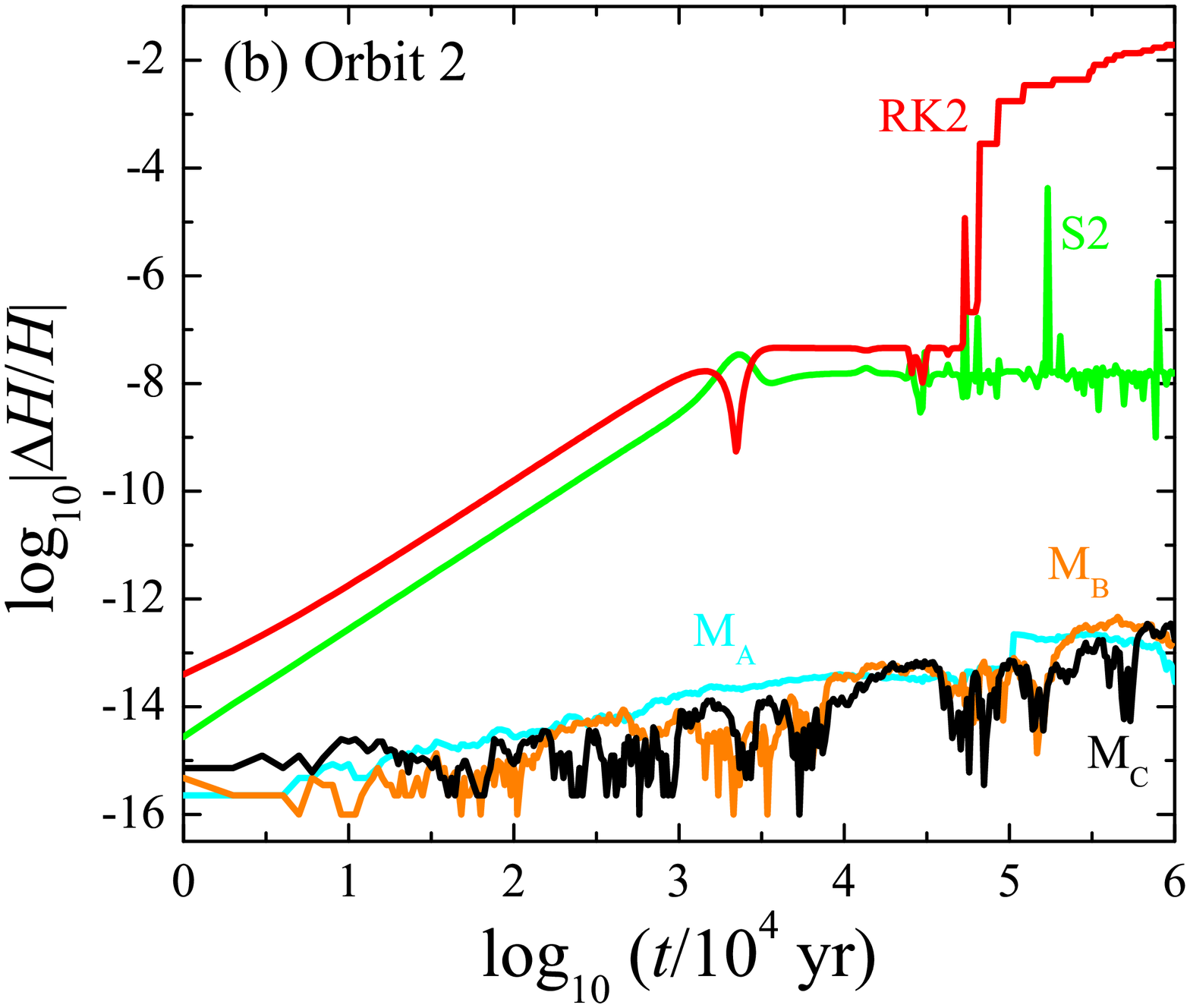}
\includegraphics[scale=0.3]{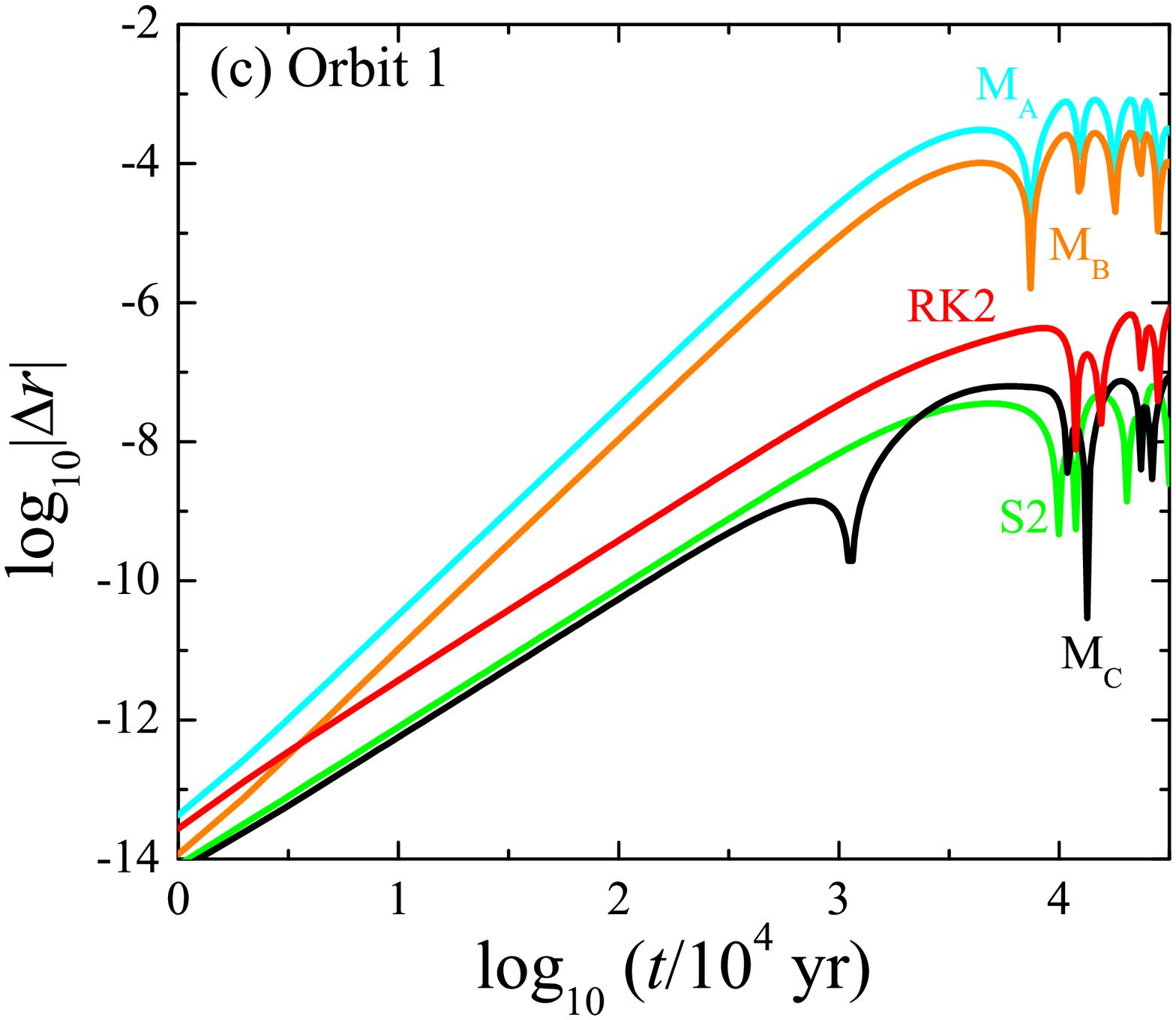}
\includegraphics[scale=0.3]{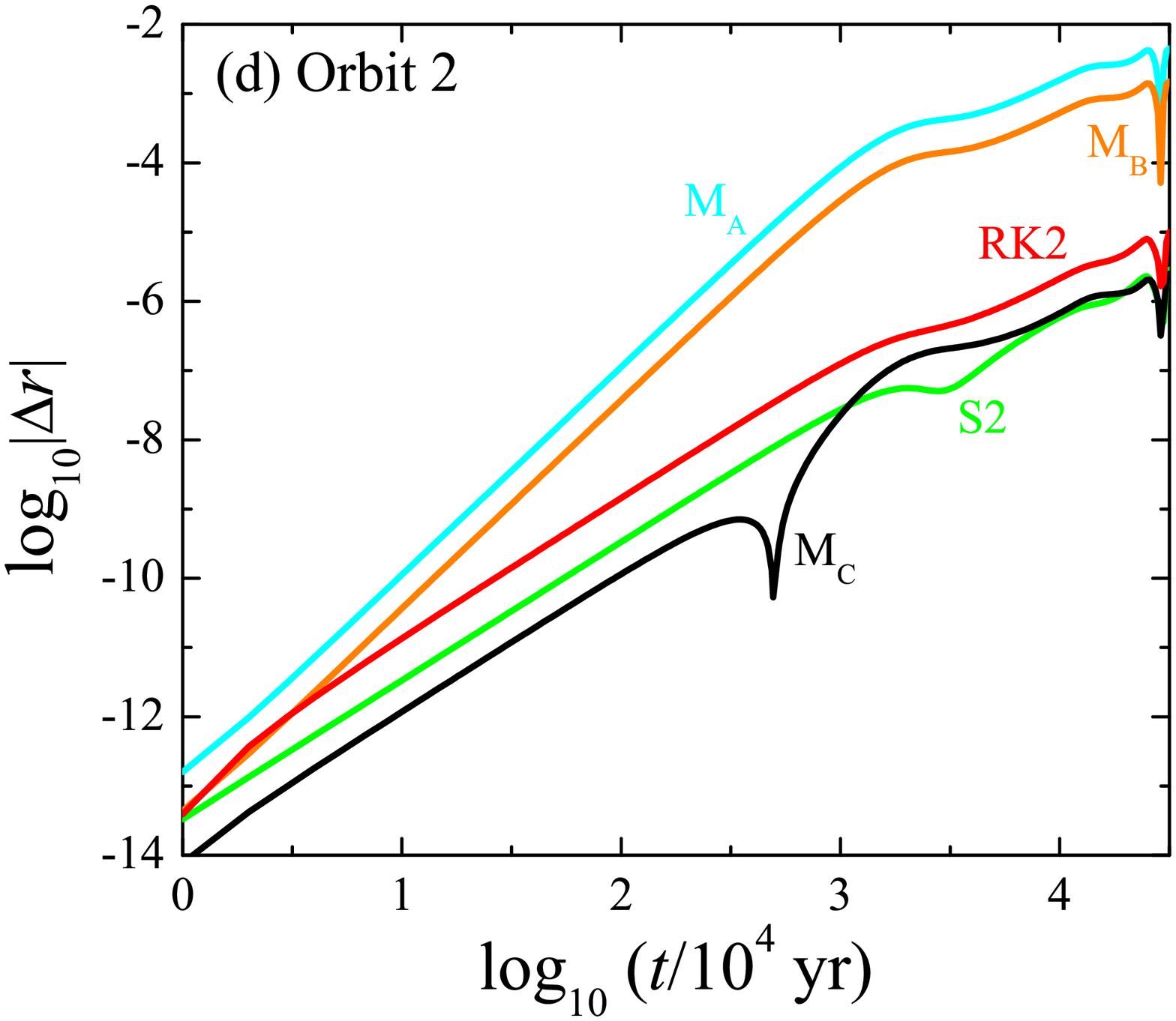}
\caption{(a) and (b): Relative Hamiltonian errors for five
methods.  The initial conditions and parameters of two orbits are
$H=450$, $b=1$, $\lambda=y=0$, $x=3$, $z=0.1$, and $p_x=p_z=0$.
The other parameters are $\alpha=1$ and $M_{n}=10$ for Orbit 1,
while $\alpha=0.1$ and $M_{n}=400$ for Orbit 2.
In the accuracy of the Hamiltonian, Method S2 is
better than Method RK2 and remains bounded; the three
energy-conversing methods $M_A$, $M_B$ and $M_C$ are almost the
consistently best. (c) and (d): Absolute position errors for the
five methods. $M_A$ and $M_B$ have almost the
same accuracies in the positions, and $M_C$ and S2 do. The
accuracy for $M_C$ is several orders of magnitude better than that
for $M_B$. This shows that $M_B$ gives a first-order accuracy to
the numerical solutions, and $M_C$ yields a second-order
accuracy. }} \label{fig1}
\end{figure*}

\begin{figure*}%[tbph]
\center{
\includegraphics[scale=0.21]{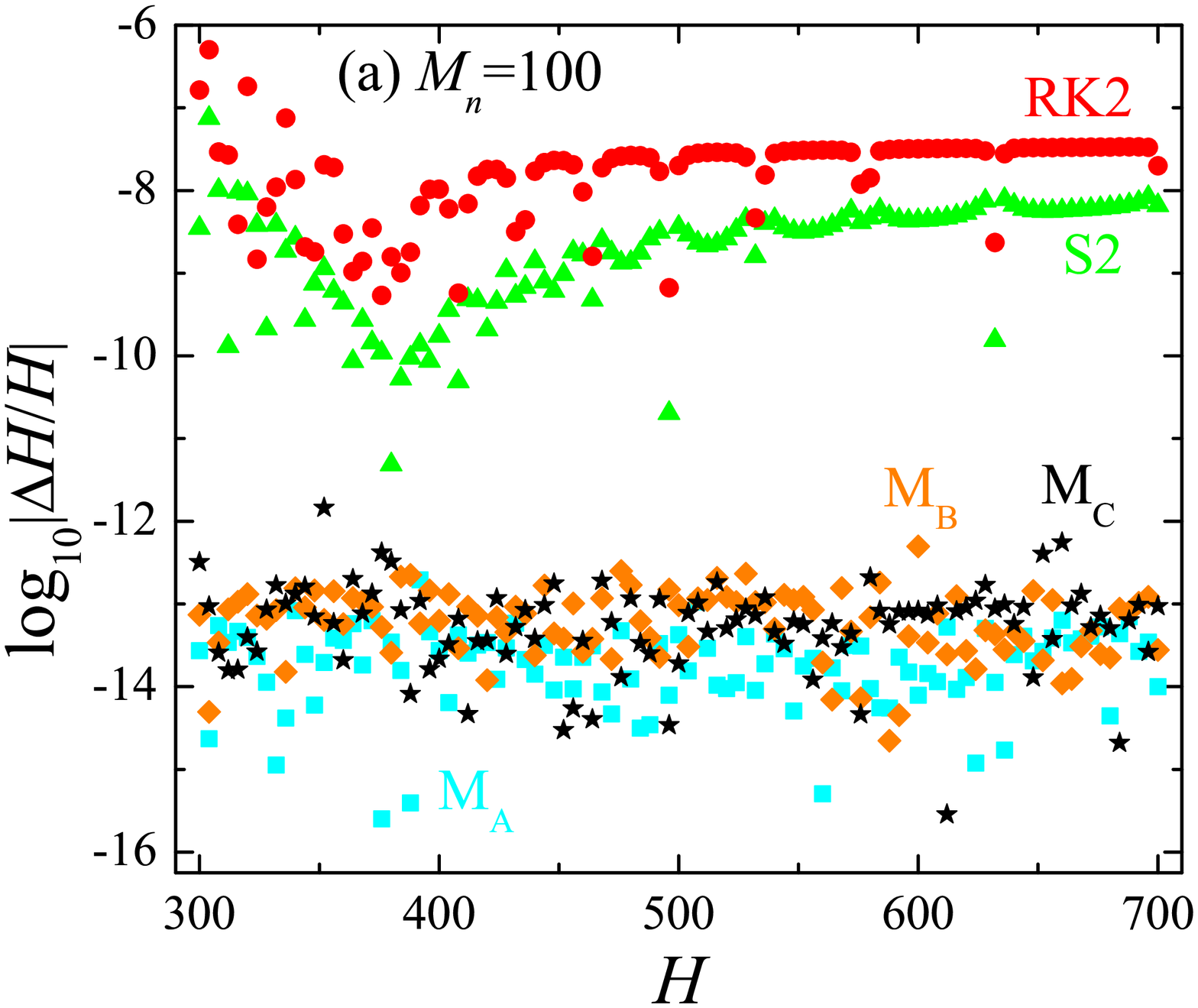}
\includegraphics[scale=0.21]{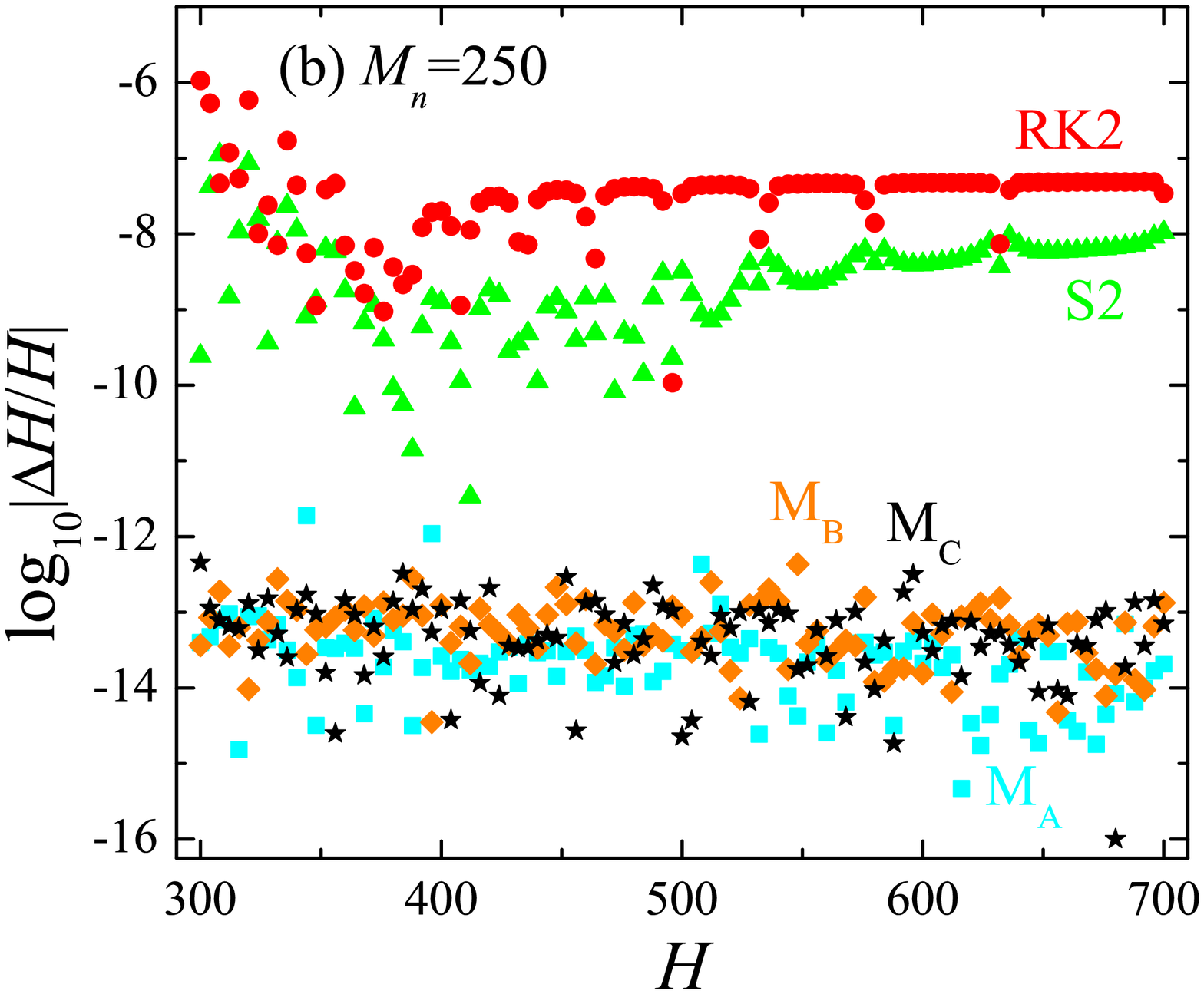}
\includegraphics[scale=0.21]{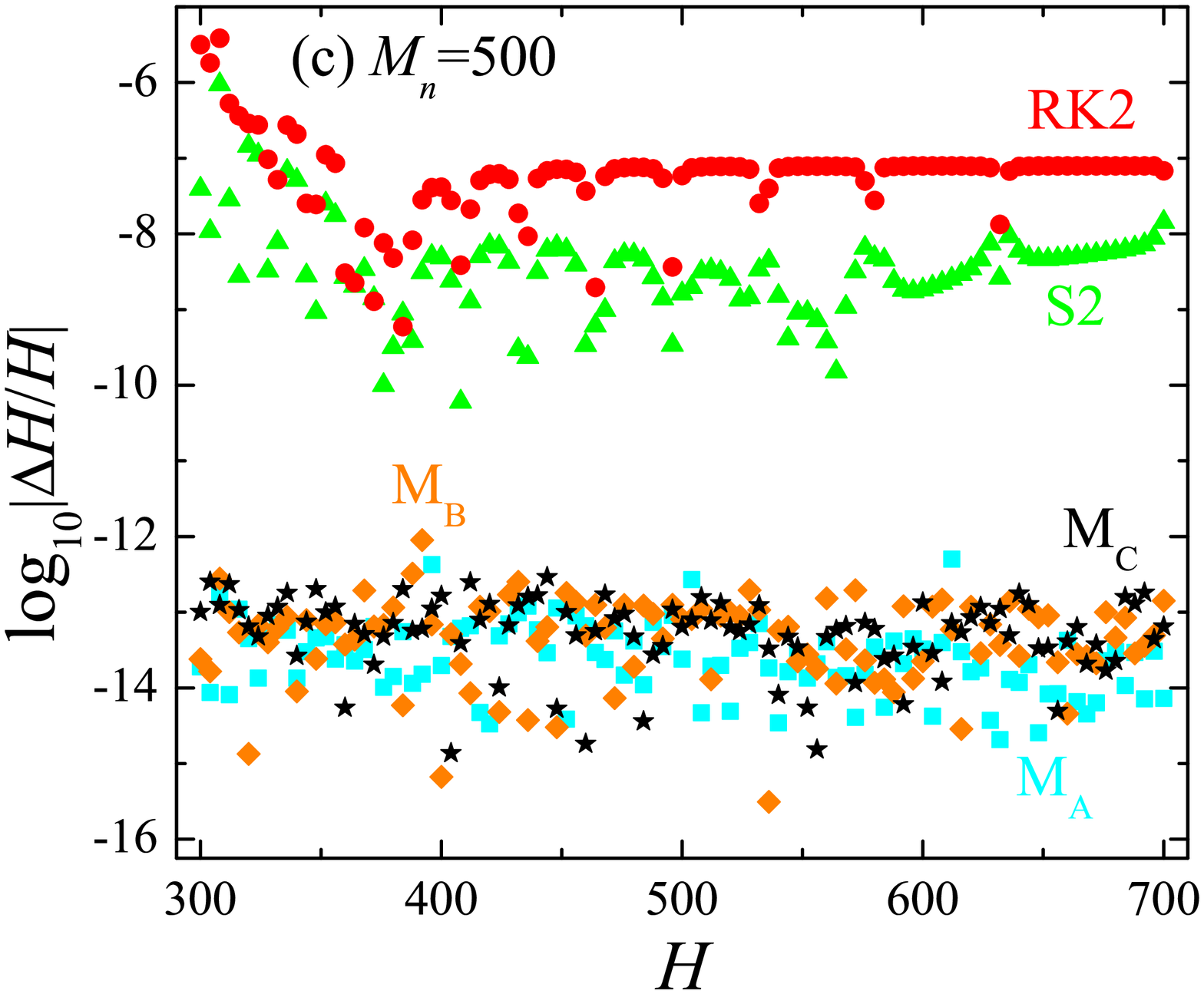}
\includegraphics[scale=0.21]{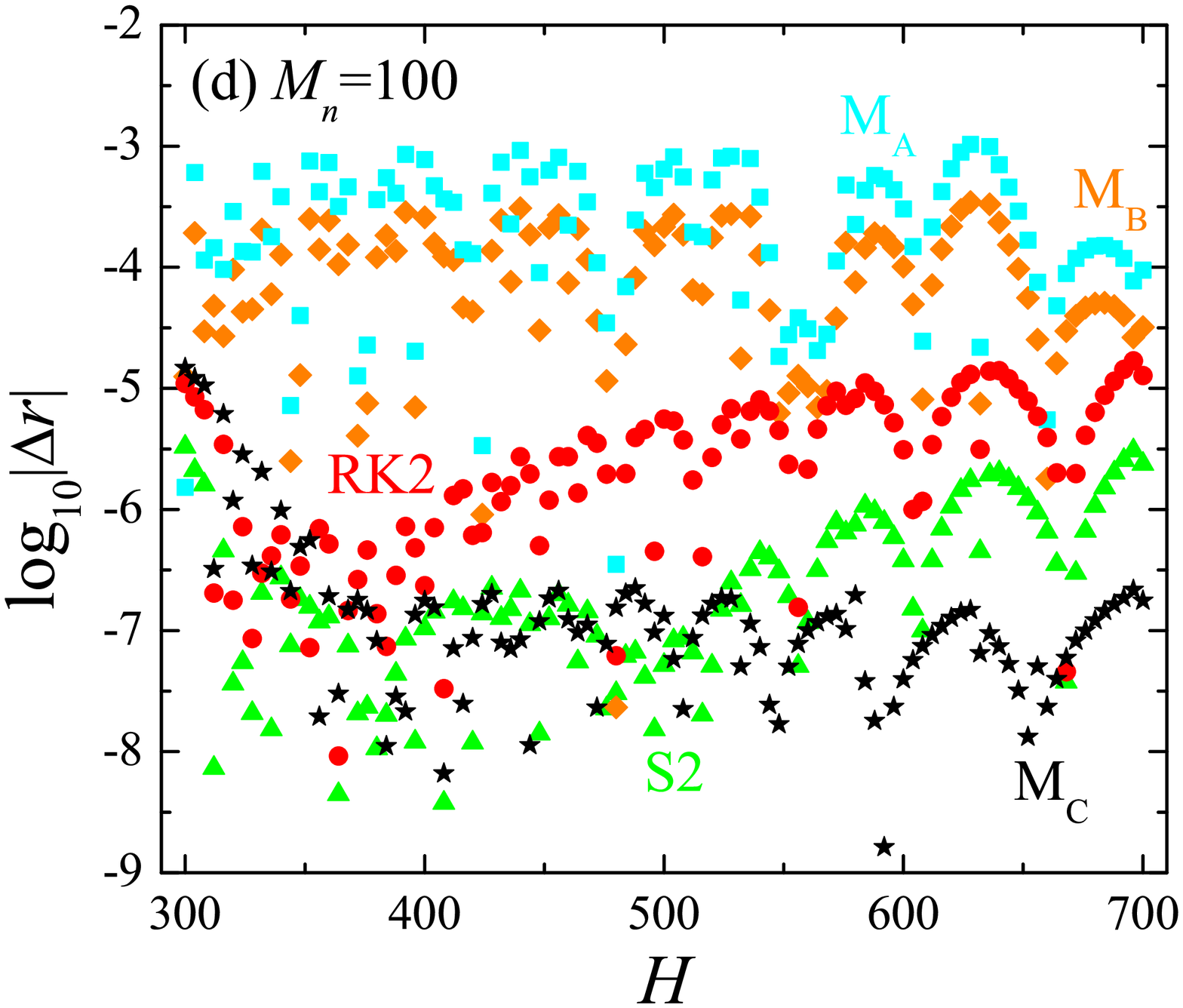}
\includegraphics[scale=0.21]{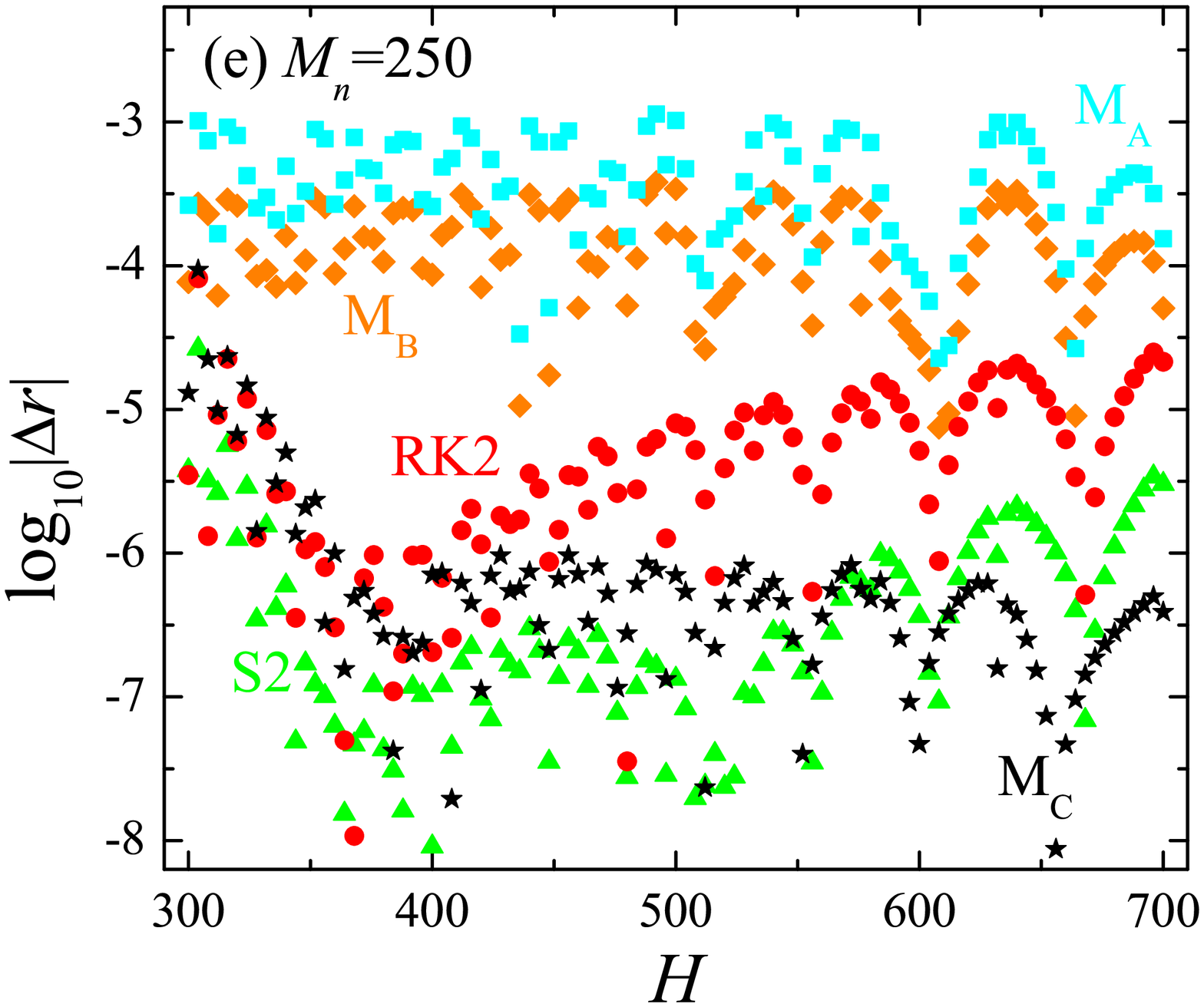}
\includegraphics[scale=0.21]{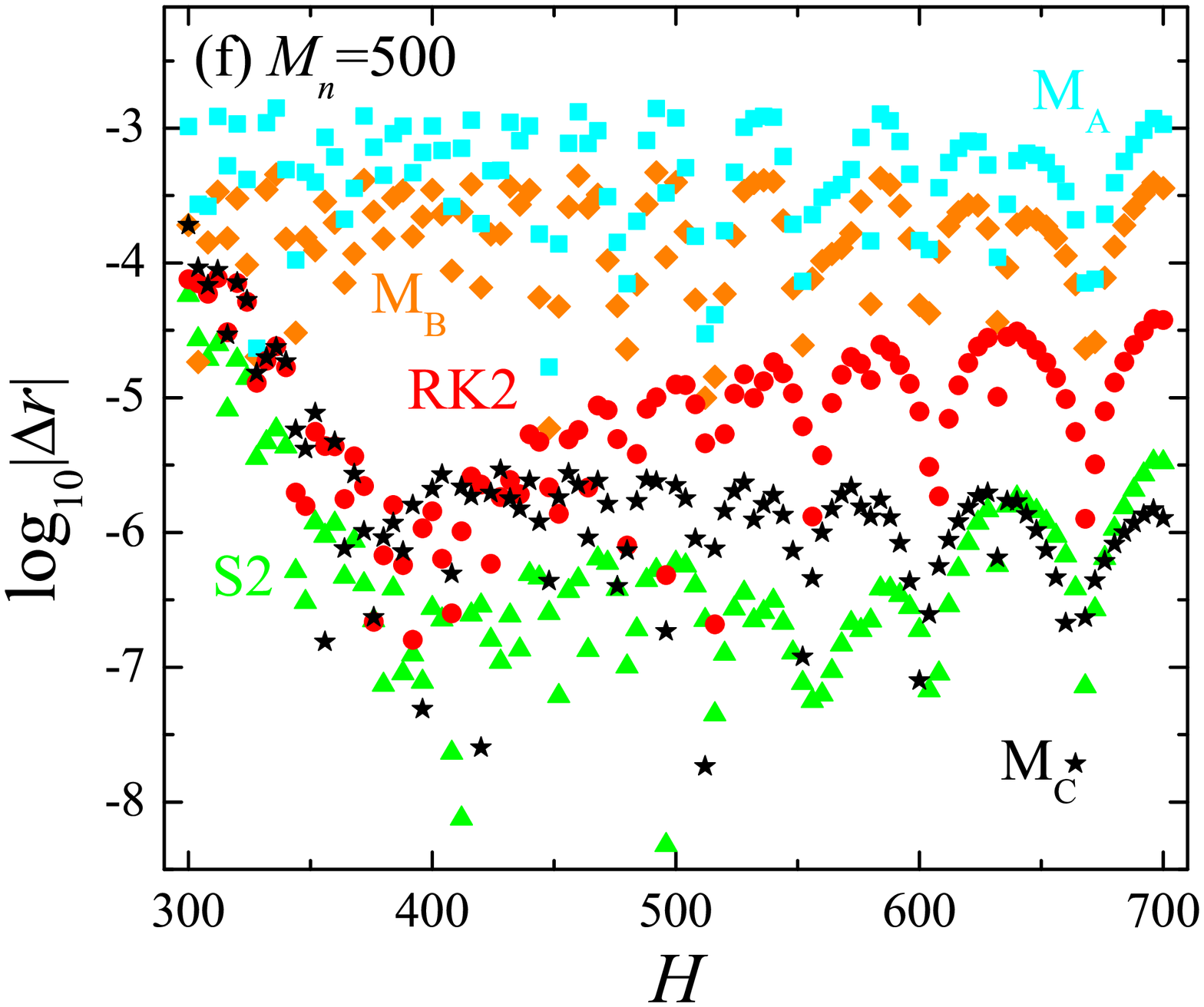}
\caption{(a)-(c): Dependence of relative Hamiltonian errors on the
Hamiltonian $H$ for three values of mass parameter $M_{n}$.
(d)-(f): Dependence of absolute position errors on the
Hamiltonian. The other parameters are $\lambda=0.001$ and
$\alpha=b=1$, and the initial conditions are those of Orbit 1.
Each error is obtained after $10^{5}$ integration steps.
$M_A$, $M_B$ and $M_C$ have almost the
consistently best performance in the Hamiltonian errors, and S2 is
better than RK2. On the other hand, $M_C$ and S2 are drastically
superior to $M_A$ and $M_B$ in the accuracies of the solutions.
This shows again that $M_B$ and $M_C$ have the same performance in
the conservation of energy, but do not have in the accuracies of
the solutions.  }} \label{fig2}
\end{figure*}

\begin{figure*}%[tbph]
\center{
\includegraphics[scale=0.21]{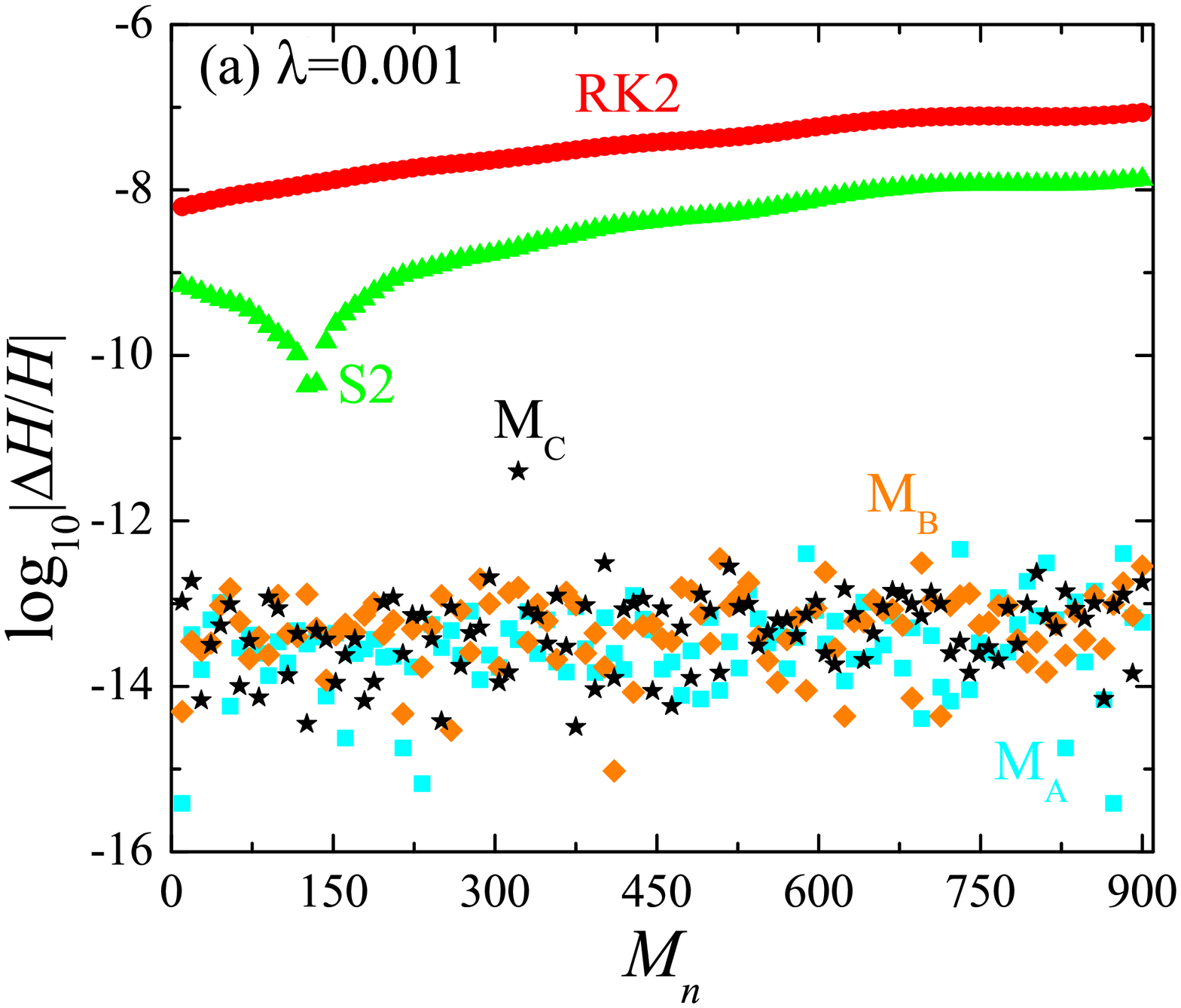}
\includegraphics[scale=0.21]{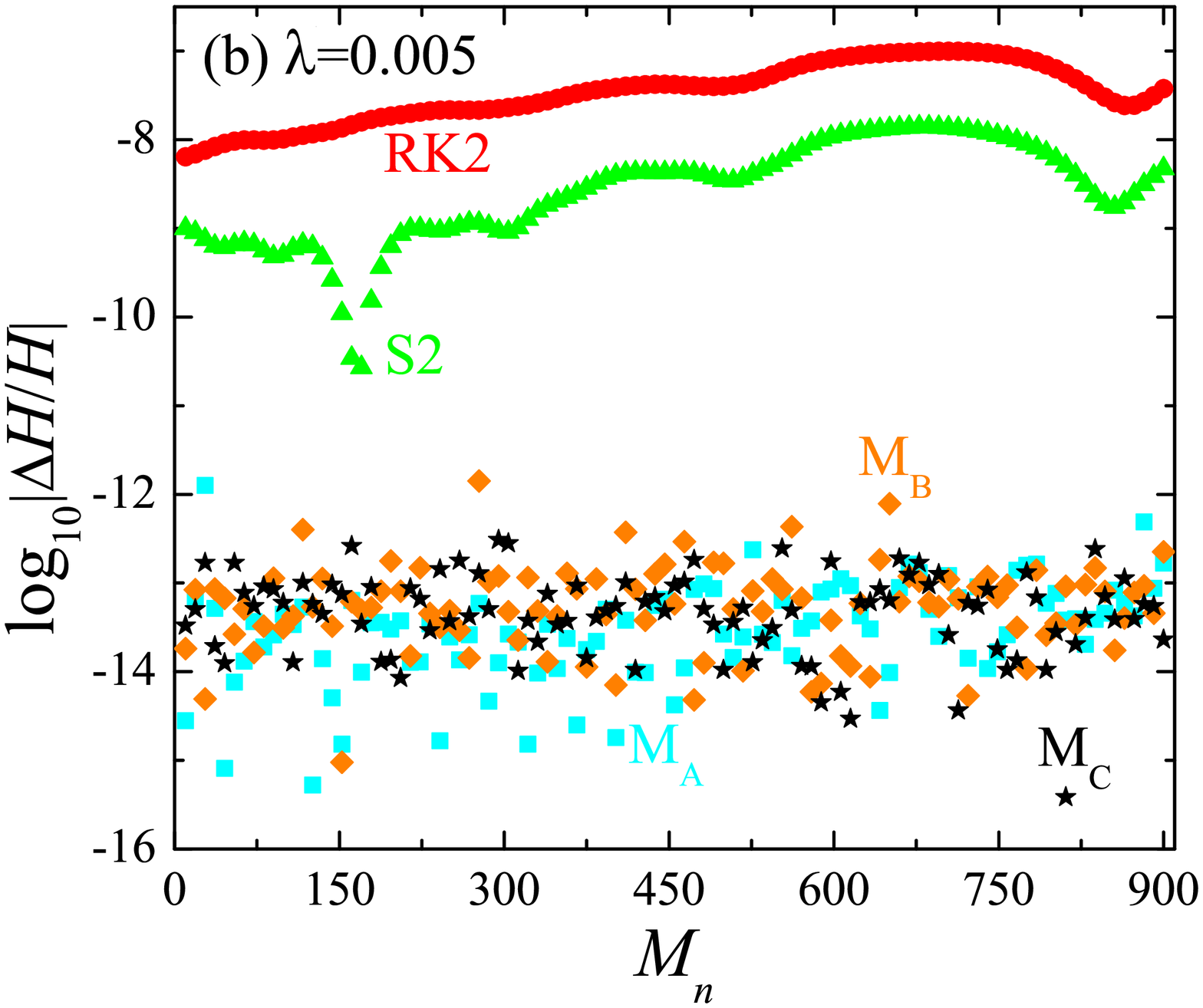}
\includegraphics[scale=0.21]{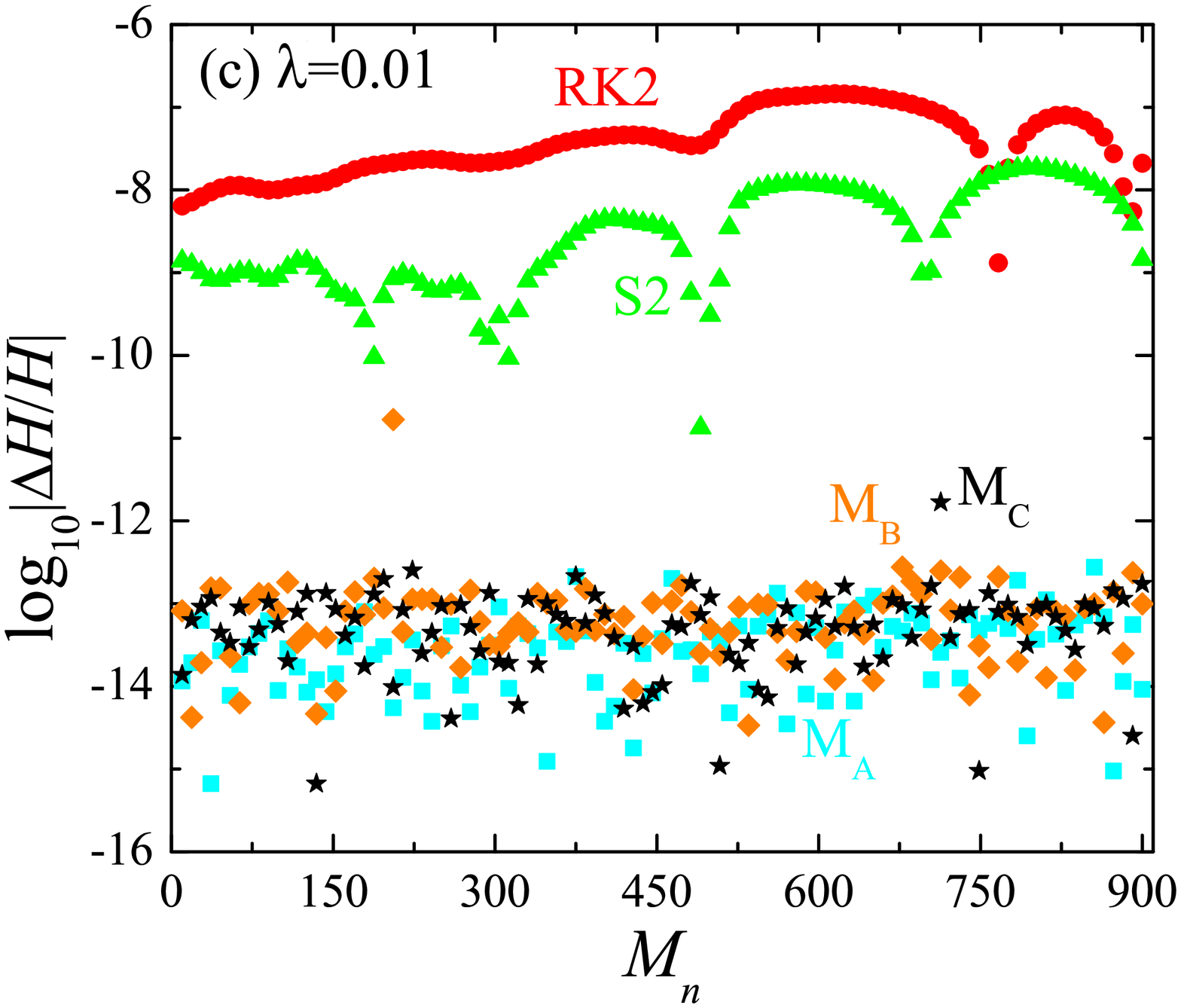}
\includegraphics[scale=0.21]{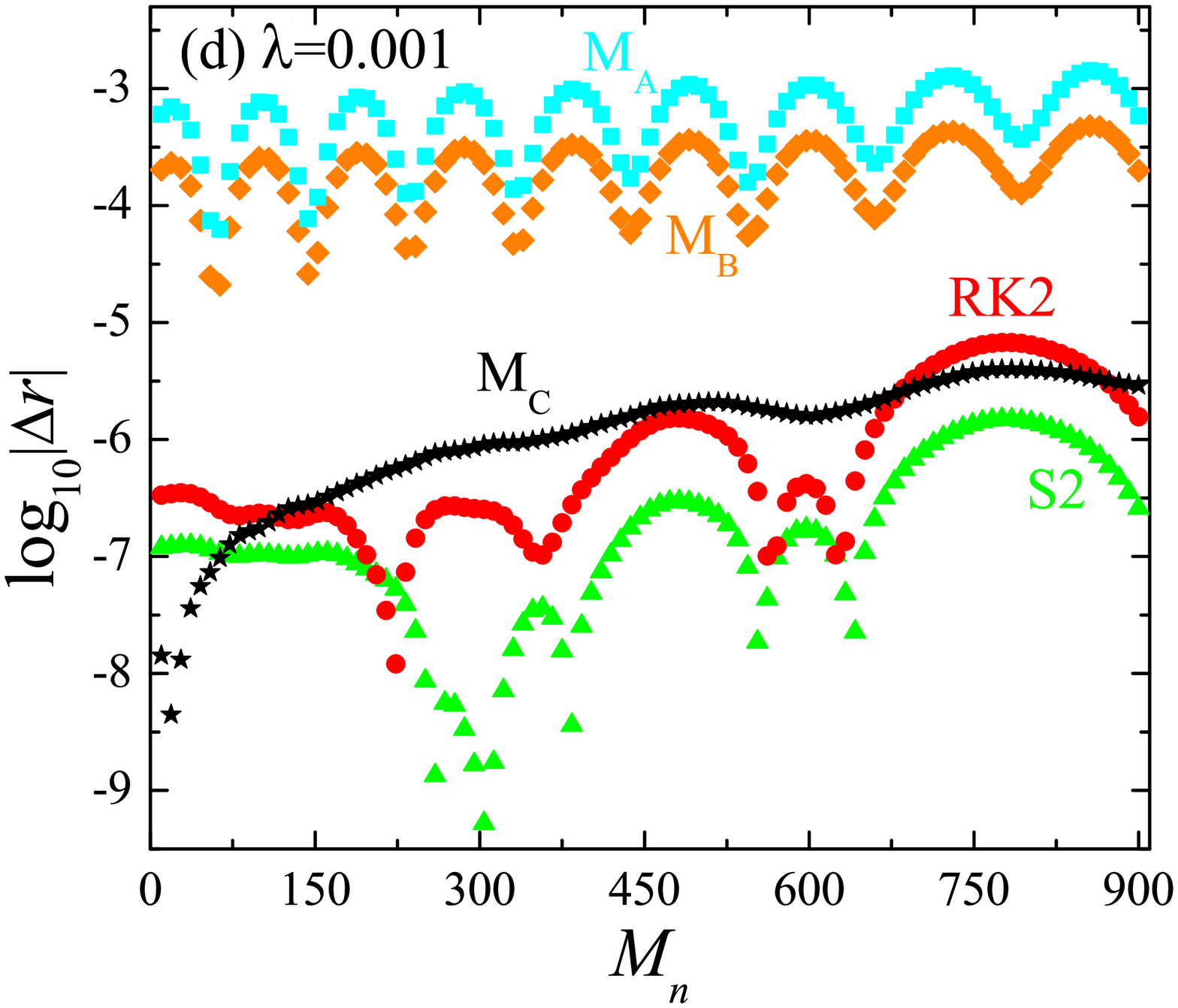}
\includegraphics[scale=0.21]{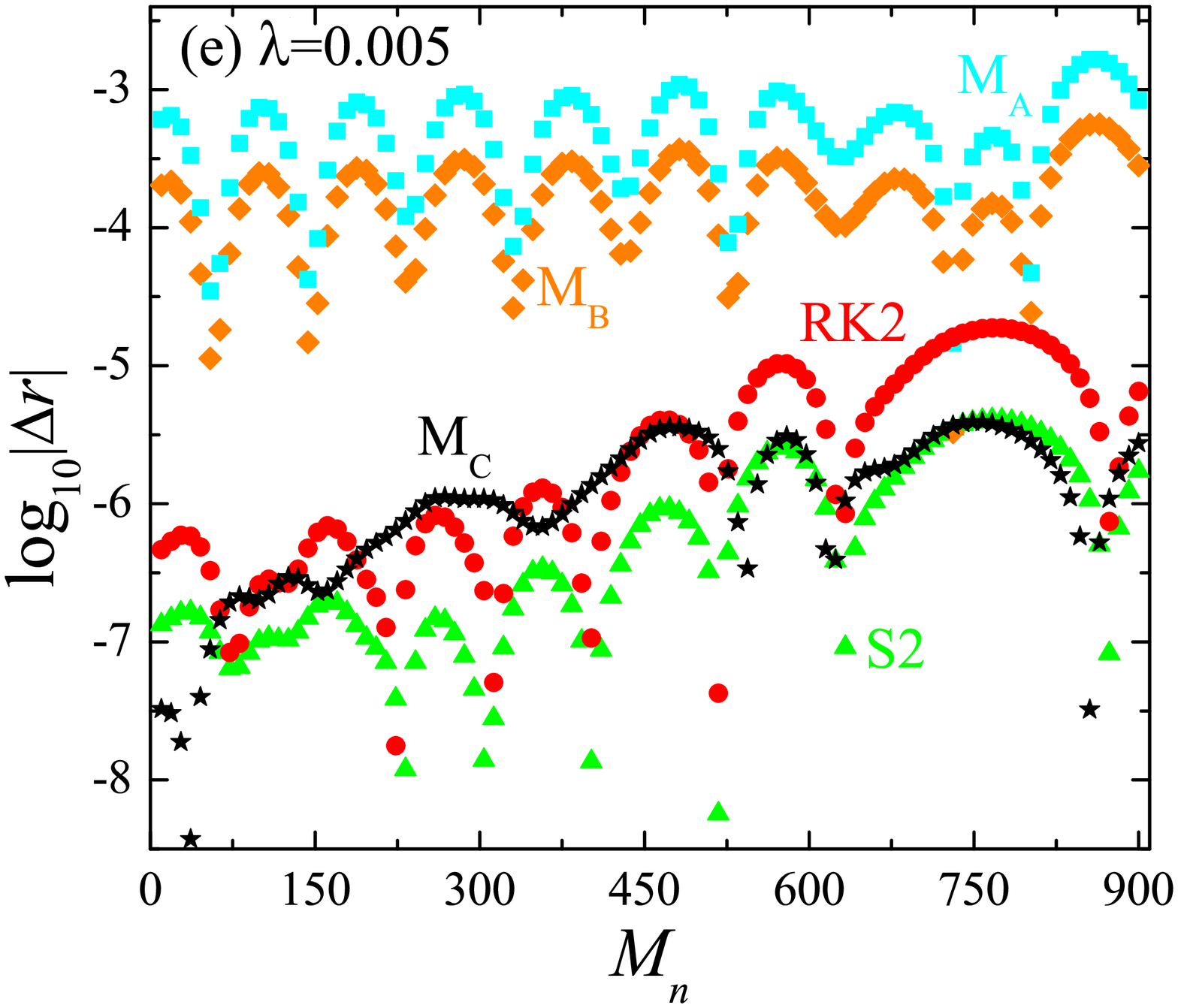}
\includegraphics[scale=0.21]{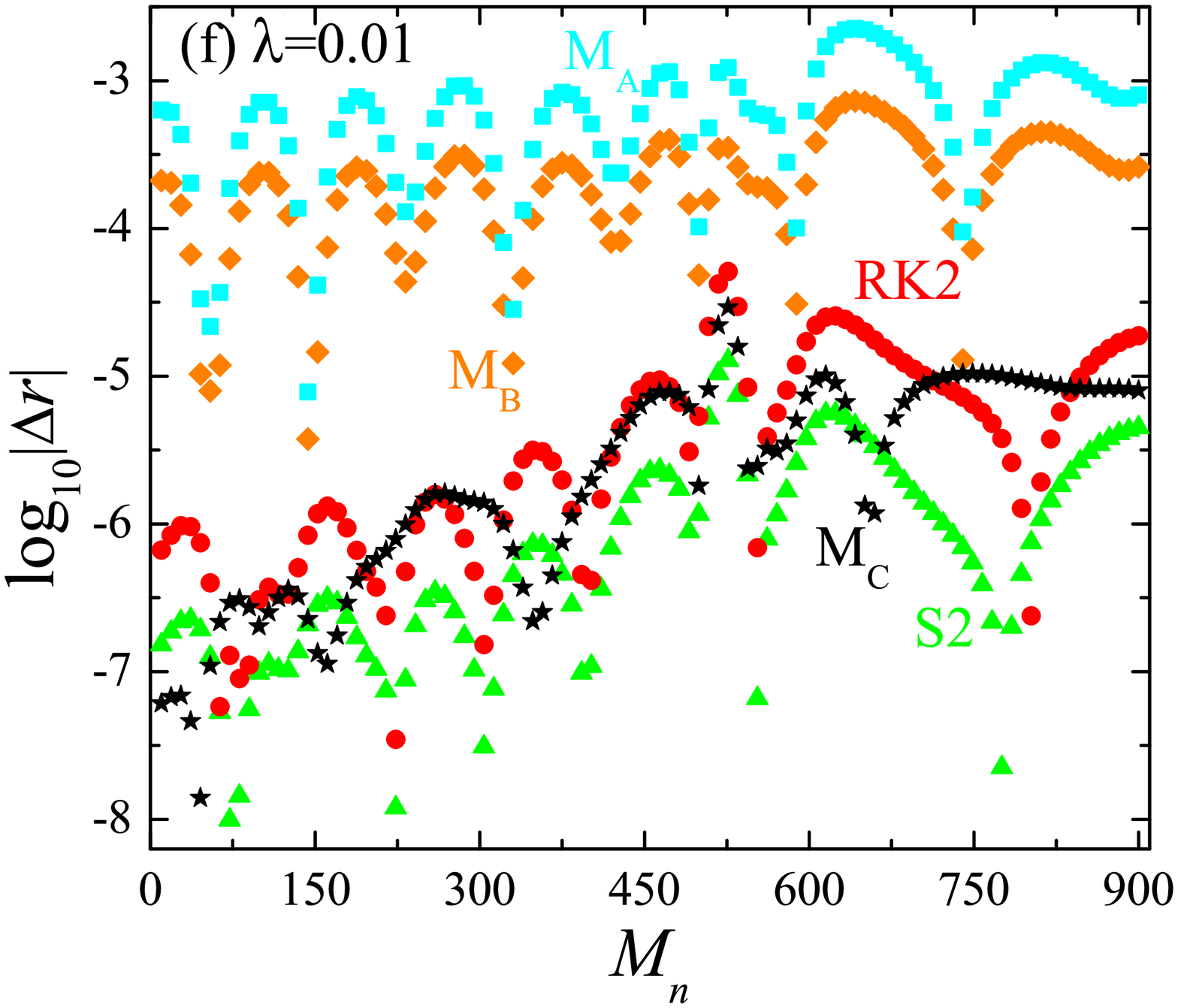}
\caption{(a)-(c): Dependence of relative Hamiltonian errors on
mass parameter $M_{n}$ for three values of parameter $\lambda$.
(d)-(f): Dependence of absolute position errors on mass parameter
$M_{n}$. The other parameters are  $H=400$ and $\alpha=b=1$, and
the initial conditions are those of Orbit 1. The
performances of $M_A$, $M_B$ and $M_C$ in the Hamiltonian errors
are approximately the same as those in Figures 1 and 2. So are the
performances of $M_B$ and $M_C$  in the accuracies of the
solutions. }} \label{fig3}
\end{figure*}

\begin{figure*}%[tbph]
\center{
\includegraphics[scale=0.3]{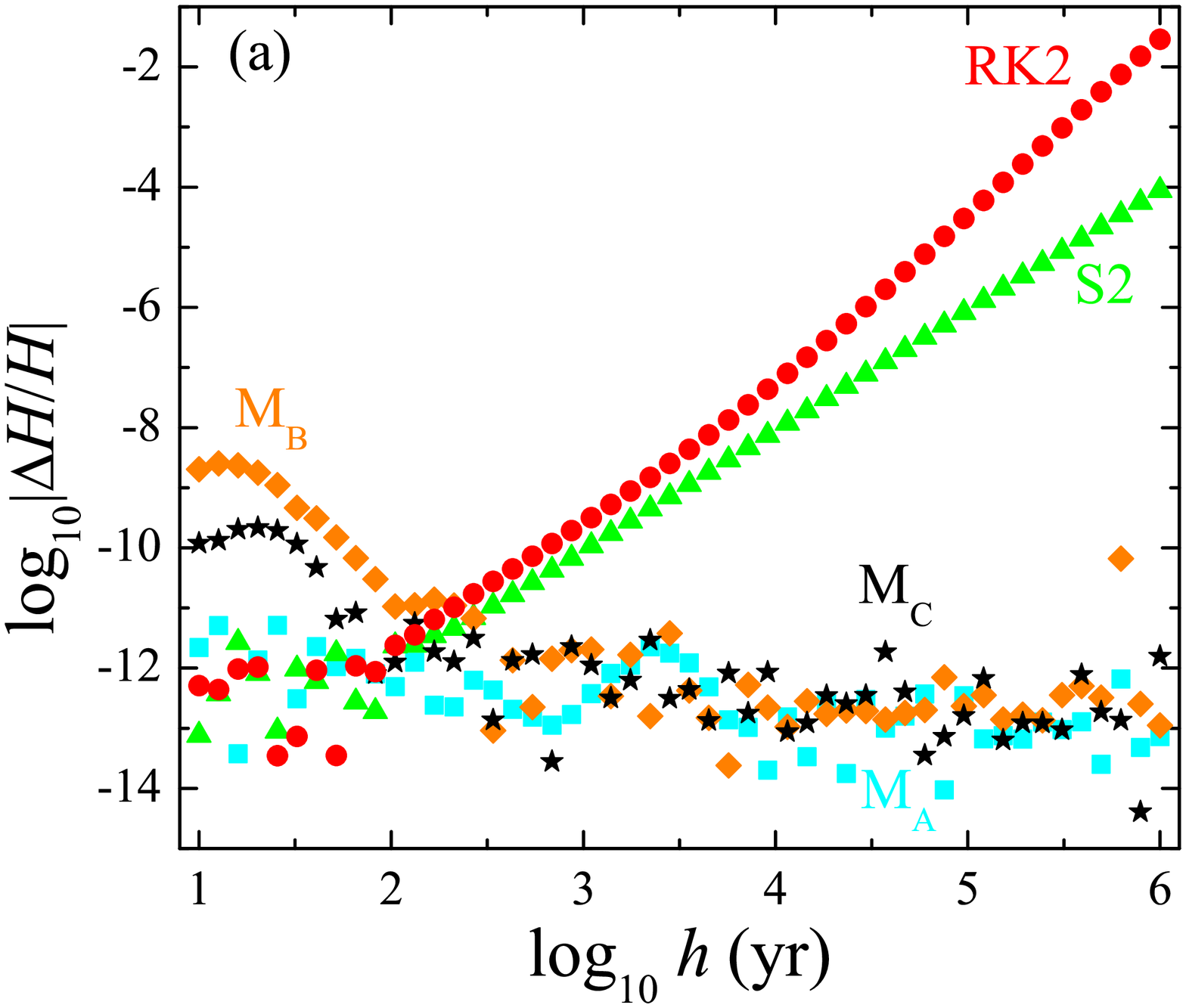}
\includegraphics[scale=0.3]{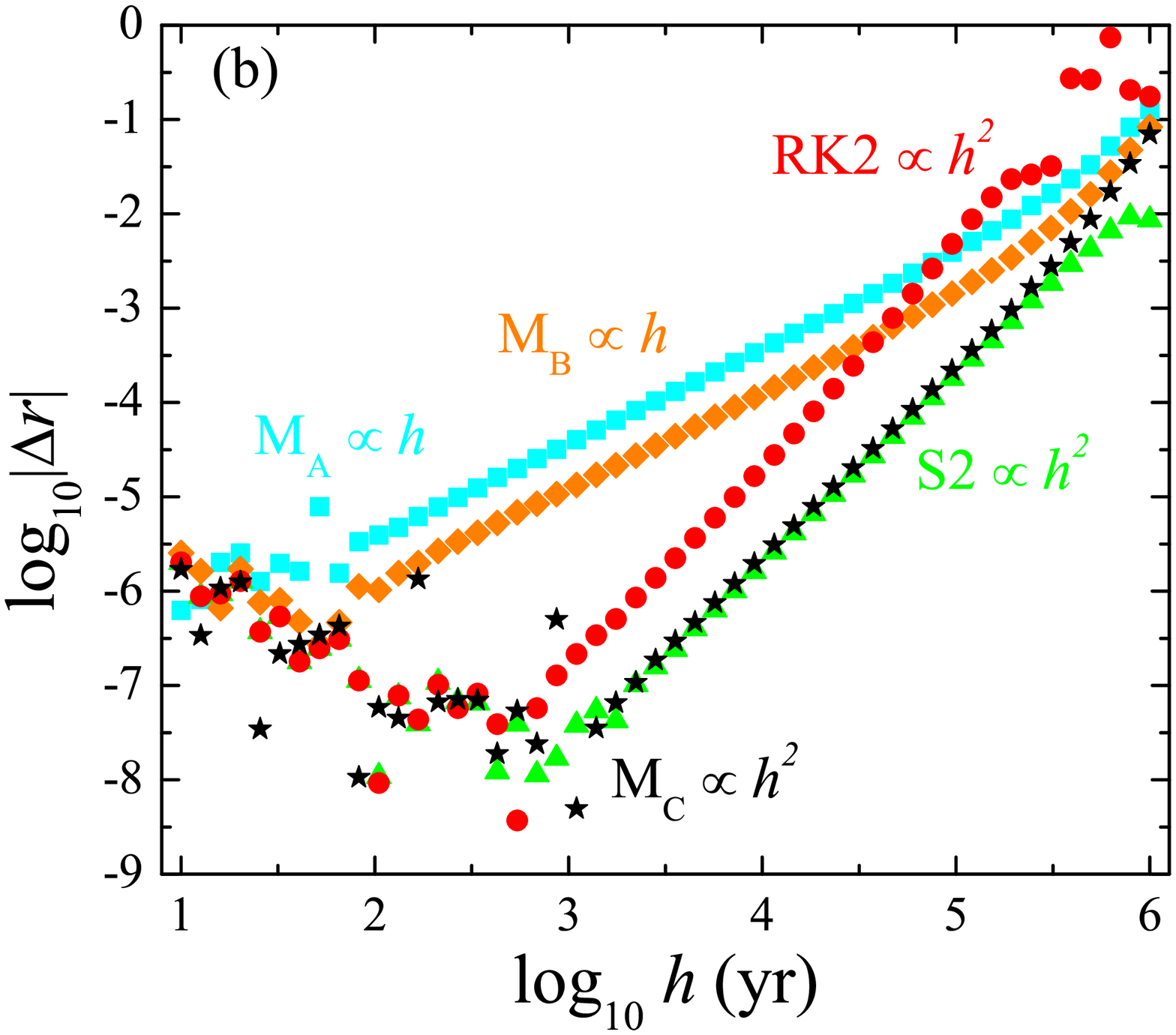}
\caption{(a): Dependence of relative Hamiltonian errors on time
steps $h$. (b): Dependence of absolute position errors on time
steps $h$. The parameters are $H=400$, $M_{n}=200$, $\alpha=1.6$,
$b=0.8$, and $\lambda=0$. The initial conditions are those of
Orbit 1. The integration lasts $10^{10}$ years for each time step.
The time steps run from 10 years to  $10^{6}$ years; equivalently,
the dimensionless time steps are from $10^{-7}$ to  $10^{-2}$.
Obviously, the position errors of $M_A$ and $M_B$
grow linearly with the time-step $h$, while those of $M_C$, RK2,
and S2 grow with $h^{2}$. This sufficiently shows that $M_A$ and
$M_B$ are only accurate to a first-order accuracy to the
solutions, and $M_C$, RK2, and S2 are accurate to a second-order
accuracy.}} \label{fig4}
\end{figure*}

\begin{figure*}%[tbph]
\center{
\includegraphics[scale=0.3]{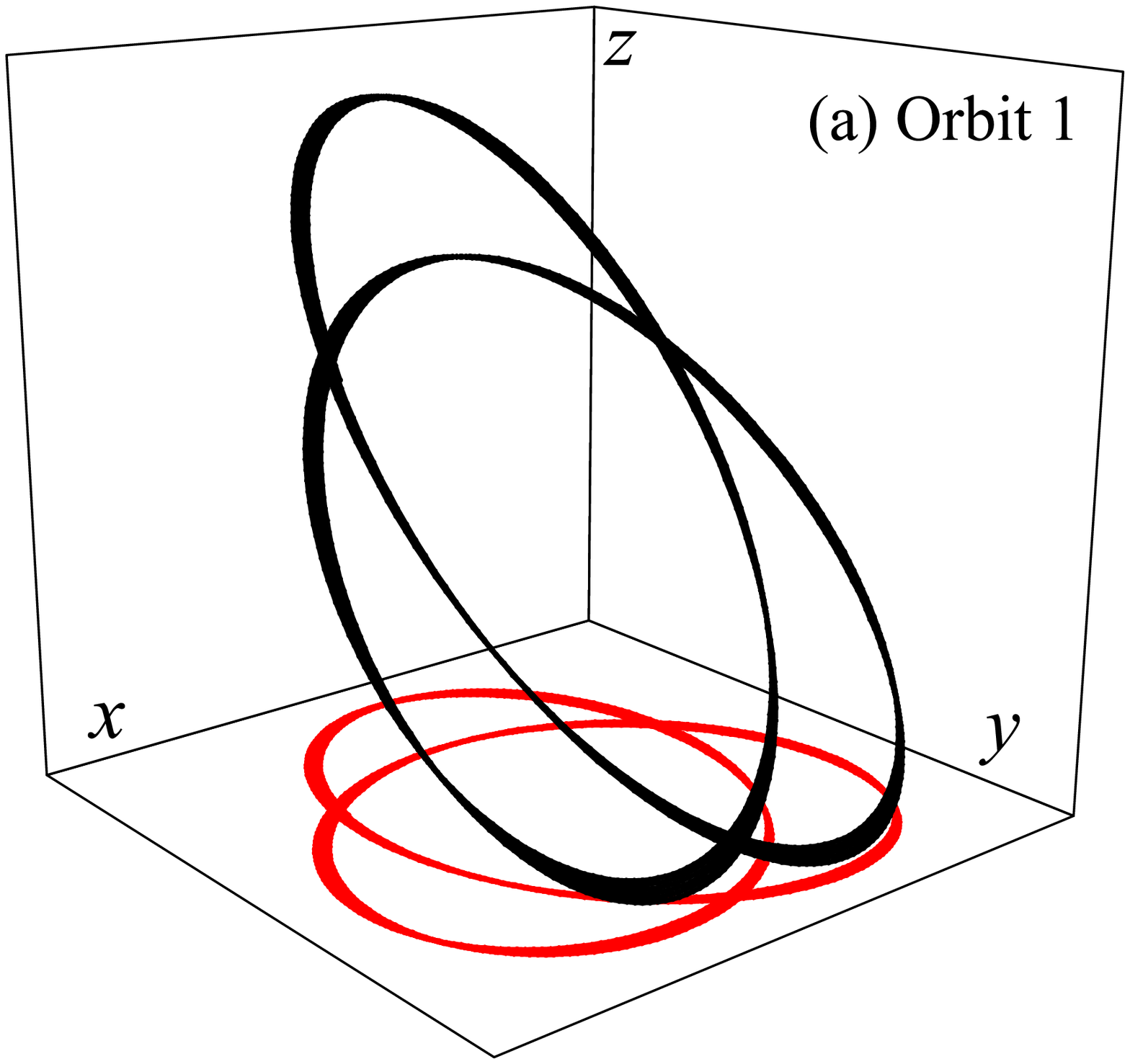}
\includegraphics[scale=0.3]{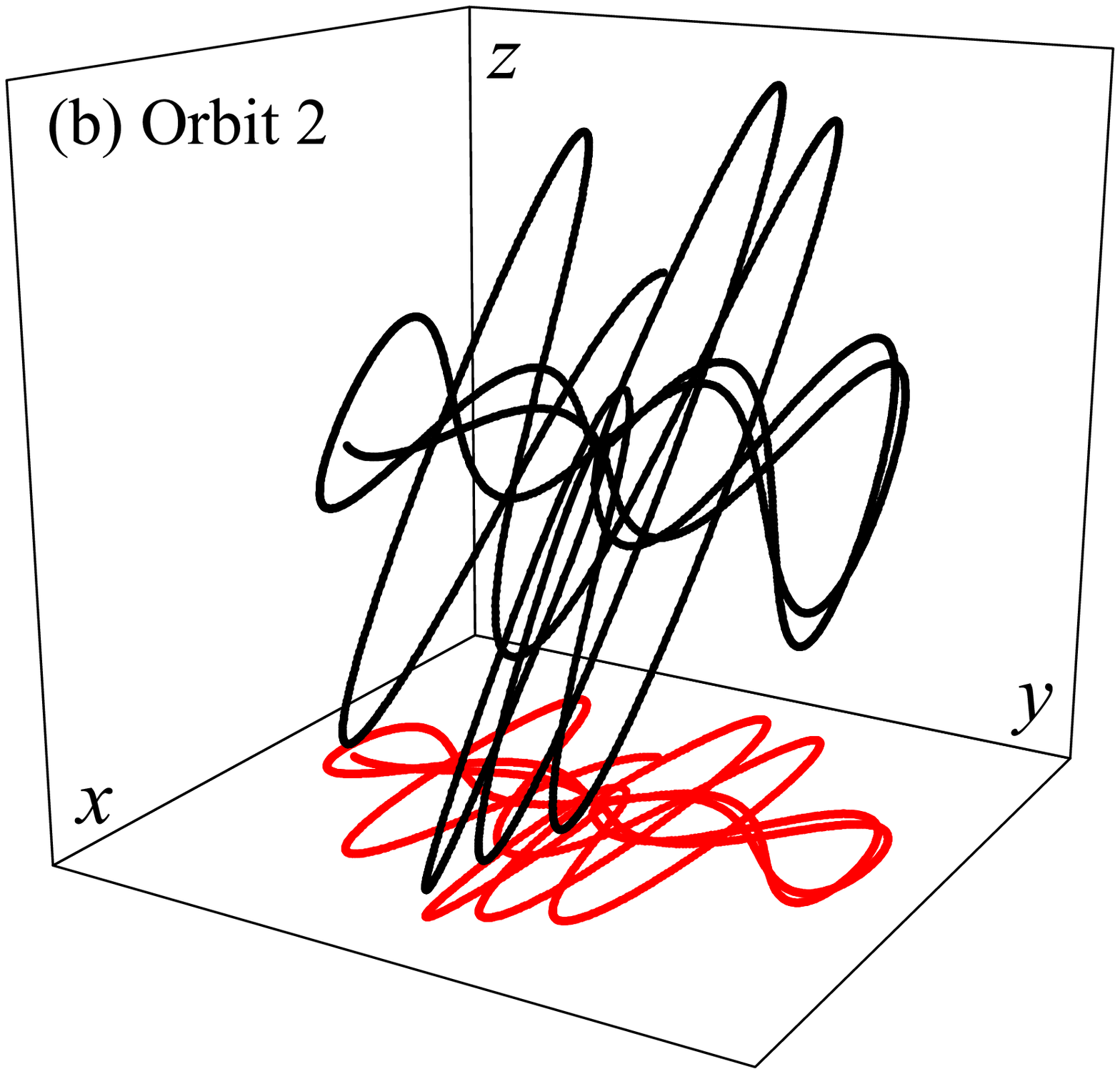}
\includegraphics[scale=0.3]{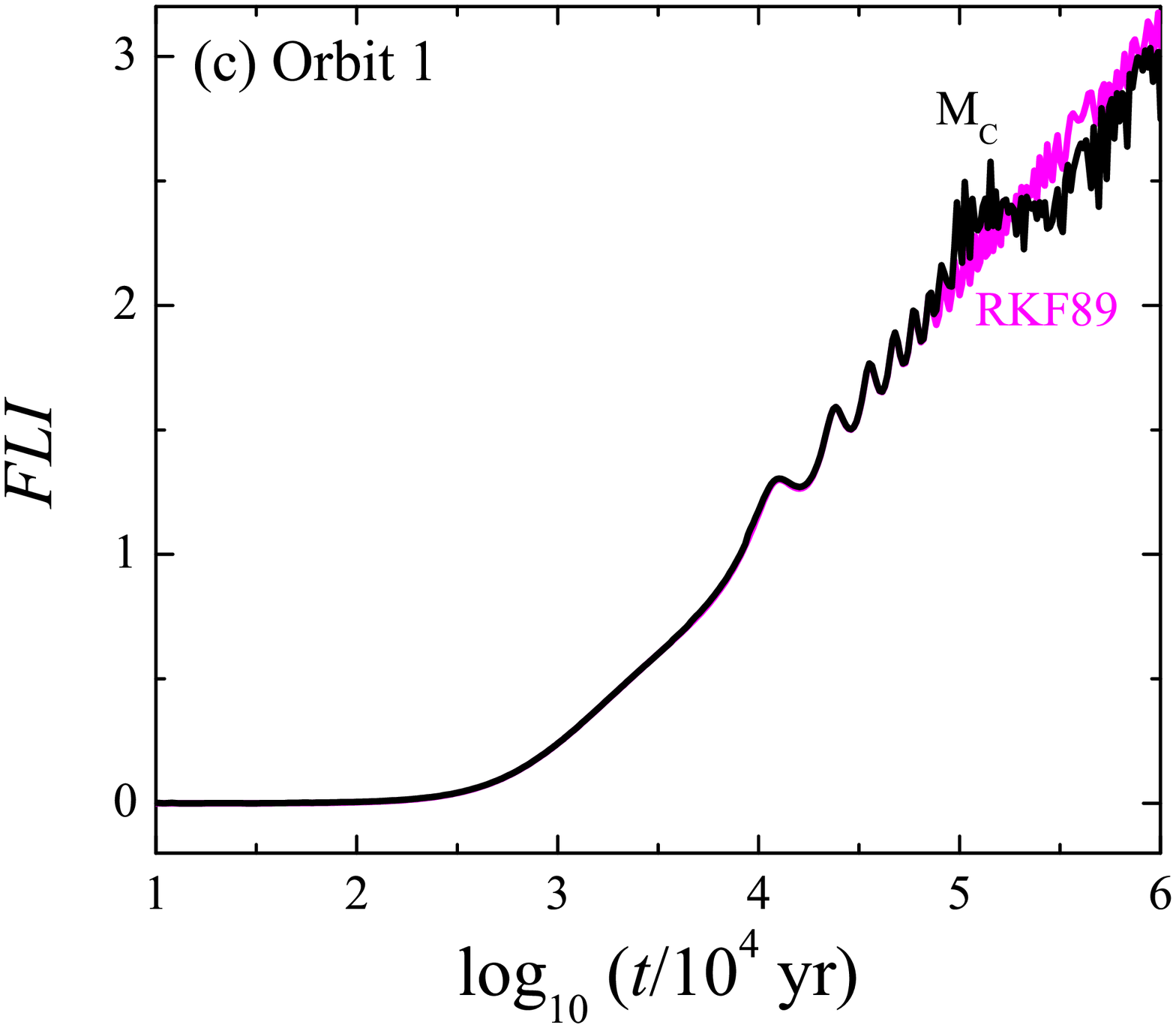}
\includegraphics[scale=0.3]{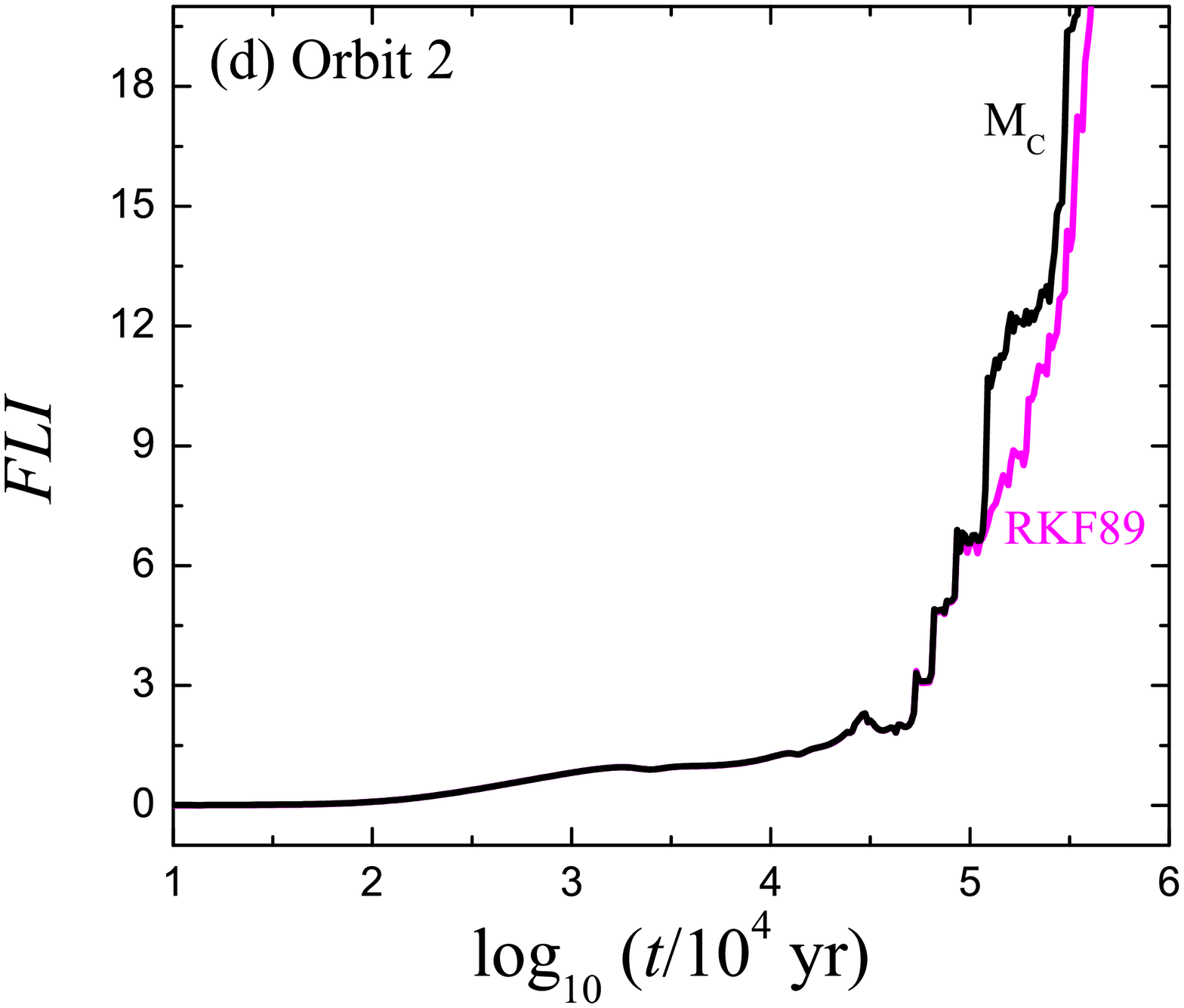}
\caption{(a) and (b): Two orbits in Figure 1 are shown in the
three-dimensional space. The red curves represent
the projection of the trajectories on the $xoy$-plane. (c) and
(d): Fast Lyapunov indicators (FLIs) for Orbits 1 and 2.
Methods $M_C$ and RKF89 give almost the same
values of the FLIs. The FLI for Orbit 1 is much smaller than that
for Orbit 2. Orbit 1 is regular, whereas Orbit 2 is chaotic.  }}
\label{fig5}
\end{figure*}

\begin{figure*}%[tbph]
\center{
\includegraphics[scale=0.21]{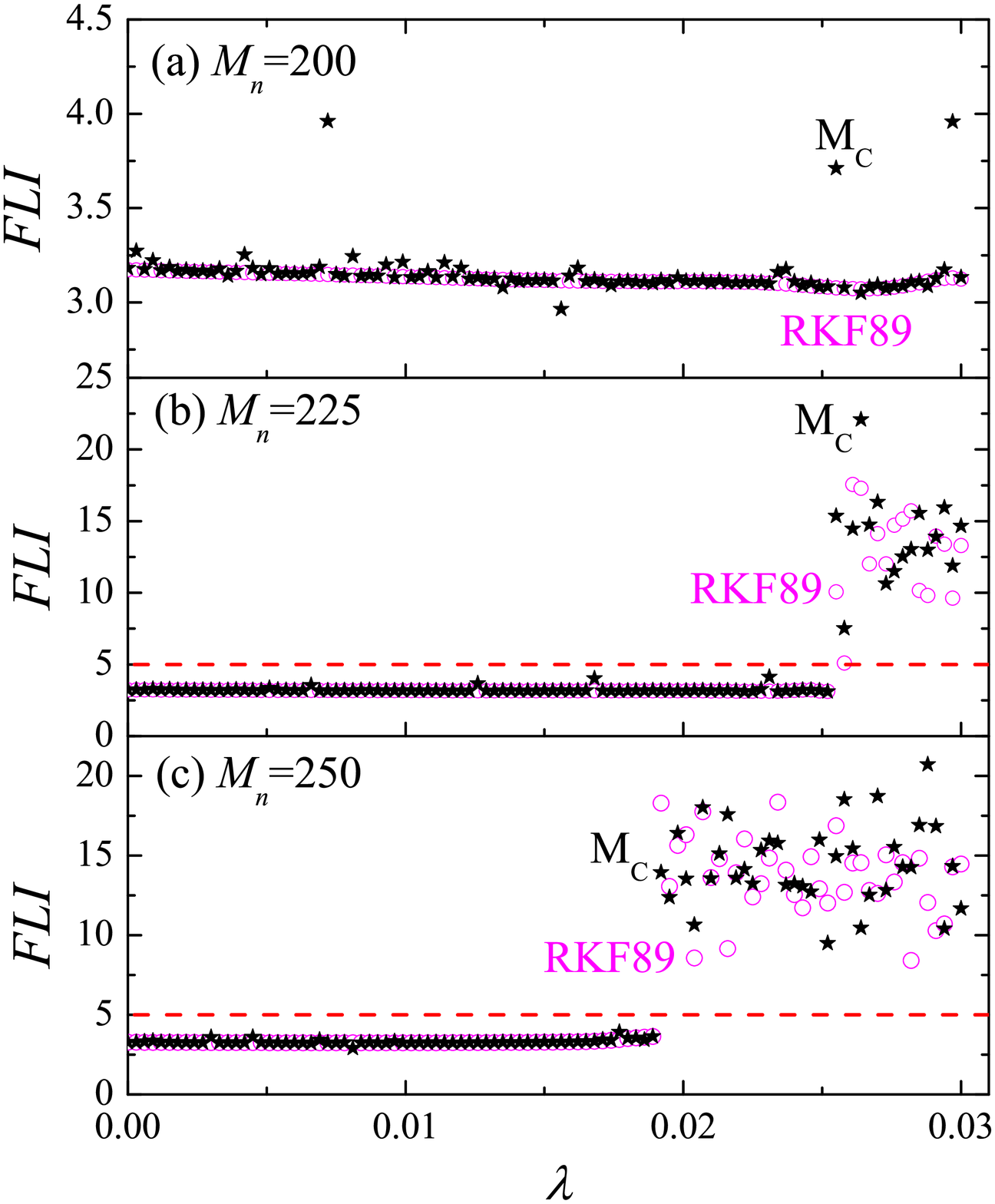}
\includegraphics[scale=0.21]{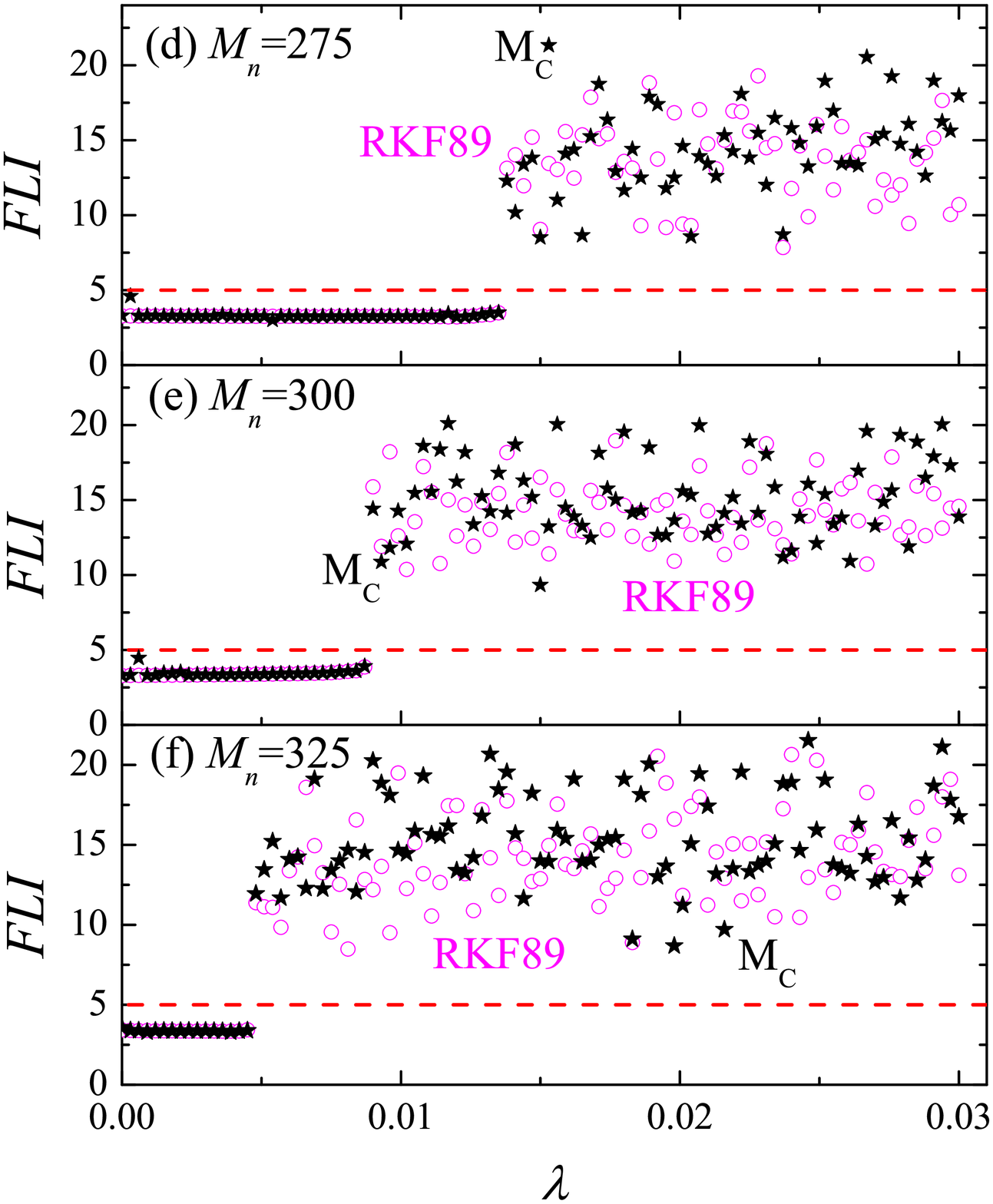}
\includegraphics[scale=0.21]{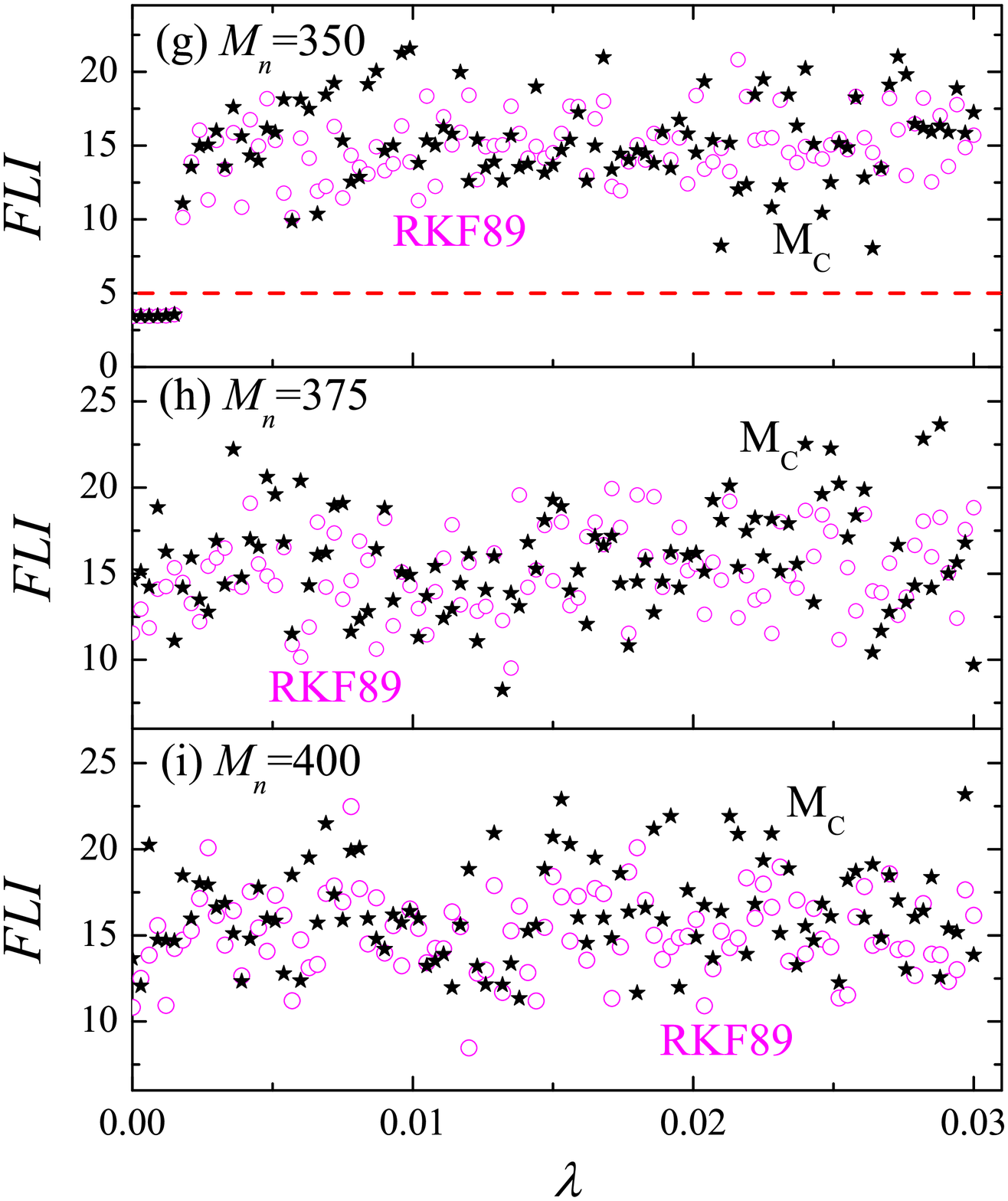}
\caption{Dependence of FLI on parameter $\lambda$ under the
circumstance of different galaxy masses $M_{n}$. $H=400$,
$\alpha=1.6$, and $b=0.8$. Each of the FLIs is obtained after $3
\times 10^{5}$ integration steps.  The FLIs larger than 5 indicate
the chaoticity; the FLIs less than 5 indicate the regularity.
For $M_{n}=200$ in (a), all values of
$\lambda\in[0, 0.03]$ correspond to order. For $M_{n}=225$ in (b),
chaos occurs when $\lambda\geq 0.0255$. For $M_{n}=250$, $275$,
$300$, $325$, and $350$ in (c)-(g), the critical values $\lambda$
for inducing chaos are 0.0192, 0.0138, 0.009, 0.0048, and 0.0018,
respectively. For $M_{n}=375$, and $400$, all values of
$\lambda\in[0, 0.03]$ correspond to chaos. These facts show that
chaos easily occurs when $M_{n}$ increases, but the critical value
of $\lambda$ for the occurrence of chaos decreases with an
increase of $M_{n}$.}} \label{fig6}
\end{figure*}

\begin{figure*}%[tbph]
\center{
\includegraphics[scale=0.3]{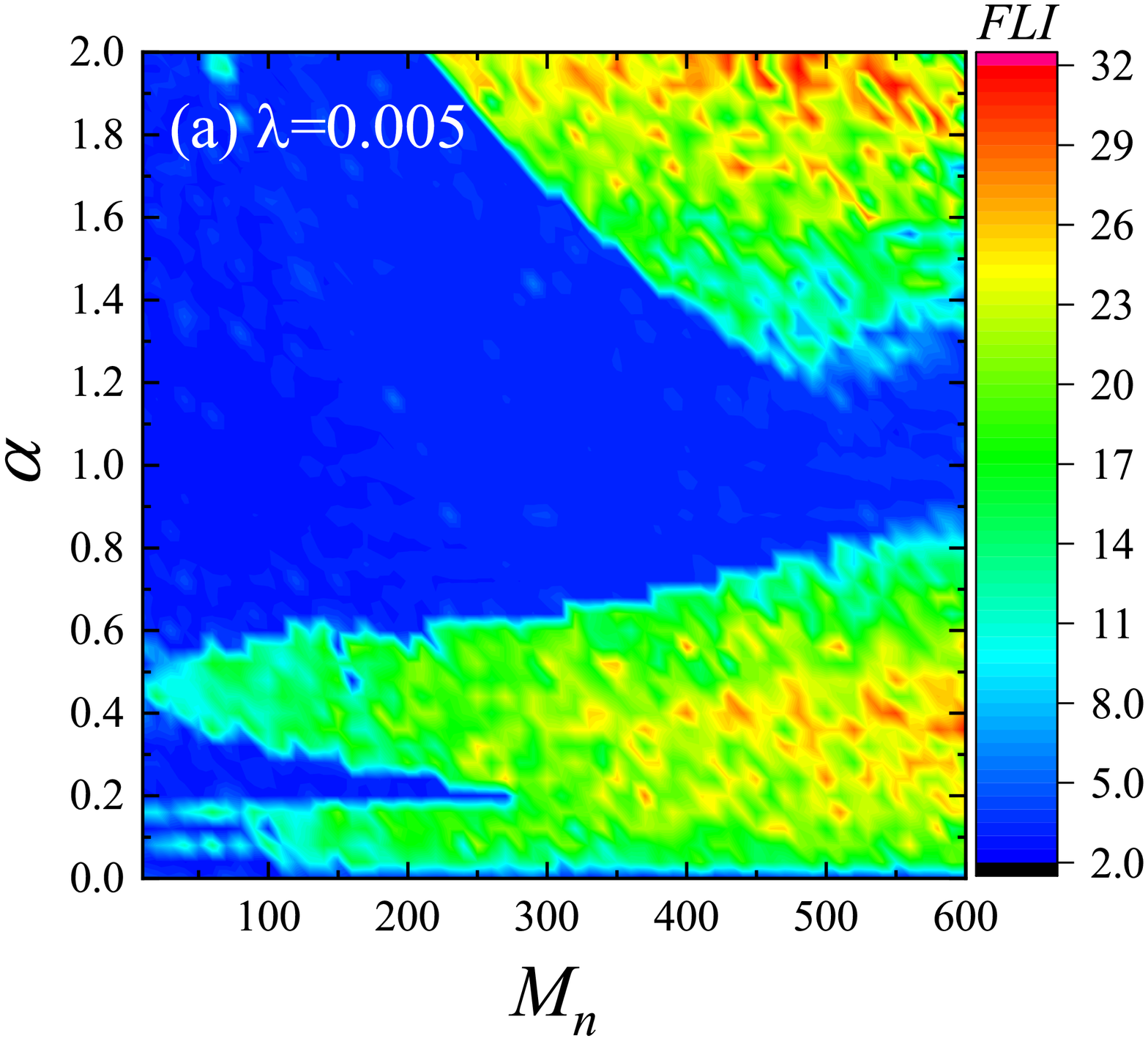}
\includegraphics[scale=0.3]{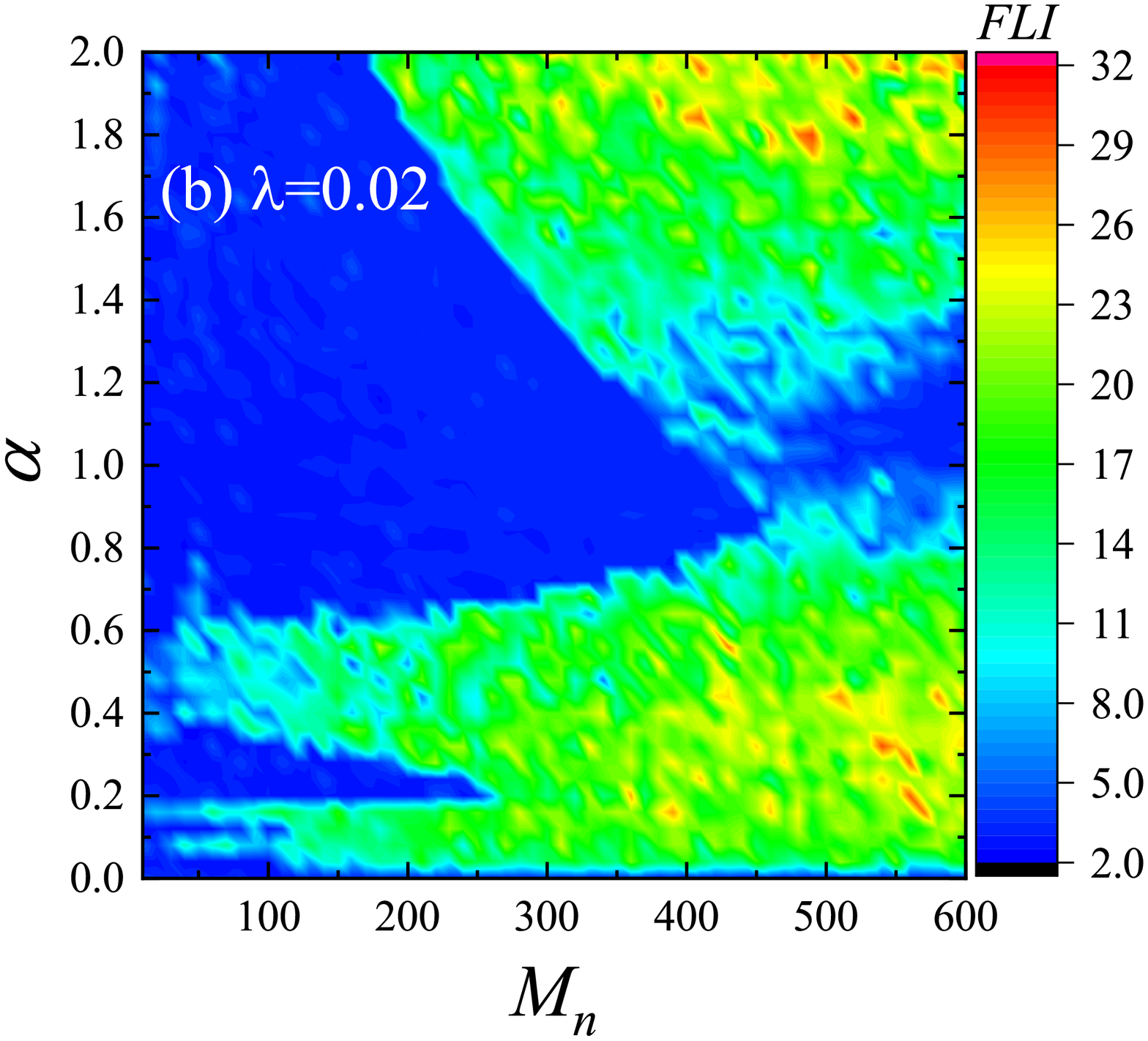}
\includegraphics[scale=0.3]{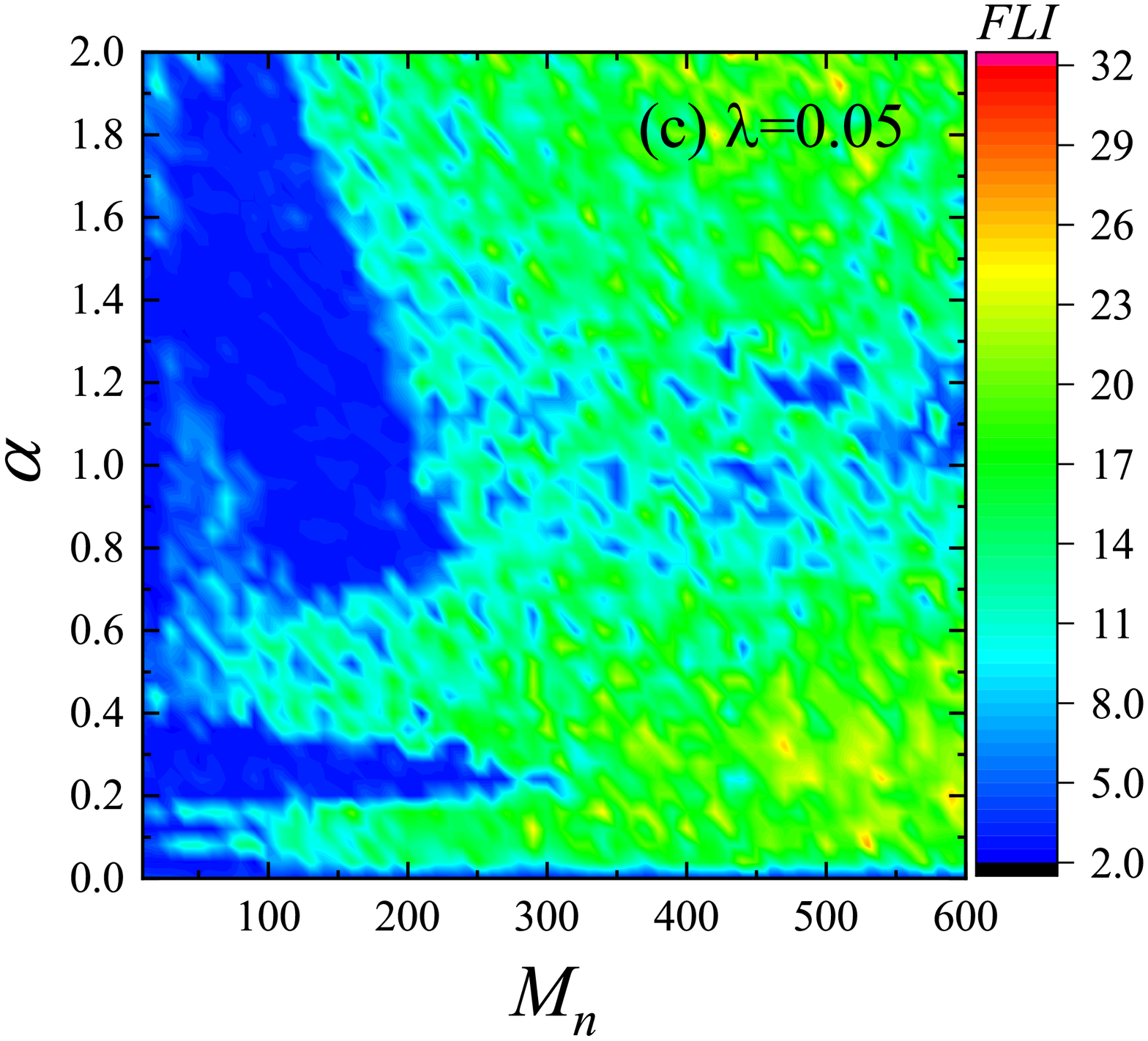}
\includegraphics[scale=0.3]{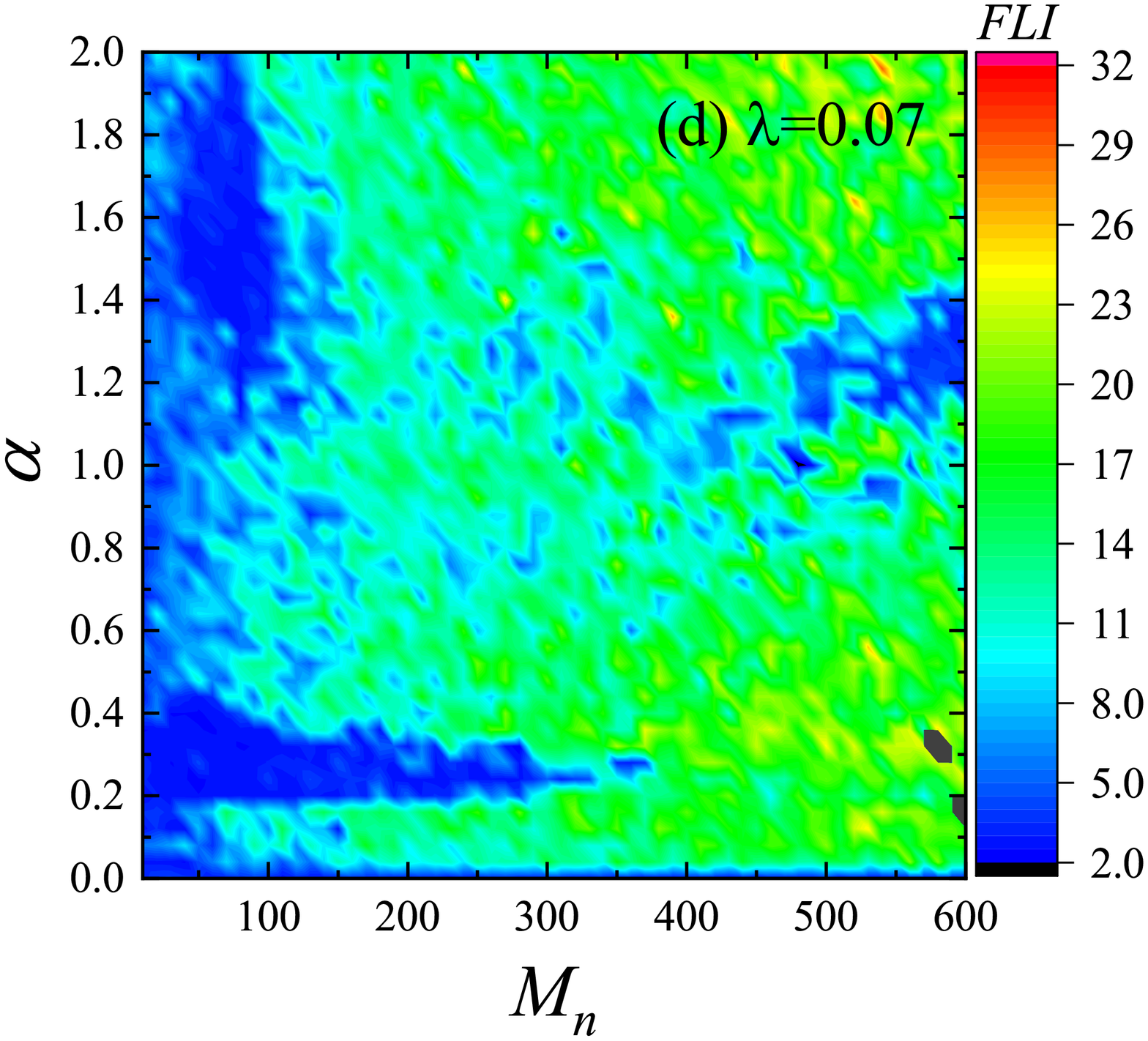}
\caption{Finding chaos by using the FLIs to scan a two-dimensional
space of parameters $\alpha$ and $M_{n}$. Other parameters are
$H=400$ and $b=1$. Chaotic regions in the
parameter space of $\alpha$ and $M_{n}$ get larger as $M_{n}$ and
$\lambda$ increase. }} \label{fig7}
\end{figure*}

\begin{figure*}%[tbph]
\center{
\includegraphics[scale=0.25]{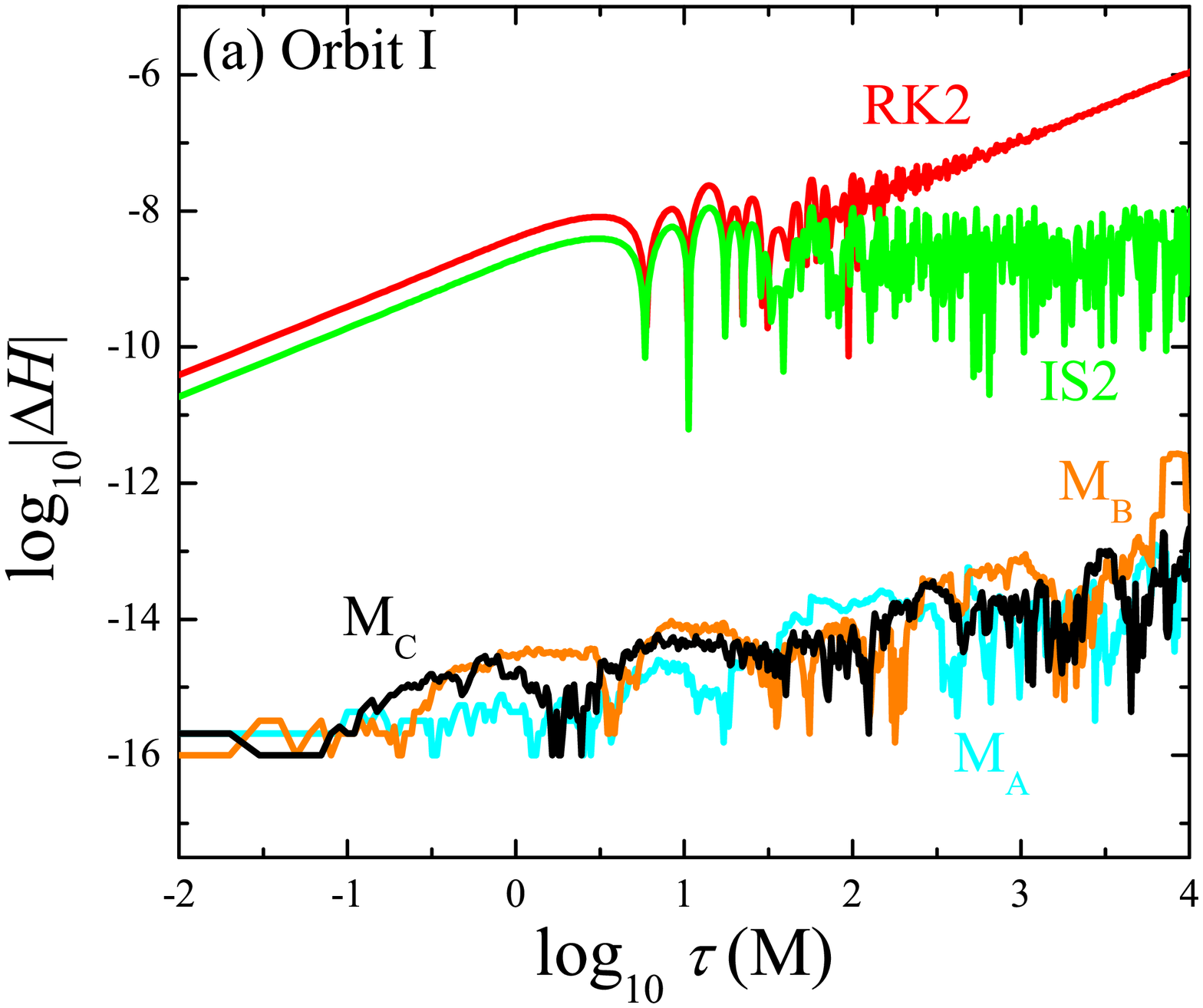}
\includegraphics[scale=0.25]{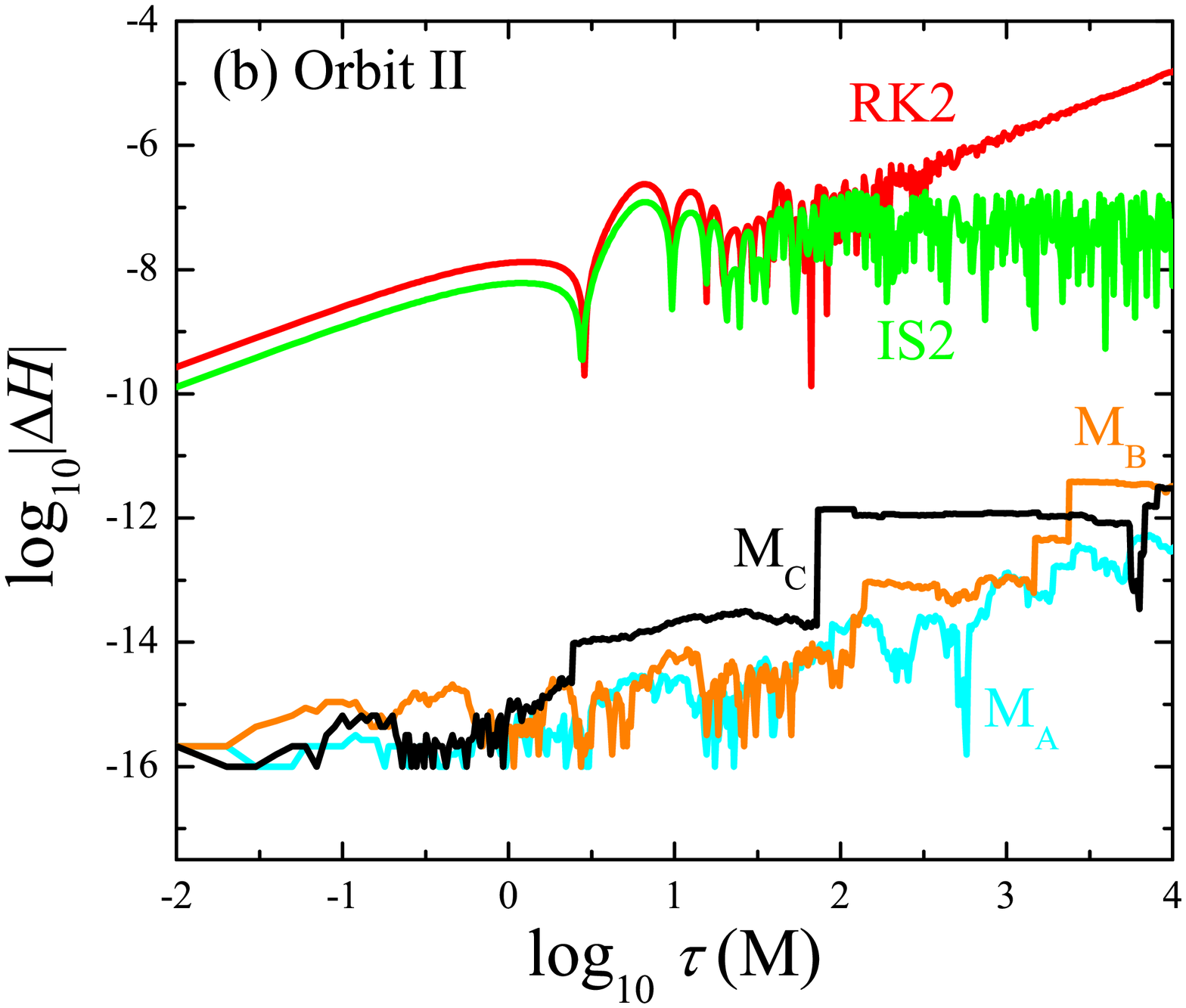}
\includegraphics[scale=0.25]{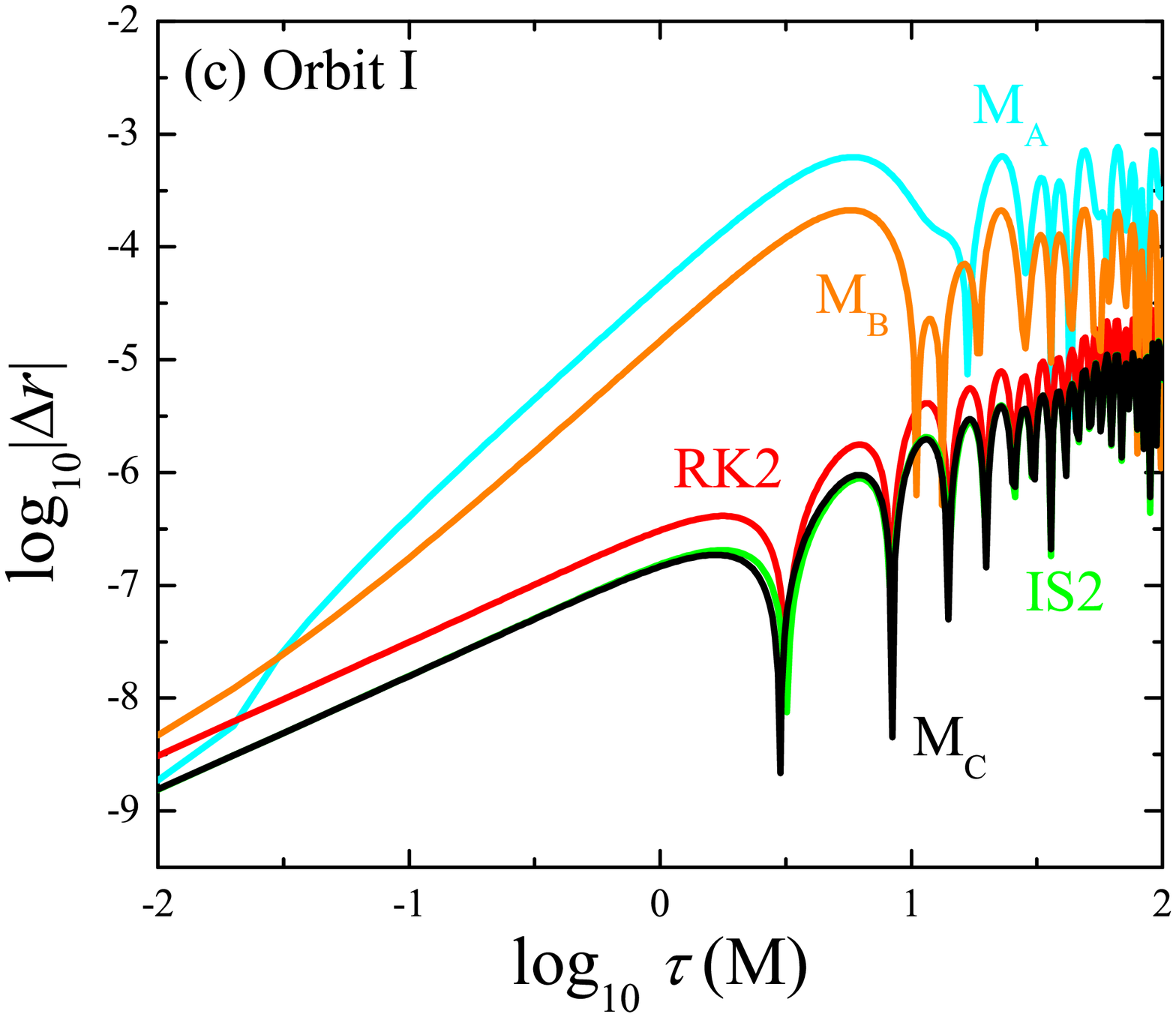}
\includegraphics[scale=0.25]{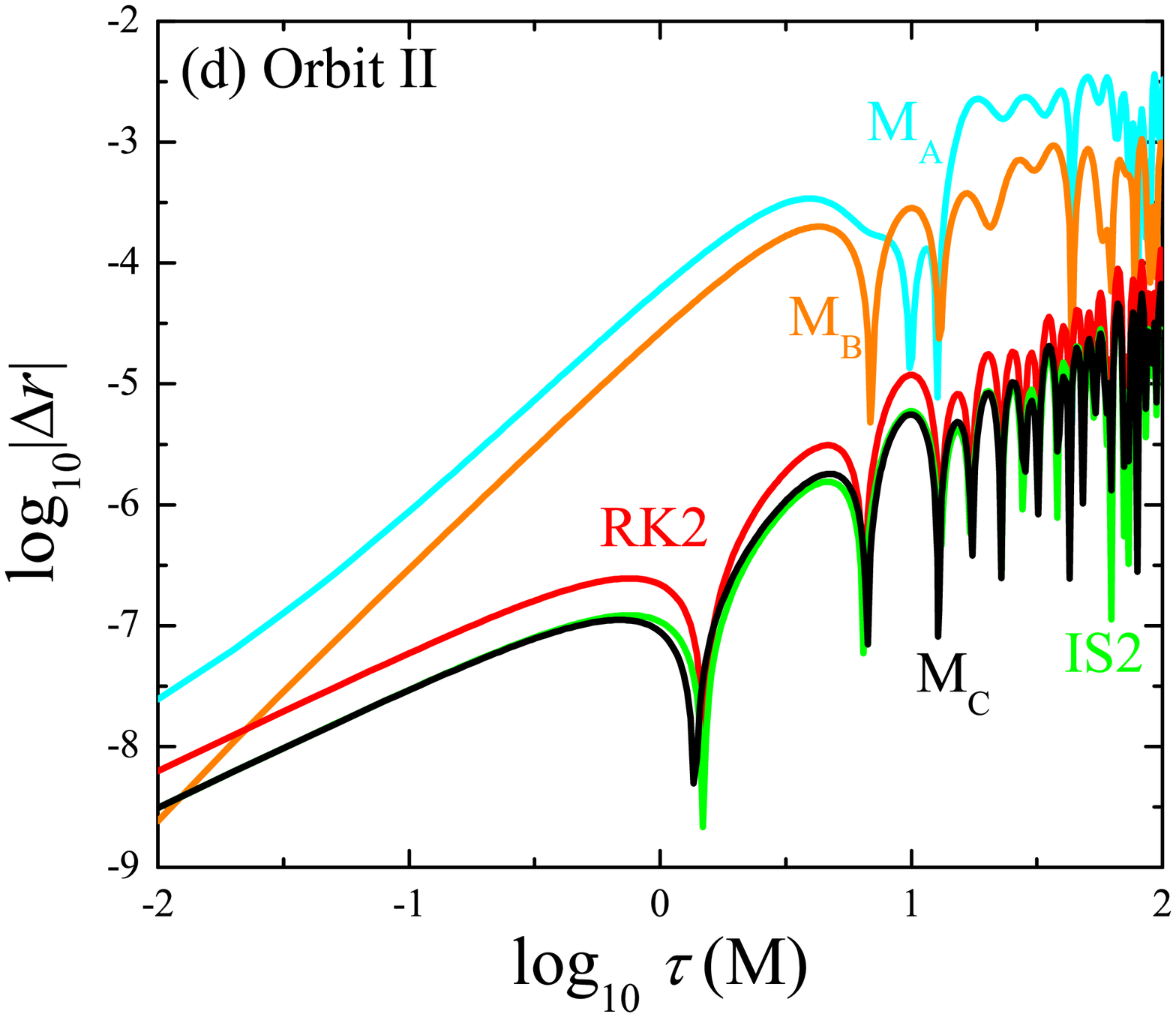}
\caption{Errors in the Hamiltonian and solutions in the magnetized
black hole background. The numerical performances
of the five algorithms in the errors of the Hamiltonian and
solutions are similar to those in Figure 1. }} \label{fig8}
\end{figure*}

\begin{figure*}%[tbph]
\center{
\includegraphics[scale=0.3]{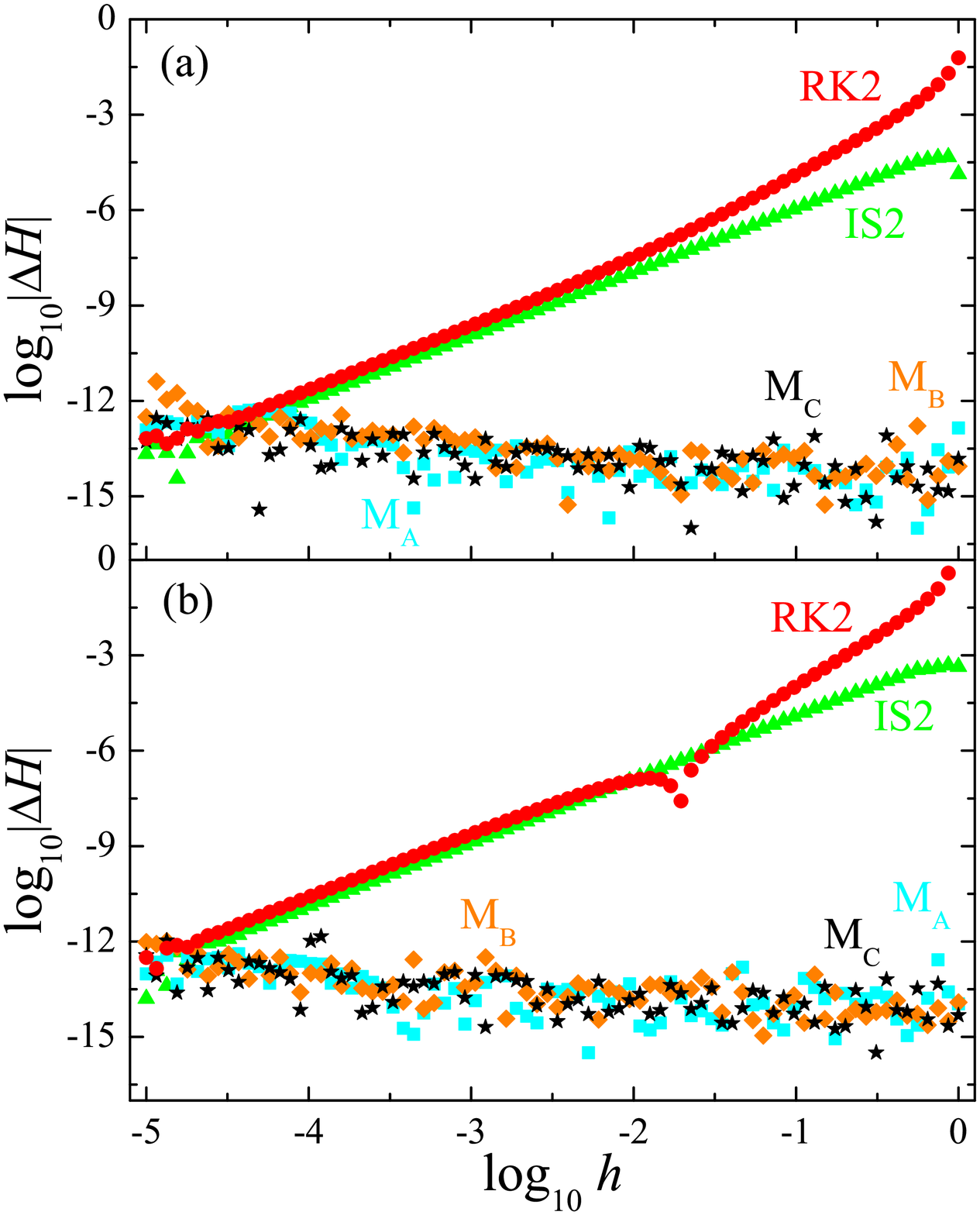}
\includegraphics[scale=0.3]{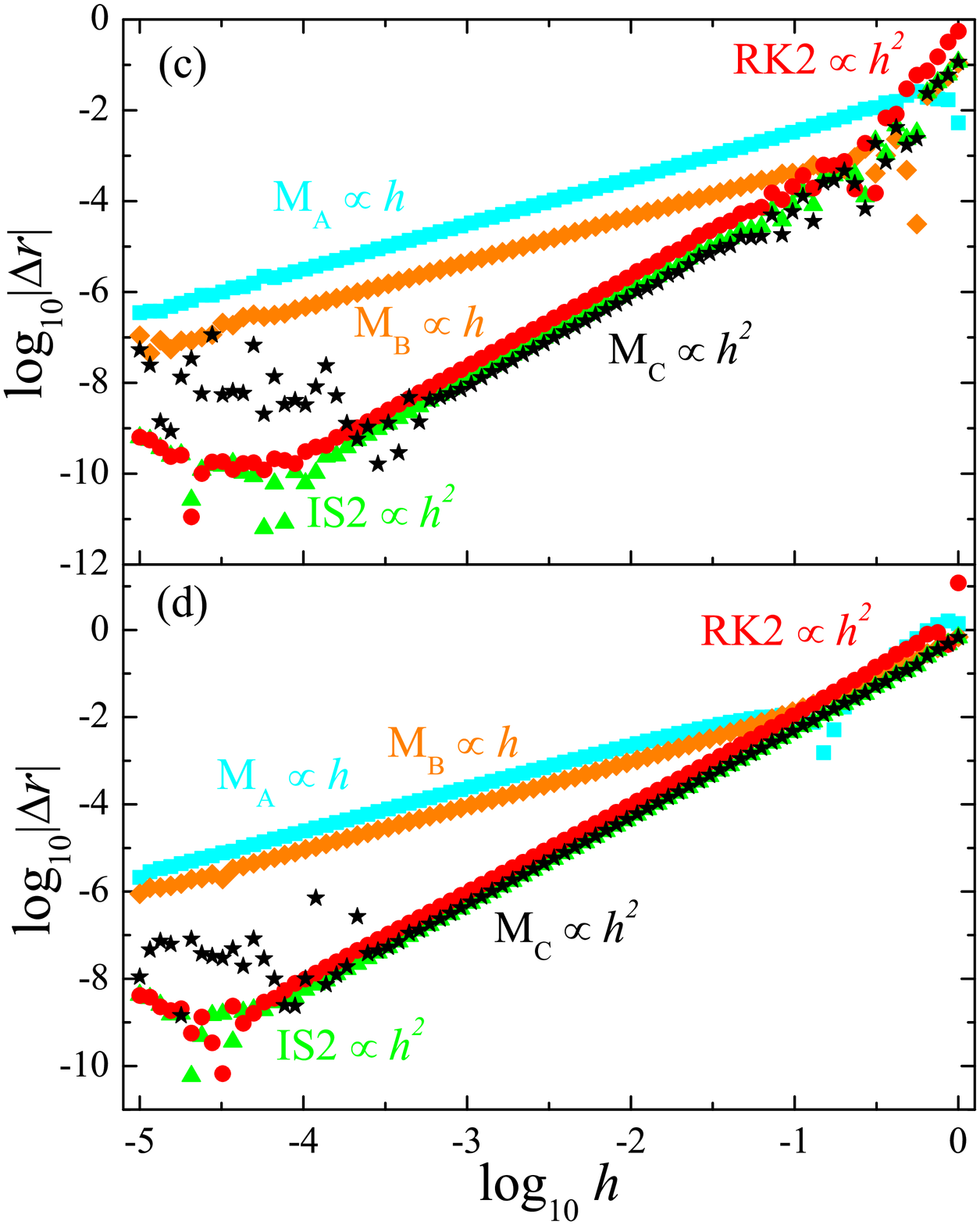}
\caption{Relations between the time steps $h$ and the errors in
the Hamiltonian and solutions. (a) and (c): Orbit I. (b) and (d):
Orbit II. Each error is obtained after the integration time
$\tau=100$. Panels (c) and (d) show that the
position errors of $M_A$ and $M_B$ grow linearly with $h$, while
those of $M_C$, RK2, and S2 grow with $h^{2}$. This implies that
$M_A$ and $M_B$ are first-order schemes, and $M_C$, RK2, and S2
are second-order schemes. The result is consistent with that in
Figure 4.}} \label{fig9}
\end{figure*}

\begin{figure*}%[tbph]
\center{
\includegraphics[scale=0.3]{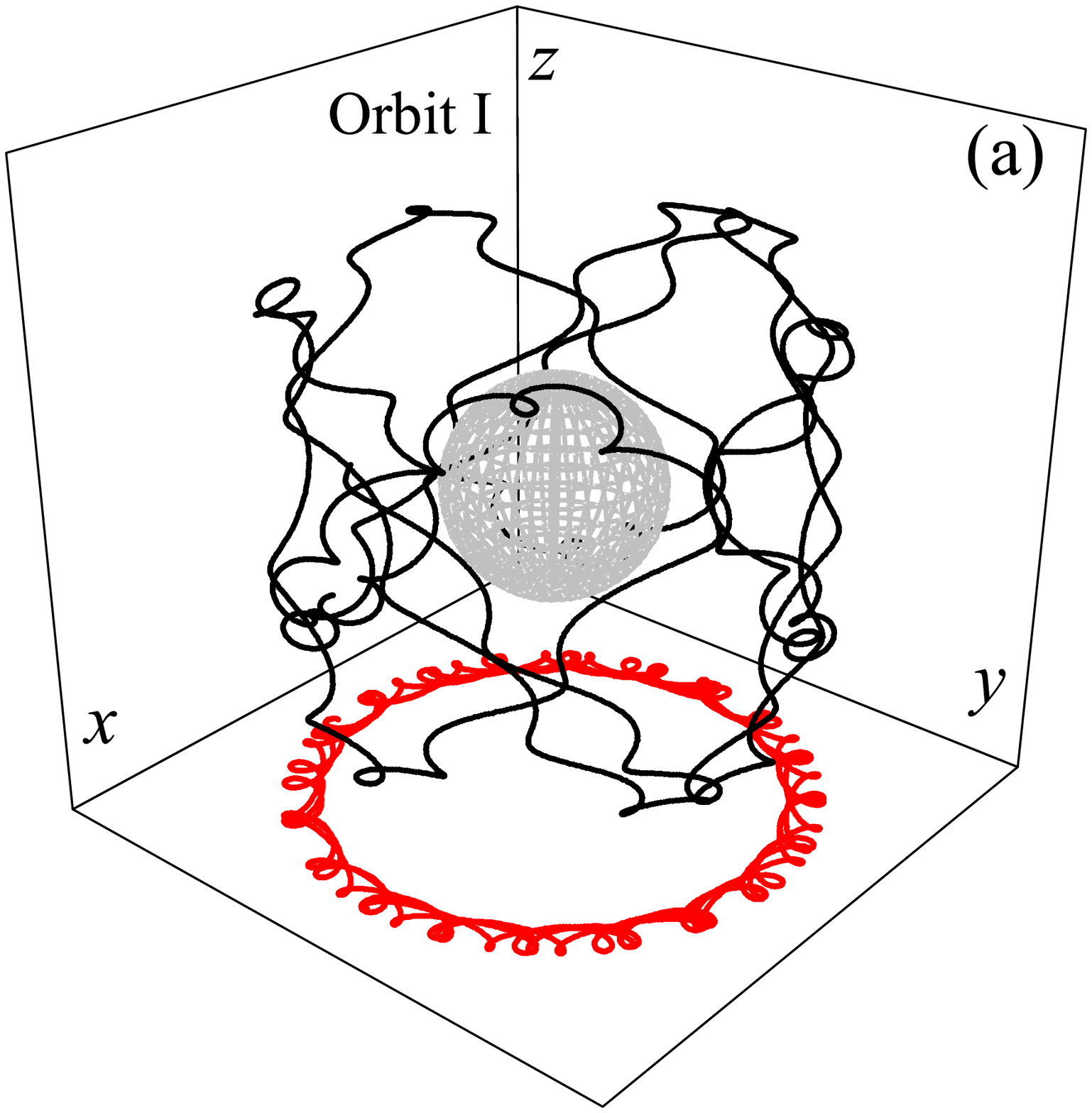}
\includegraphics[scale=0.3]{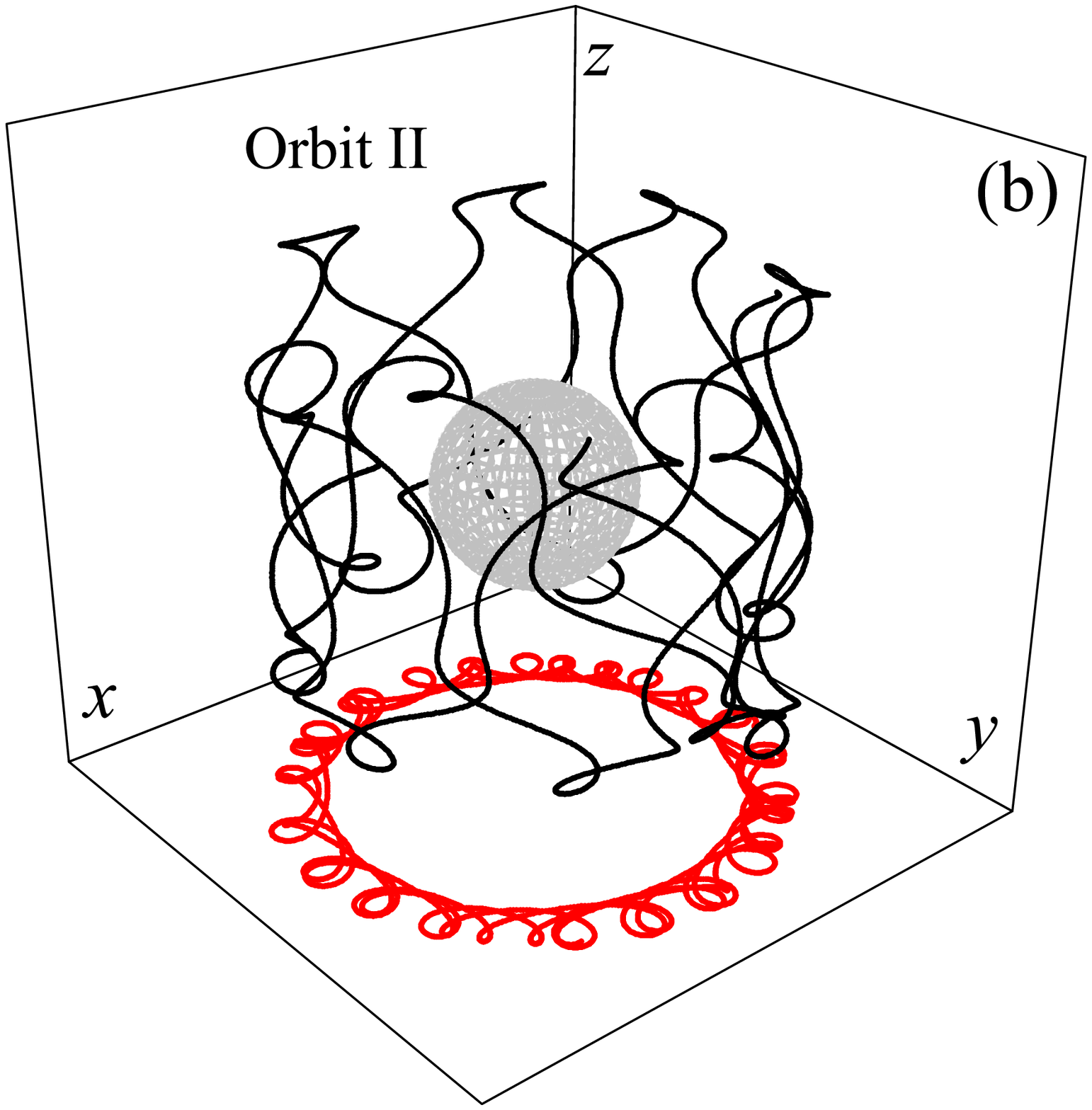}
\includegraphics[scale=0.3]{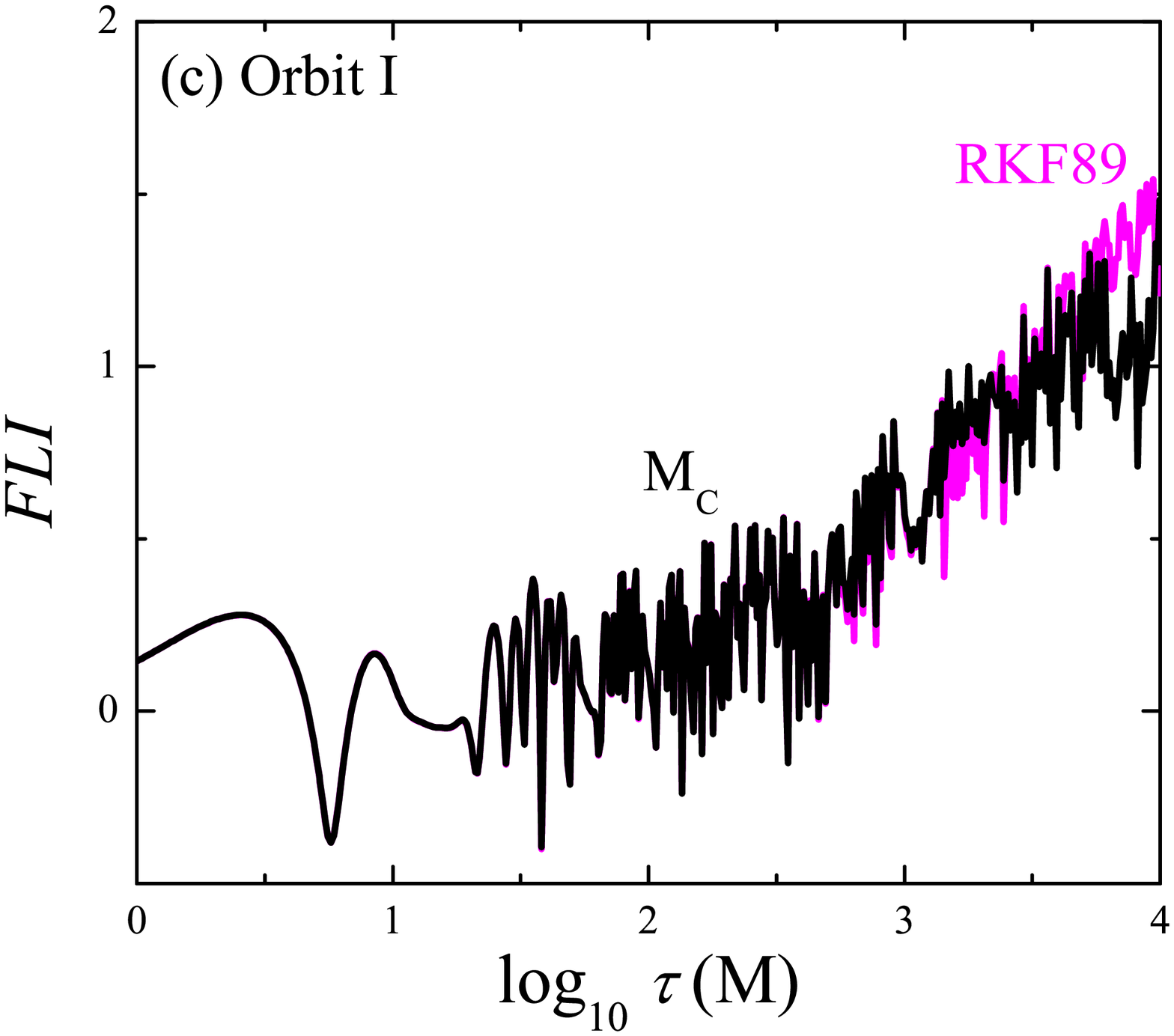}
\includegraphics[scale=0.3]{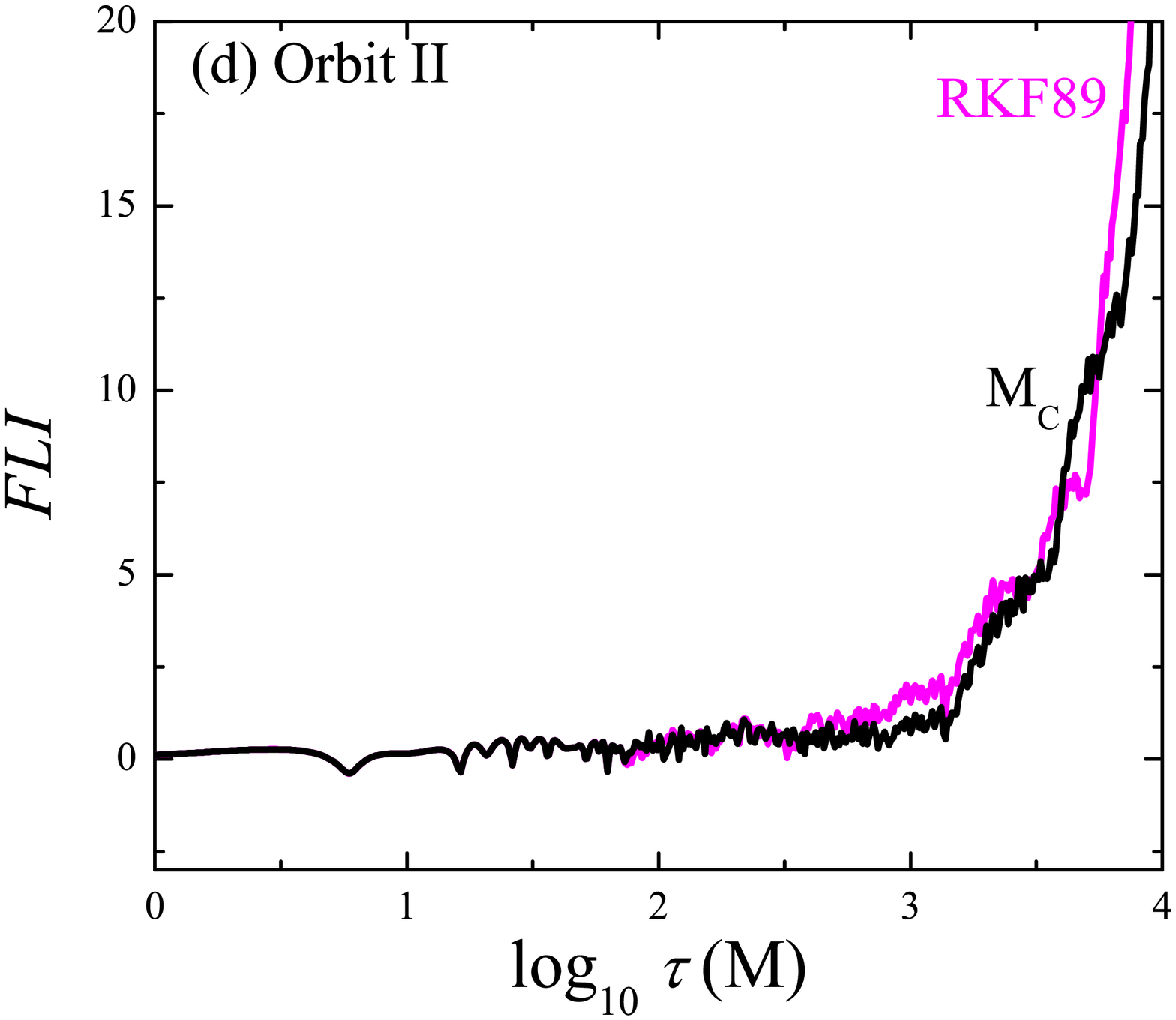}
\caption{(a) and (b): Two orbits in Figure 8 are plotted in the
three-dimensional space. The red curves represent
the projection of the trajectories on the $xoy$-plane. (c) and
(d): Growth of fast Lyapunov indicators (FLIs) with proper time
$\tau$ for Orbits I and II. Orbit I is ordered,
but Orbit II is chaotic. }} \label{fig10}
\end{figure*}

\begin{figure*}%[tbph]
\center{
\includegraphics[scale=0.21]{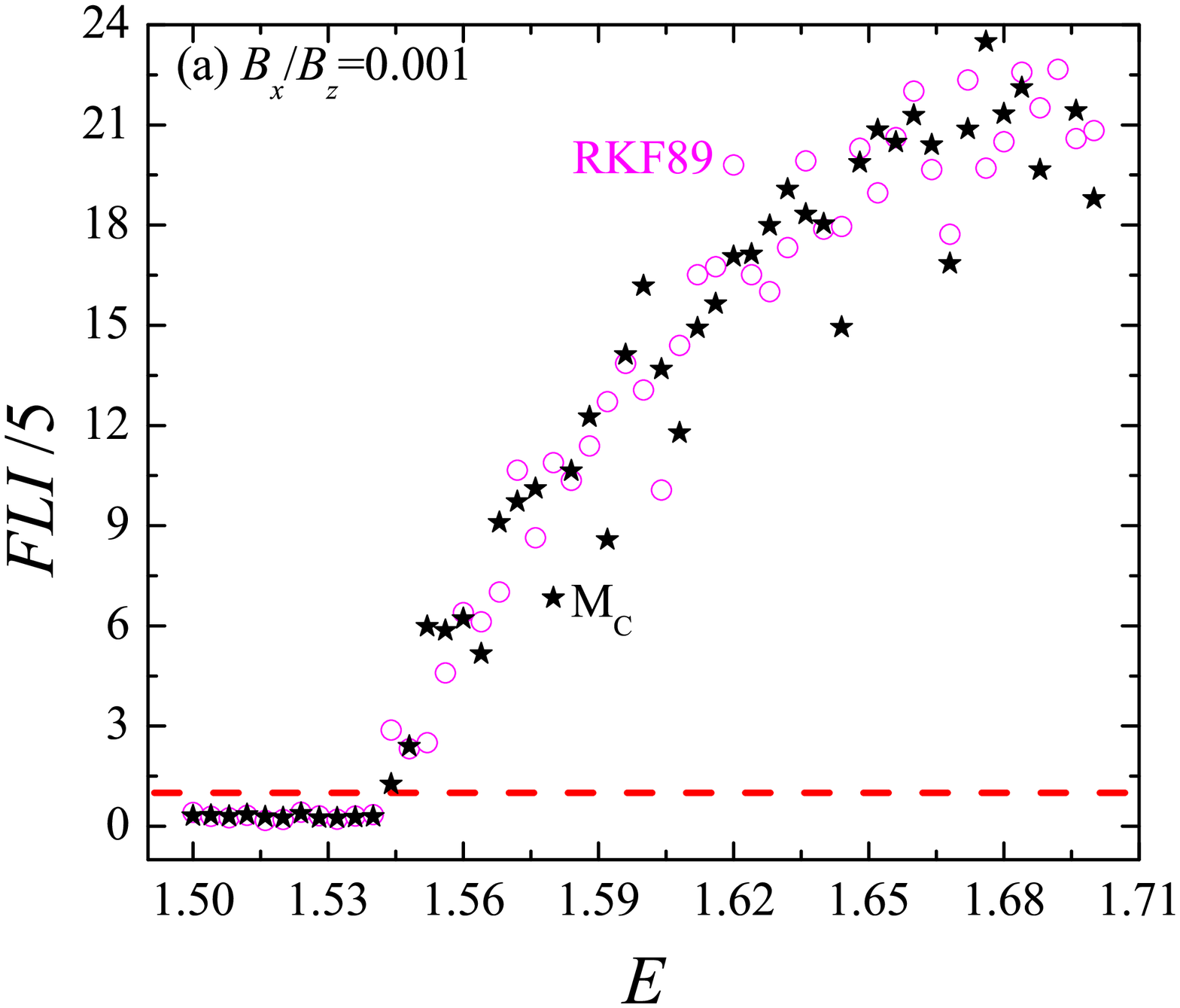}
\includegraphics[scale=0.21]{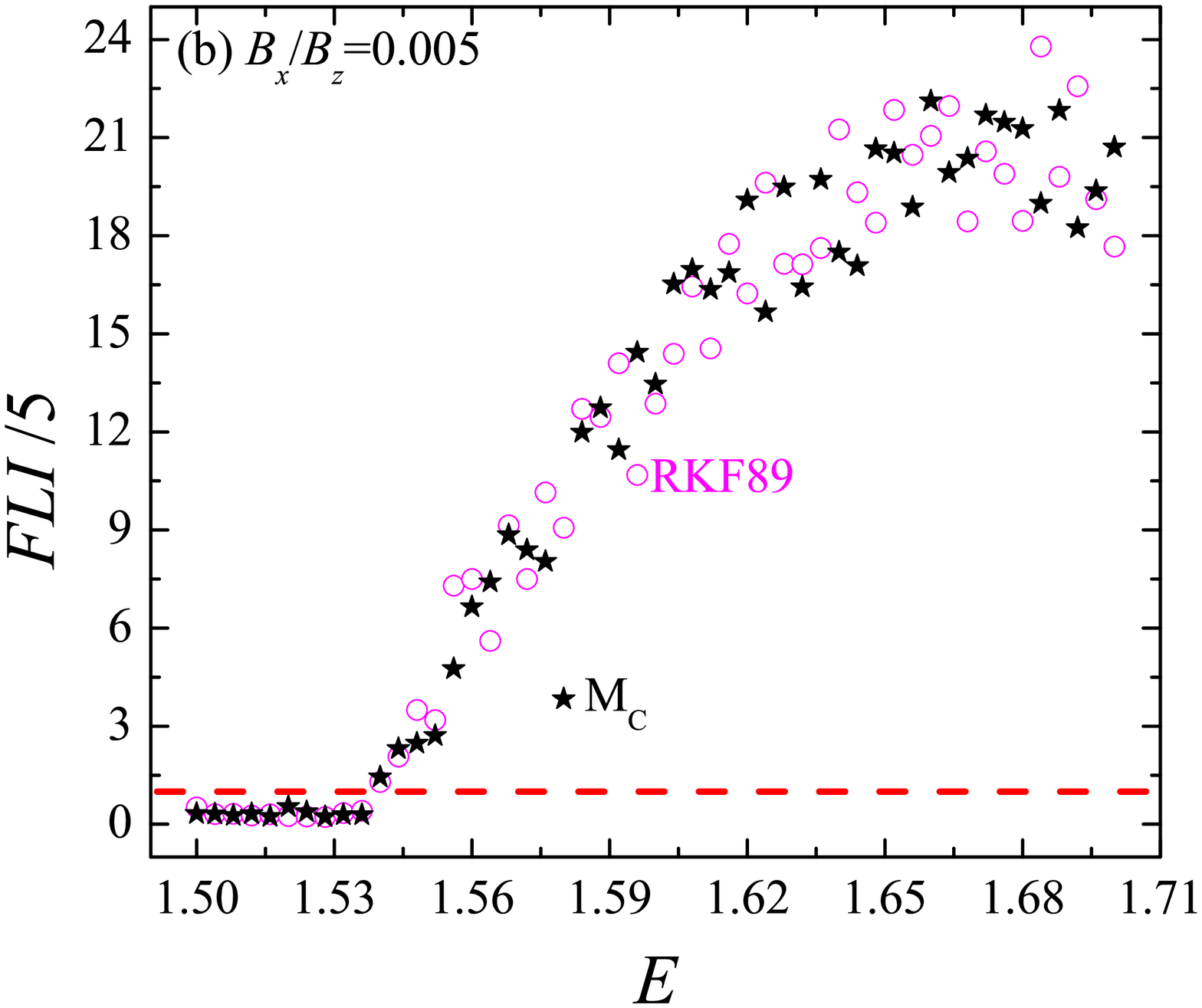}
\includegraphics[scale=0.21]{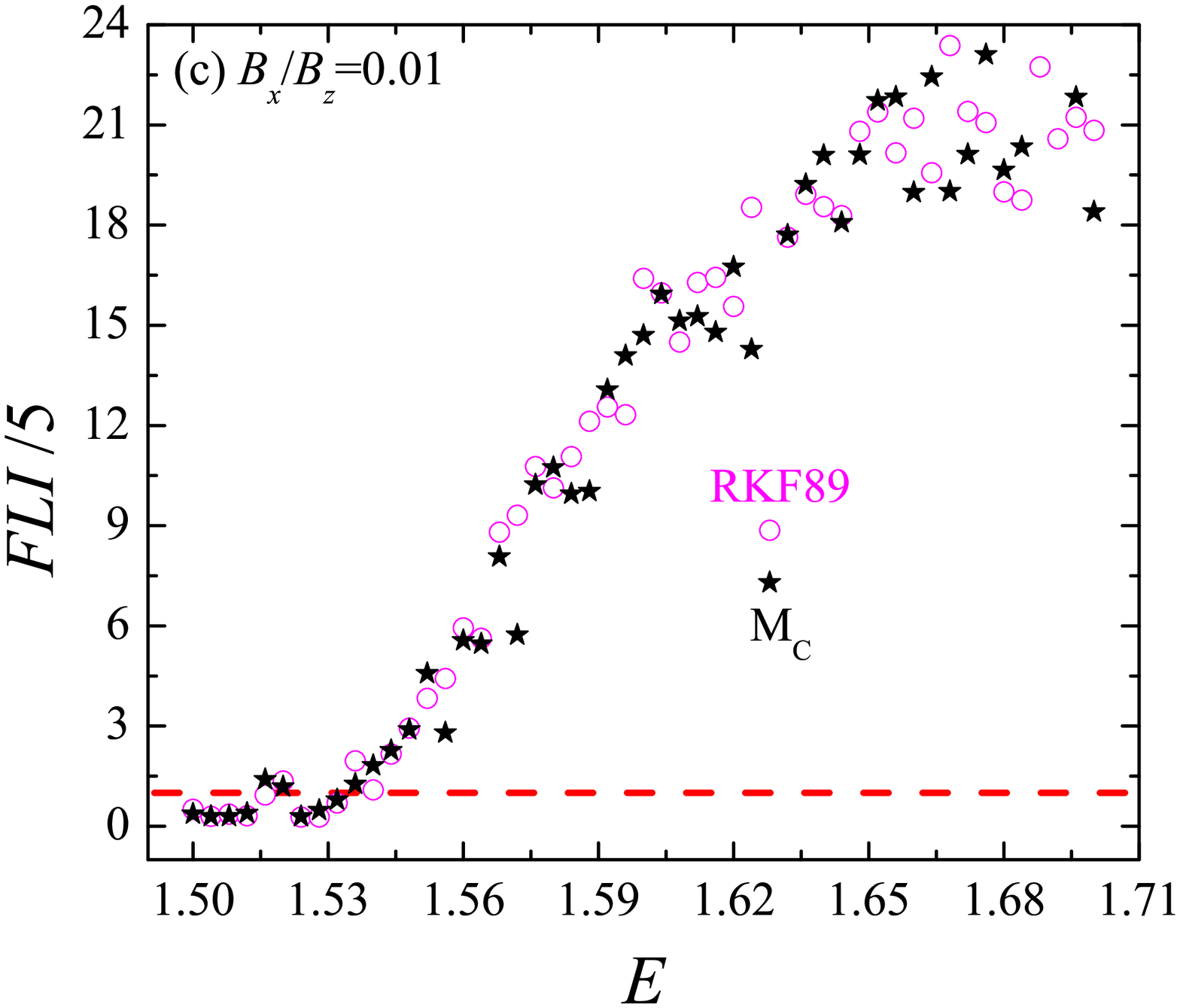}
\includegraphics[scale=0.21]{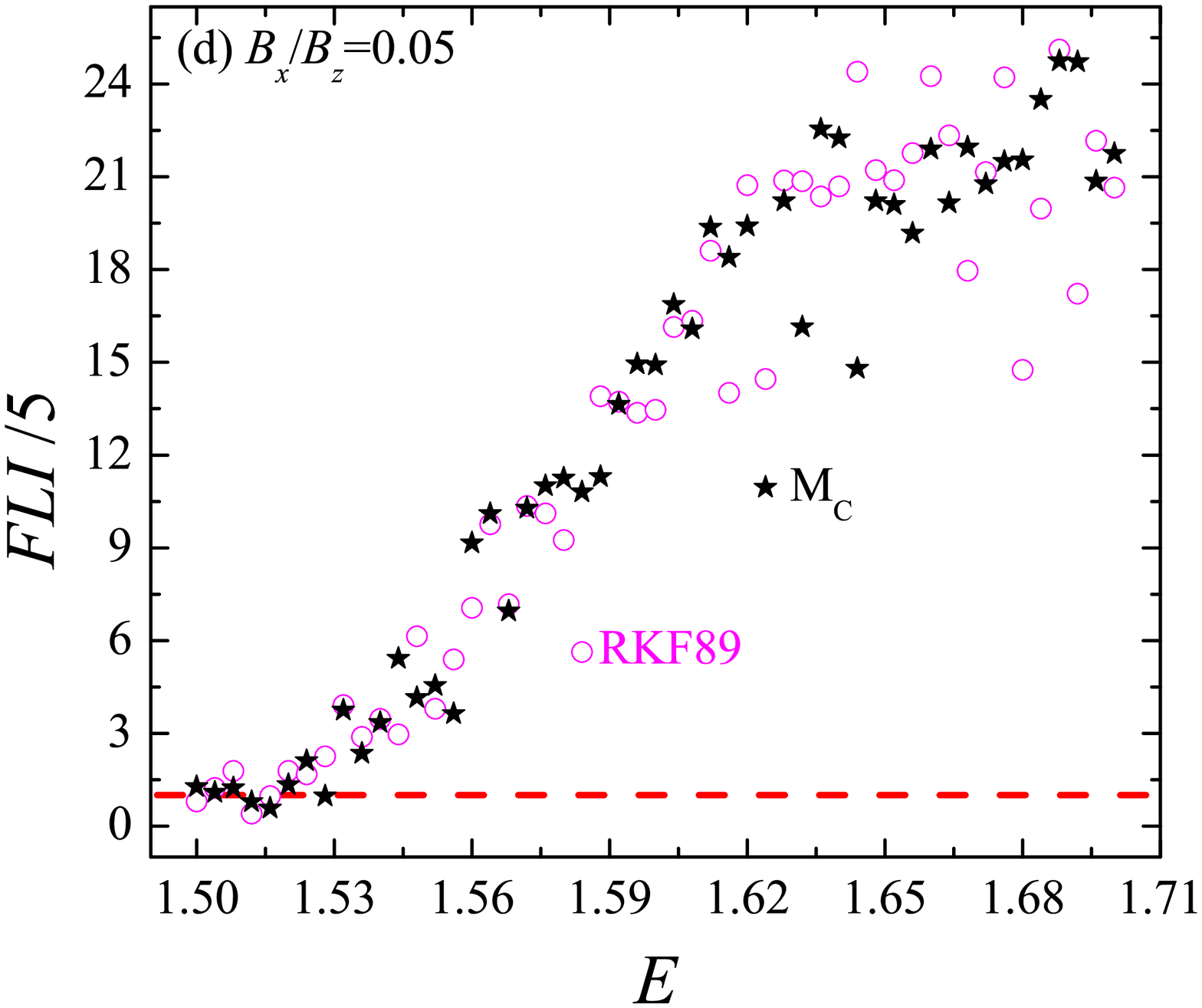}
\includegraphics[scale=0.21]{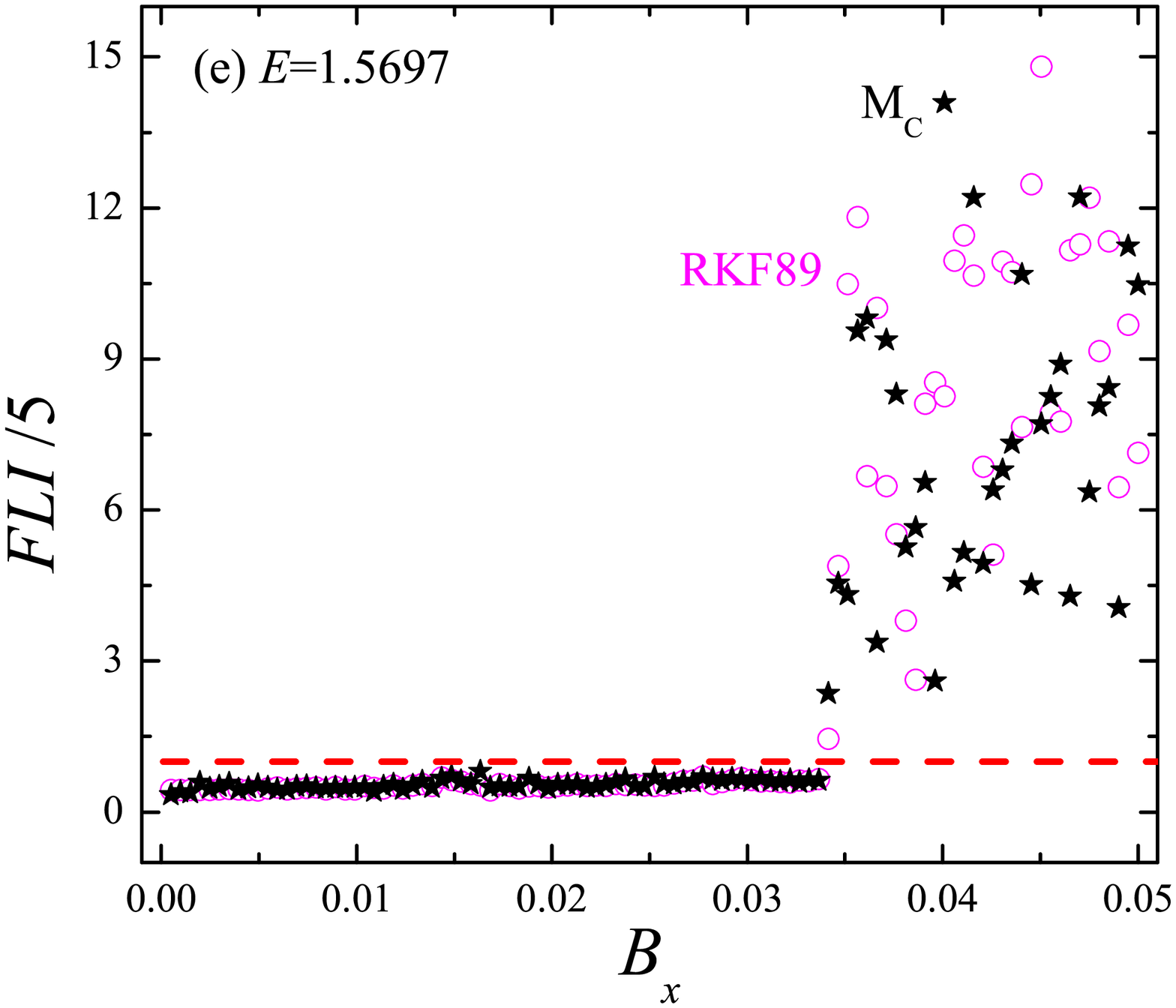}
\includegraphics[scale=0.21]{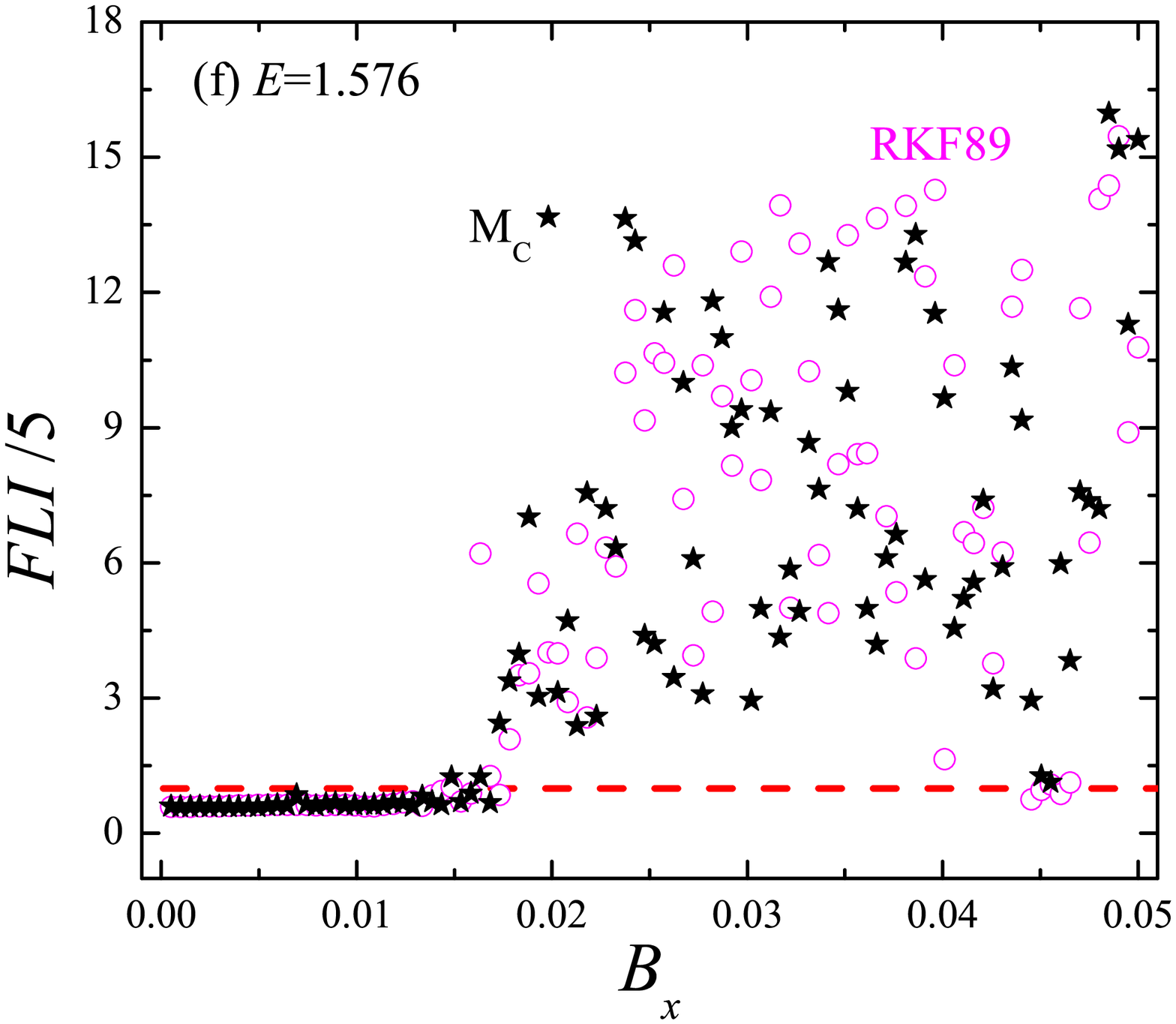}
\caption{(a)-(d): Dependence of FLI on energy
$E$. The initial conditions and other parameters are $a=0.75$,
$L=6$, $Q=1$, $B_z=1$, $r=4$, $\theta=1$, $\varphi=p_{r}=0$.  Each
of the FLIs is obtained after $5 \times 10^{5}$ integration steps.
The plotted FLIs are a fifth of the real FLIs. The plotted FLIs
larger than 1 indicate the chaoticity; the plotted FLIs less than
1 indicate the regularity. Chaos occurs for $E\geq 1.544$ in (a),
$E\geq 1.54$ in (b), $E\geq 1.536$ in (c), and $E\geq 1.52$ in
(d). This indicates that a smaller energy easily induces chaos for
a larger value of $B_x$. In other words, the onset of chaos
becomes easier as $E$ and $B_x$ get larger. (e) and (f):
Dependence of FLIs on magnetic field parameter $B_x$. The
parameters and initial conditions different from those in (a)-(d)
are $a=0.8$, $L=5$, $r=3.66$, and $\theta=\pi/2$. Chaos occurs for
$B_x \geq 0.03416$ in (e), and $B_x\geq 0.01486$ in (f). This
means that an increase of $B_x$ easily leads to the onset of
chaos. }} \label{fig11}
\end{figure*}

\begin{figure*}%[tbph]
\center{
\includegraphics[scale=0.21]{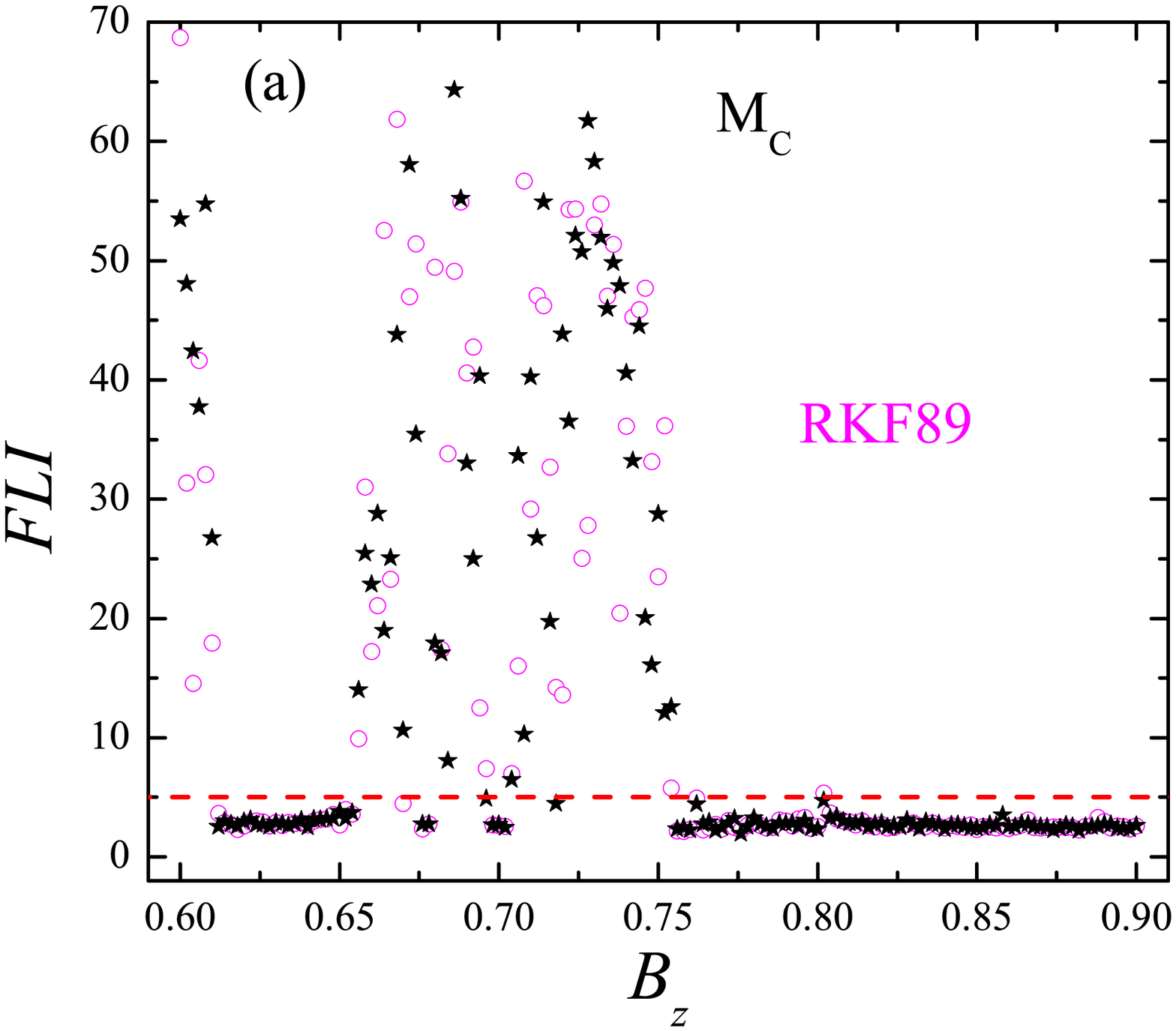}
\includegraphics[scale=0.21]{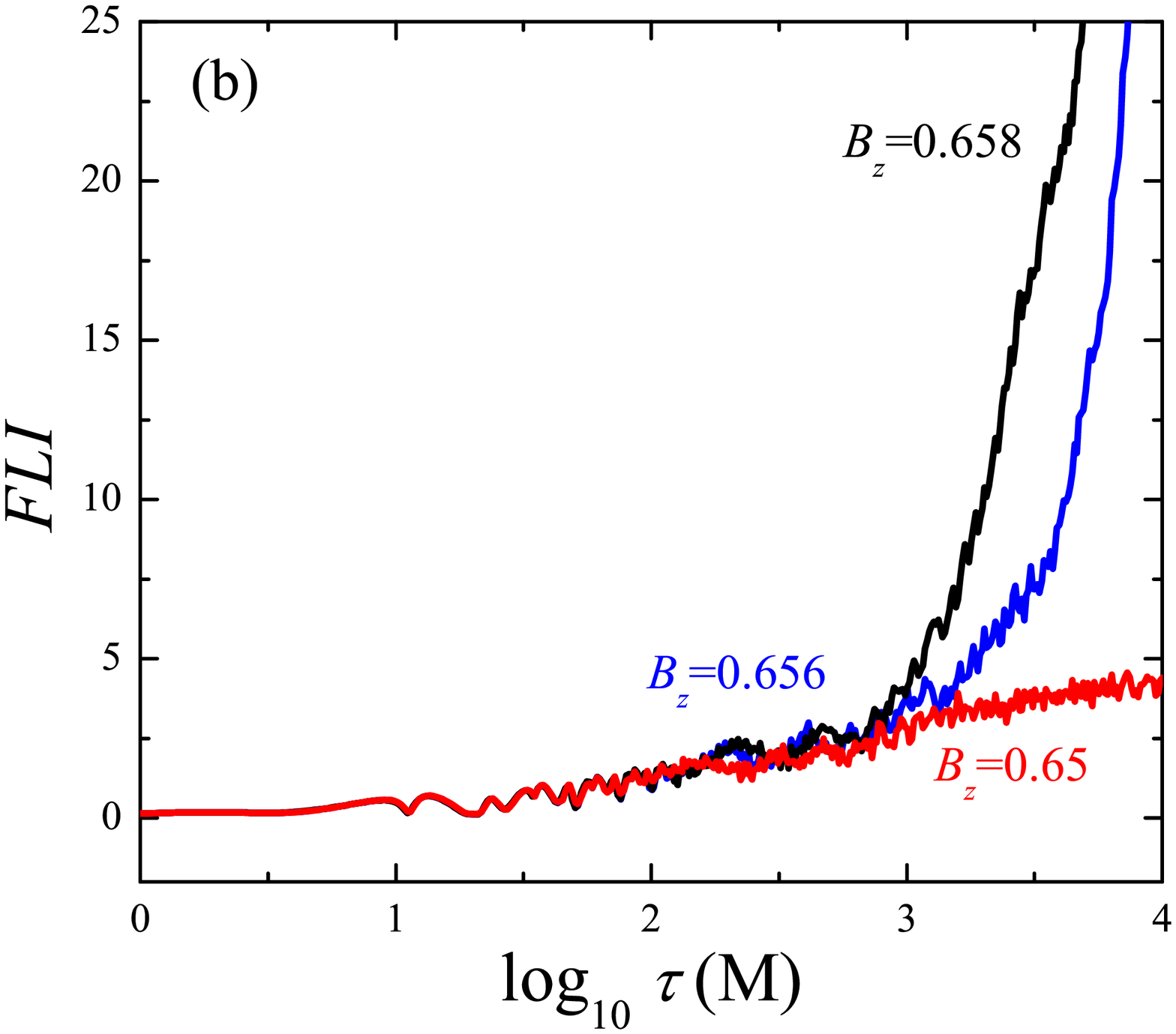}
\includegraphics[scale=0.21]{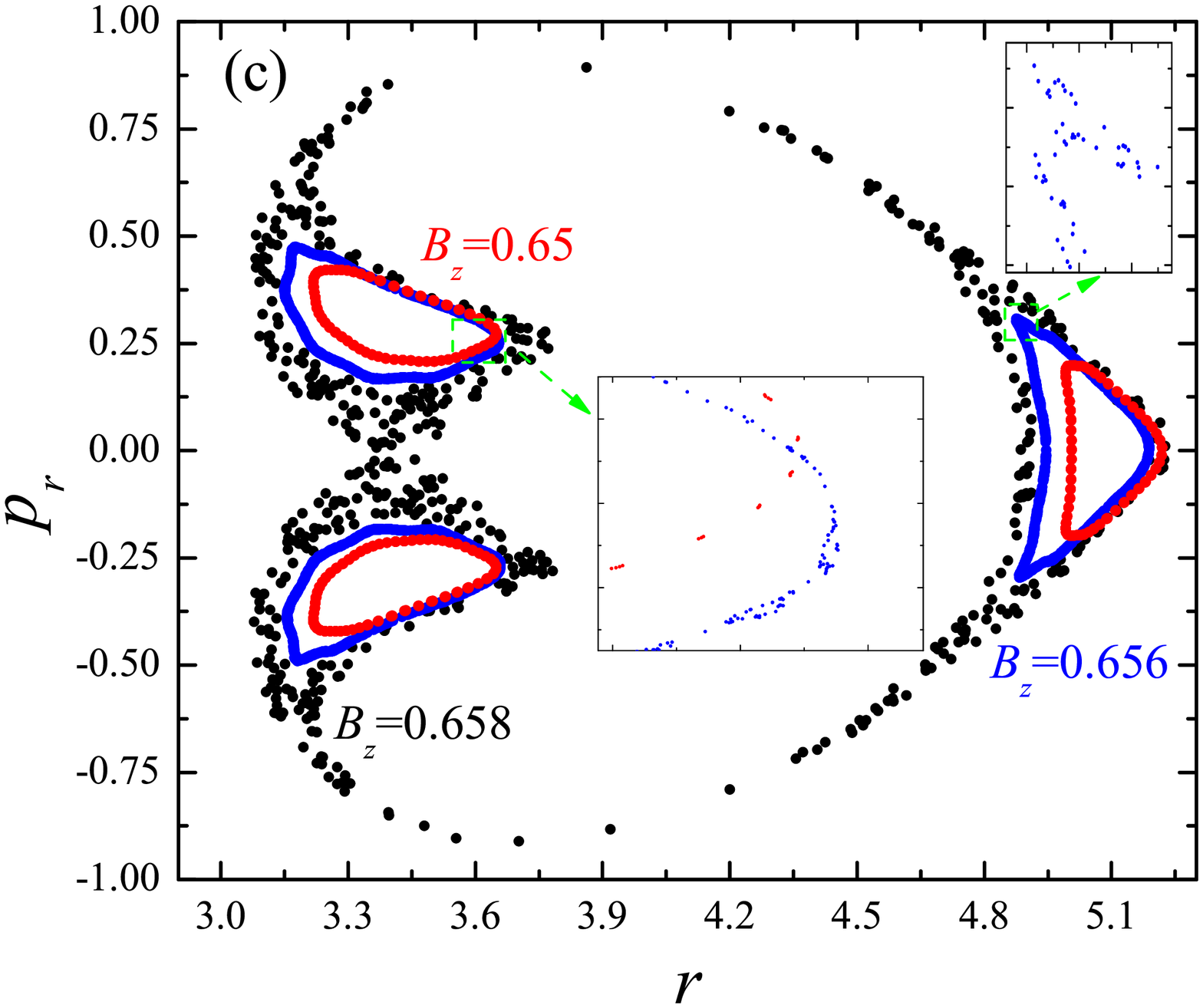}
\caption{(a): Dependence of FLIs on magnetic
field parameter $B_z$. The other parameters and initial conditions
are $B_{x}=0$, $a=0.8$, $E=1.48$, $L=5$, $Q=1$, $r=3.5$,
$\theta=1$, and $\varphi=0$. Each of the FLIs is obtained after $5
\times 10^{5}$ integration steps. The red dashed line is the
boundary between the FLIs of regularity and chaoticity. The values
of $B_z$ are [0.612, 0.654] and [0.756, 0.9] for the regular case, and [0.6, 0.61], [0.656, 0.659] for the chaotic case.
(b): Growth of FLIs with proper time $\tau$ for three values of
$B_{z}$: 0.65, 0.656 and 0.658. The FLIs increase when $B_{z}$
runs from 0.65 to 0.658. (c): Poincar\'{e}-sections/maps at plane
$\theta=\pi/2$ with $p_{\theta}>0$ for the three values of
$B_{z}$. The three values of $B_{z}$ 0.65, 0.656 and 0.658
correspond to regularity, weak chaoticity and strong chaoticity,
respectively. For the case of $B_{x}=0$, the angular momentum $L$
is a constant. Therefore, the motions are restricted to a
4-dimensional phase space $r$, $\theta$, $p_r$ and $p_{\theta}$. }} \label{fig12}
\end{figure*}

\begin{figure*}%[tbph]
\center{
\includegraphics[scale=0.21]{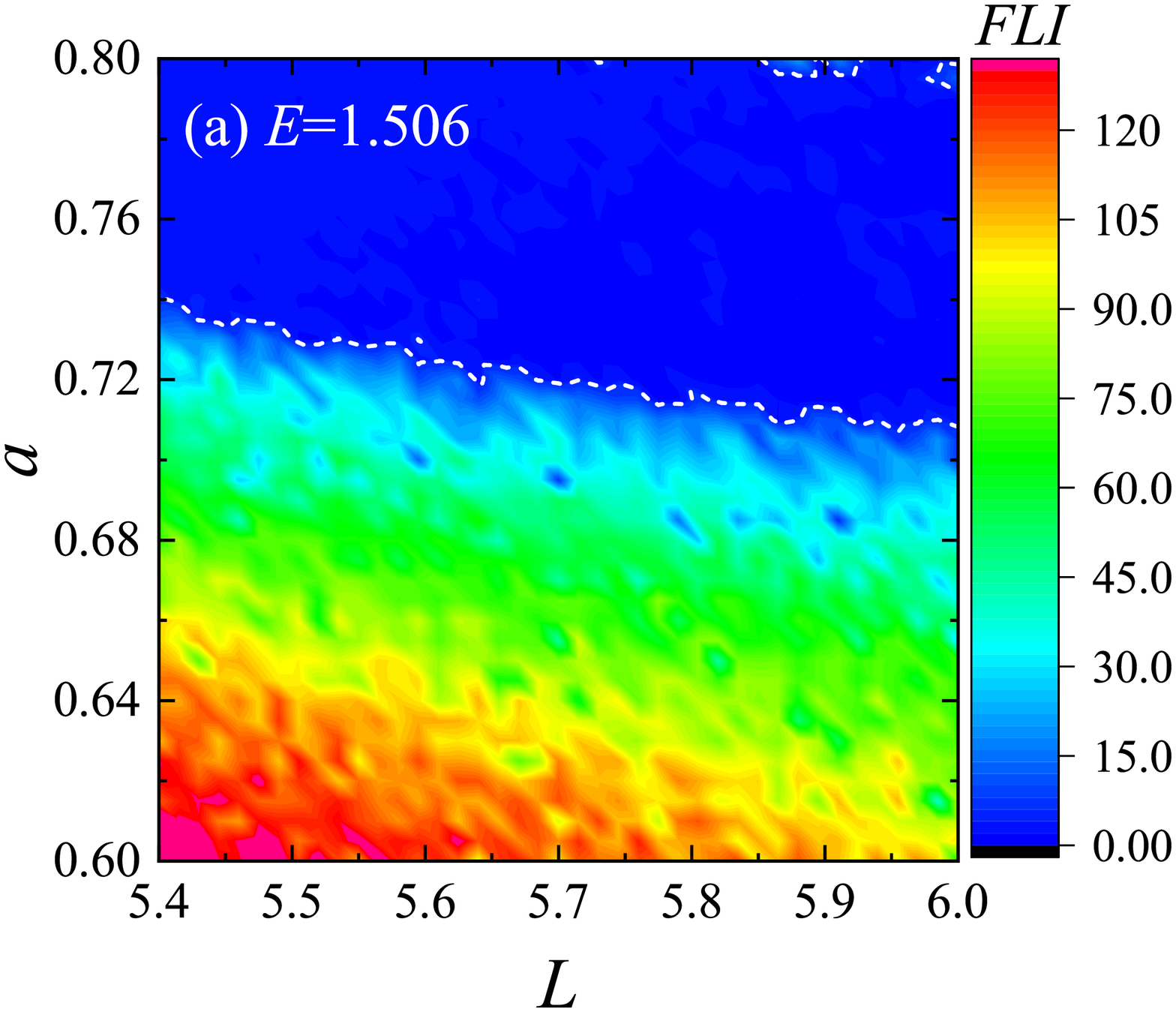}
\includegraphics[scale=0.21]{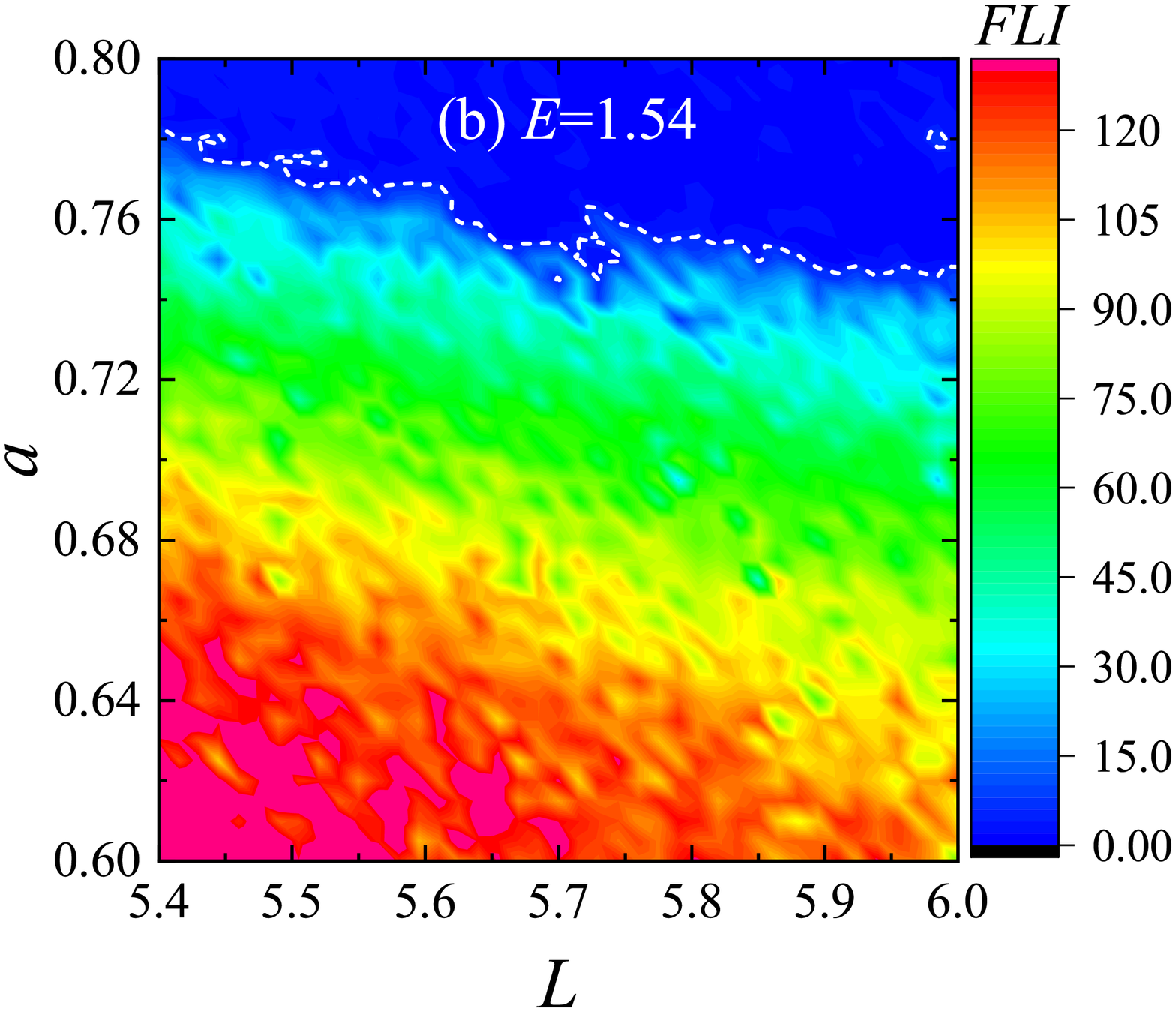}
\includegraphics[scale=0.21]{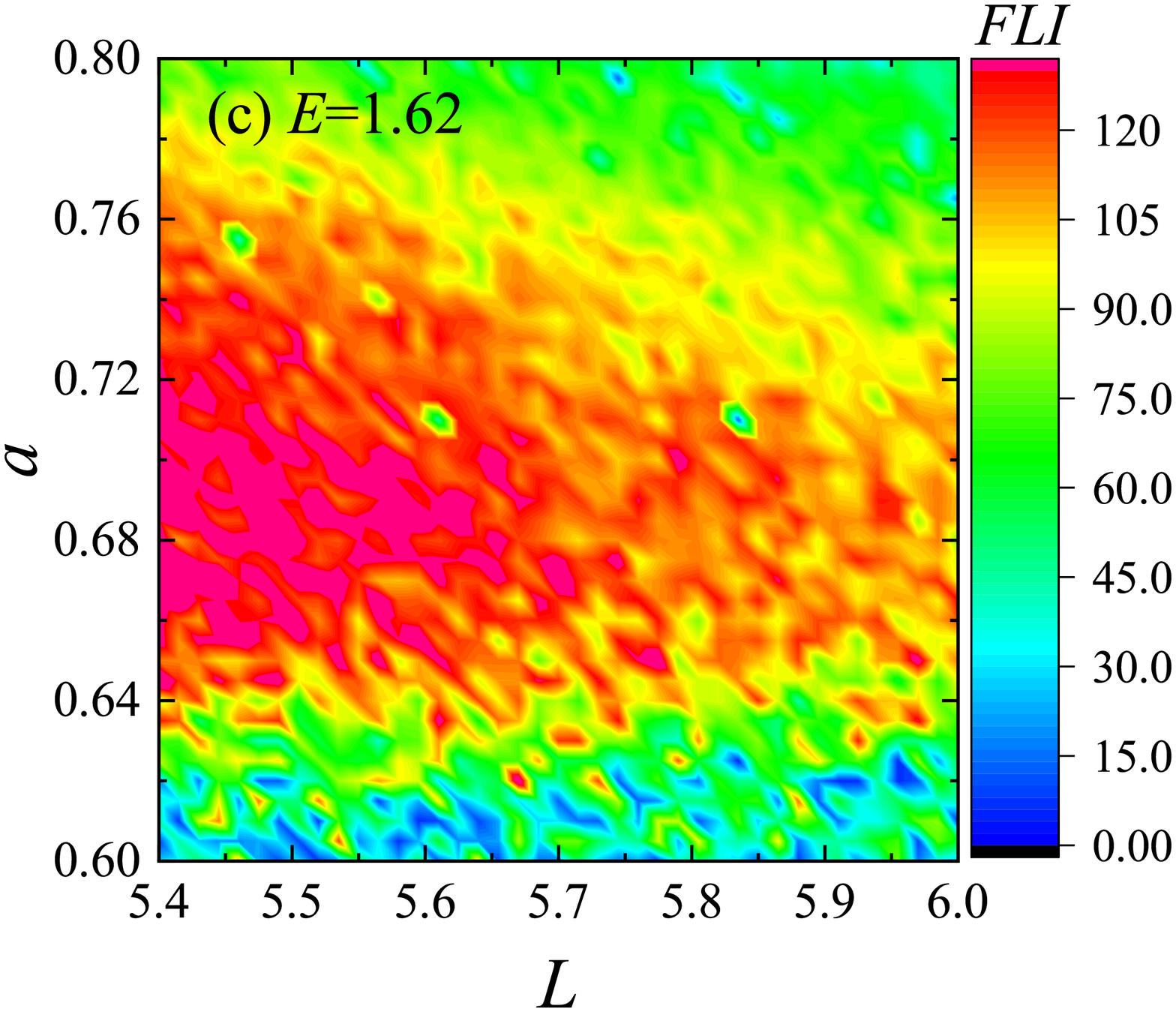}
\caption{Finding chaos by using the FLIs to scan a two-dimensional
space of parameter $a$ and the initial angular momentum $L$.
$B_z=1$ and $B_x=0.001$. The white dashed line
corresponds to $FLI=5$, which is the boundary between the chaotic
and ordered regions. Strong chaos exists for smaller values of $a$
and $L$ in (a) and (b), but it does for $a\in [0.64, 0.77]$ in
(c). }} \label{fig13}
\end{figure*}


\begin{thebibliography}{99}

\bibitem[Avdyushev (2003)]{r11} Avdyushev, V, A. 2003, CeMDA, 383, 409
\bibitem[Bacchini et al. (2018)]{r30} Bacchini, F., Ripperda, B., Chen, A, Y., $\&$ Sironi, L. 2018, ApJS, 237, 6 (arXiv 1801. 02378 [gr-pc])
\bibitem[Bacchini et al. (2019)]{r29} Bacchini, F., Ripperda, B., Porth, O., $\&$ Sironi, L. 2019, ApJS, 240, 40 (arXiv 1810. 00842 [astro-ph.HE])
\bibitem[Bi$\check{\textrm{c}}$\'{a}k $\&$ Jani$\check{\textrm{s}}$ (1985)]{r44} Bi$\check{\textrm{c}}$\'{a}k, J., $\&$ Jani$\check{\textrm{s}}$, V. 1985, MNRAS, 212, 899
\bibitem[Bronzwaer et al. (2018)]{r5} Bronzwaer, T., Davelaar, J.,
Younsi, Z., Mo\'{s}cibrodzka, M., Falcke, H., Kramer, M., $\&$
Rezzolla, L. 2018, A$\&$A, 613, A2
\bibitem[Bronzwaer et al. (2020)]{r6} Bronzwaer, T., Younsi, Z.,
Davelaar, J., $\&$ Falcke, H. 2020, A$\&$A, 641, A126
\bibitem[Brown (2006)]{r13} Brown, J, D. 2006, PhRvD, 73, 024001
\bibitem[Carlberg $\&$ Innanen (1987)]{r1} Carlberg, R, G., $\&$ Innanen, K, A. 1987, AJ, 94, 666
\bibitem[Caranicolas (1984)]{r2} Caranicolas, N, D. 1984, Celestial. Mech., 33, 209
\bibitem[Caranicolas (1993)]{r3} Caranicolas, N, D. 1993, A$\&$A, 267, 368
\bibitem[Chorin et al. (1978)]{r25} Chorin, A. J., Huges, T. J. R., McCracken, M. F., $\&$ Marsden, J. E. 1978, Comm. Pure and Appl. Math., 31, 205
\bibitem[Christian $\&$ Chan (2021)]{r24} Christian, P., $\&$ Chan, C. 2021, ApJ, 909, 67
\bibitem[Deng et al. (2020)]{r24b} Deng, C., Wu, X., $\&$ Liang, E. 2020, MNRAS, 496, 2946
\bibitem[Feng $\&$ Qin (2009)]{r26} Feng, K., $\&$ Qin, M. Z. 2009,
\emph{Symplectic Geometric Algorithms for Hamiltonian Systems}
(Zhejiang Science and Technology Publishing House, Hangzhou China,
Springer, New York)
\bibitem[Fukushima (2003)]{r12a} Fukushima, T. 2003, AJ, 126, 1097
\bibitem[Hairer (2006)]{r12} Hairer, E., Lubich, C., $\&$ Wanner, G. 2006, Geometric Numerical Integration:
Structure-Preserving Algorithms for Ordinary Differential
Equations (2nd ed.; Berlin: Springer)
\bibitem[Hu et al. (2019)]{r31} Hu, S, Y., Wu, X., Huang, G, Q., $\&$ Liang, E, W. 2019, ApJ, 887, 191 (arXiv 1910. 10353 [gr-pc])
\bibitem[Hu et al. (2021)]{r32} Hu, S, Y., Wu, X., $\&$ Liang, E, W. 2021, ApJS, 253, 55 (arXiv 2102. 08000 [gr-pc])
\bibitem[Huang et al. (2016)]{r35} Huang, L., Wu, X., $\&$ Ma, D, Z., 2016, EPJC, 76, 488
\bibitem[Itoh $\&$ Abe (1988)]{r28} Itoh, T., $\&$ Abe, K. 1988, JCoPh, 76, 85
\bibitem[Kop\'{a}\v{c}ek et al. (2010)]{r16} Kop\'{a}\v{c}ek, O., Karas, V., Kov\'{a}\v{r}, J., $\&$ Stuchl\'{\i}k, Z. 2010, ApJ, 722, 1240
\bibitem[Kop\'{a}\v{c}ek $\&$ Karas (2014)]{r41} Kop\'{a}\v{c}ek, O., $\&$ Karas, V. 2014, ApJ, 787, 117
\bibitem[Li $\&$ Wu (2017)]{r23} Li, D., $\&$ Wu, X. 2017, MNRAS, 469, 3031
\bibitem[Liu et al. (2016)]{r21} Liu, L., Wu, X., Huang, G., $\&$ Liu, F. 2016, MNRAS, 459, 1968
\bibitem[Lubich et al. (2010)]{r17} Lubich, C., Walther, B., $\&$ Br\"{u}gmann, B. 2010, PhRvD, 81, 104025
\bibitem[Luo et al. (2017)]{r22} Luo, J, J., Wu, X., Huang, G, Q., $\&$ Liu, F. 2017, ApJ, 834, 64
\bibitem[Mei et al. (2013a)]{r19} Mei, L, J., Wu, X., $\&$ Liu, F, Y. 2013a, EPJC, 73, 2413
\bibitem[Mei et al. (2013b)]{r20} Mei, L, J., Ju, M, J., Wu, X., $\&$ Liu, S. 2013b, MNRAS, 435, 2246
\bibitem[Misner et al. (1973)]{r42} Misner, C., Thorne, K., $\&$ Wheeler, J. 1973, Gravitation (San Francisco, CA: Freeman)
\bibitem[Nacozy (1971)]{r42b} Nacozy, P. E. 1971, Ap$\&$SS, 14, 40
\bibitem[Pan et al. (2021)]{r42c} Pan, G., Wu, X., $\&$ Liang, E. 2021, PhRvD, accepted
\bibitem[Preto $\&$ Saha (2009)]{r15} Preto, M., $\&$ Saha, P. 2009, ApJ, 703, 1743
\bibitem[Qin (1987)]{r27} Qin, M, Z. 1987, JCM, 5, 203
\bibitem[Seyrich $\&$ Lukes-Gerakopoulos (2012)]{r14} Seyrich, J., $\&$ Lukes-Gerakopoulos, G. 2012, PhRvD, 86, 124013
\bibitem[Sun et al. (2021)]{r46} Sun, W., Wang, Y., Liu, F, Y., $\&$ Wu, X. 2021, submitted to
EPJC
\bibitem[Takahashi $\&$ Koyama (2009)]{r45} Takahashi, M., $\&$ Koyama, H. 2009, ApJ, 693, 472
\bibitem[Wald (1974)]{r43} Wald, R. 1974, PhRvD, 10, 1680
\bibitem[Wang et al. (2018)]{r7a}  Wang, S. C., Huang, G. Q., $\&$ Wu, X. 2018, AJ, 155, 67
\bibitem[Wang et al. (2021a)]{r7} Wang, Y., Sun, W., Liu, F, Y., $\&$ Wu, X. 2021a, ApJ, 907,
66 (Paper I)
\bibitem[Wang et al. (2021b)]{r8} Wang, Y., Sun, W., Liu, F, Y., $\&$ Wu, X. 2021b, ApJ, 909,
22 (Paper II)
\bibitem[Wang et al. (2021c)]{r9} Wang, Y., Sun, W., Liu, F, Y., $\&$ Wu, X. 2021c, ApJS, 254,
8 (Paper III)
\bibitem[Wang et al. (2016)]{r9b} Wang, S. C., Wu, X., $\&$ Liu, F. Y. 2016, MNRAS, 463, 1352
\bibitem[Wisdom $\&$ Holman (1991)]{25} Wisdom, J., $\&$ Holman, M. 1991, AJ, 102, 1528
\bibitem[Wu et al. (2007)]{r40a} Wu, X., Huang, T. Y., Wan, X. S.,  $\&$ Zhang, H. 2007, AJ, 133, 2643
\bibitem[Wu et al. (2006)]{r40} Wu, X., Huang, T., $\&$ Zhang, H. 2006, PhRvD, 74, 083001
\bibitem[Wu $\&$ Xie (2010)]{r33} Wu, X., $\&$ Xie, Y. 2010, PhRvD, 81, 084045
\bibitem[Wu et al. (2015)]{r34} Wu, X., Mei, L., Huang, G., $\&$ Liu, S. 2015, PhRvD, 91, 024042
\bibitem[Wu et al. (2021)]{r10} Wu, X., Wang, Y., Sun, W., $\&$ Liu, F, Y. 2021, ApJ, 914,
63 (Paper IV)
\bibitem[Zhong et al. (2010)]{r18} Zhong, S., Wu, X., Liu, S., $\&$ Deng, X. 2010, PhRvD, 82, 124040
\bibitem[Zotos (2011)]{r4} Zotos, E. 2011, NewA, 16, 391
\bibitem[Zotos (2012a)]{r36} Zotos, E. 2012a, NewA., 17, 576
\bibitem[Zotos (2012b)]{r37} Zotos, E. 2012b, ApJ, 750, 56
\bibitem[Zotos (2013)]{r38} Zotos, E. 2013, PASA, 30, 12
\bibitem[Zotos (2014)]{r39} Zotos, E. 2014, Astron. Nachr., 335, 886

\end{thebibliography}
\end{document}